\documentclass[11pt]{article}
\usepackage{mathrsfs}
\usepackage{amsfonts}
\usepackage{amsfonts}
\usepackage{mathrsfs}
\usepackage{amsmath}
\usepackage{amssymb}
\usepackage{hyperref,color}

\numberwithin{equation}{section}

\author{\bf Yu Hou$^1,$ ~\bf Peng Zhao$^1,$  ~\bf Engui Fan$^1$\footnote{Corresponding
author and  e-mail address:
      faneg@fudan.edu.cn} ~and~ Zhijun Qiao$^2$  }
\date{   \small{$^1$ School of Mathematical Science, Fudan University, Shanghai
200433, P.R. China\\
$^2$ Department of Mathematics, University of Texas-Pan American,
Edinburg, TX 78539, U.S.A}}
\title{\bf The algebro-geometric solutions for  Degasperis-Procesi hierarchy }
\begin{document}
\maketitle
\begin{abstract}

Though completely integrable Camassa-Holm (CH) equation and Degasperis-Procesi (DP) equation  are cast in the same  peakon family,
they possess the  second- and third-order  Lax  operators, respectively.  From the viewpoint of algebro-geometrical study,
this difference lies in hyper-elliptic and  non-hyper-elliptic curves. The non-hyper-elliptic curves lead to great difficulty
in the construction of algebro-geometric solutions of the DP equation.
In this paper,  we derive the
DP hierarchy with the help of Lenard recursion operators. Based on the characteristic
polynomial of a Lax matrix for the DP hierarchy,  we introduce a third order
algebraic curve $\mathcal{K}_{r-2}$ with
genus $r-2$, from
which  the associated Baker-Akhiezer functions, meromorphic function and Dubrovin-type equations are  established.  Furthermore,  the theory
of algebraic curve is applied to  derive  explicit representations of the theta function
 for the Baker-Akhiezer functions and the meromorphic
function.  In particular,   the algebro-geometric solutions are
obtained for all  equations in  the whole  DP hierarchy.

\end{abstract}

 \section{ Introduction}
  The Degasperis-Procesi (DP) equation
    \begin{equation}\label{1.1}
      u_t-u_{txx} + 4uu_x-3u_xu_{xx}-uu_{xxx}=0,
    \end{equation}
was first discovered in a search
for asymptotically integrable PDEs \cite{2}. It arose as a model
equation in the study of  the two-dimensional water waves
propagating in an irrotational flow over a flat bed \cite{18},
\cite{28}, \cite{55}. Given the intricate structure of the full
governing equations for water waves, it is natural to seek simpler
approximate model equations in various physical regimes. The DP
equation may be derived in the moderate amplitude regime:
introducing the wave-amplitude parameter $\varepsilon$  and the
long-wave parameter $\delta$. In this regime we assume that $\delta
\ll 1$ and $\varepsilon \sim \delta$. This regime is more
appropriate for the study of nonlinear waves than dispersive waves,
the stronger nonlinearity of which could allow for the occurrence of
wave-breaking. The other regime studied most is the shallow water
system for which $\delta \ll 1 $ and $\varepsilon \sim \delta^2$. In
the parameter $\delta$ range, due to a balance between nonlinearity
and dispersion, various integrable systems like the Korteweg-de
Vries (KdV) equation arose as approximations to the governing
equations.
 However, among the models of moderate amplitude regime, only the CH equation and the DP
  equation are integrable  in the peakon family \cite{3} in the sense that they  admit a bi-Hamiltonian structure and a Lax pair.
   Also, they are two integrable equations from a family,  corresponding to parameters $b=2$
 and $b=3$, respectively, of the following $b$-family
  of equations
    \begin{equation}
      u_t-u_{txx}+ (b+1)uu_x=bu_xu_{xx}+uu_{xxx},
    \end{equation}
 where $b$ is a constant.

Quasi-periodic solutions (also called algebro-geometric solutions or
finite gap solutions) of nonlinear equations were originally studied
on the KdV equation based on the inverse spectral theory and
algebro-geometric method developed by pioneers such as the authors
in Refs.\cite{1},\cite{4}-\cite{11} in the late 1970s.  This theory
has been extended to the whole hierarchies of nonlinear integrable
equations by Gesztesy and Holden using polynomial recursion method
\cite{12}-\cite{15.0}.
 As a degenerated case of algebro-geometric solution, the
   multi-soliton solution and elliptic function solution may be
   obtained \cite{4},\cite{8}, \cite{34}. It is well known that the algebro-geometric solutions of the CH
   hierarchy have been obtained with different techniques, see Gesztesy and Holden \cite{13},
   and Qiao \cite{39}.
   However,  within the authors' knowledge,
   the algebro-geometric solutions of the DP hierarchy are
   still not presented yet.

 Before turning to each section,
 it seems appropriate to
 review some related literature as usual.
 Over recent three decades soliton equations
 associated with $2 \times 2$ matrix spectral problems have widely been studied.
   Various  methods were developed  to construct   algebro-geometric
 solutions for  integrable equations   such as   KdV, mKdV,
  Kadomtsev-
Petviashvili equation, Schr$\mathrm{\ddot{o}}$dinger, CH equations, sine-Gordon,
 AKNS,  Ablowitz-Ladik lattice and Toda lattice etc
 \cite{4}-\cite{11}, \cite{12}-\cite{15.0},
 \cite{19}, \cite{20}, \cite{56}, \cite{57}.
 But it is  very
 difficult to extend these methods to soliton equations associated
 with $ 3 \times 3$ matrix spectral problems.  The main  reasons  for this
complexity  get traced back to
the associated  algebraic curve, which is the second order hyper-elliptic
 in the $2 \times 2$  matrix spectral problems while it is non-hyper-elliptic
of the third order one typically arising in the $ 3 \times 3$ case.

 In Refs.\cite{38}, Qiao proposed the DP hierarchy through the procedure of recursion
 operator and connected the DP hierarchy (including the DP equation as a special negative member)
 to finite-dimensional integrable systems and gave
 its parametric solution on a symplectic submanifold by
 using the C Neumann constraint under the nonlinearization technique.
 In Refs.\cite{31}, the $N$-soliton of the DP equation is obtained by
 Hirota's method.  In Refs.\cite{17},
 the inverse scattering method for the DP equation is studied based on a
 $ 3 \times 3$ matrix Riemann-Hilbert (RH) problem, where the
 solution of the DP equation is extracted from the large-$k$
 behavior of the solution of the RH problem.  In
 Refs.\cite{21}, \cite{22}, Dickson and Gesztesy proposed  an unified framework,
 which yields all algebro-geometric solutions of the entire
 Boussinesq (Bsq) hierarchy.  Geng et. al. further investigated  the
 algebro-geometric solutions of the modified Bsq hierarchy
in a recent paper \cite{23}.

The purpose  of this paper is to construct
 the algebro-geometric solutions for the DP  hierarchy
   which contains the DP equation (\ref{1.1})  as special  member.  The outline of the present paper is as follows.
 In section 2, based
 on the Lenard recursion operators and the stationary zero-curvature
 equation, we derive the DP hierarchy associated with a
 $ 3 \times 3$ matrix spectral problem. An algebraic curve
 $\mathcal{K}_{r-2}$ of arithmetic genus $r-2$ is introduced with
 the help of the characteristic polynomial of Lax matrix for the
 stationary DP hierarchy.

 In section 3, we study the meromorphic
 function $\phi$
 satisfying a second-order nonlinear differential equation. Moreover, the stationary DP
 equations are decomposed into a system of Dubrovin-type
 equations.

 In section 4, we present the explicit
 theta function representations for the Baker-Akhiezer function and the
 meromorphic function.
 In particular,  we give the  algebro-geometric solutions
 of the entire stationary DP hierarchy.

 In sections 5 and 6, we extend all the Baker-Akhiezer function, the meromorphic function,
 the Dubrovin-type equations, and the theta function representations dealt with
 in sections 3  and  4  to   the time-dependent cases.
 Each equation in the time-dependent  DP  hierarchy is permitted to evolve in terms of an independent time
 parameter $t_p$. We use a stationary solution of
 the $n$th equation of the DP hierarchy as an initial data  to  construct a
 time-dependent solution of  the $p$th equation of the DP hierarchy.

\section{The DP hierarchy}
    In this section, we  derive the DP hierarchy and the corresponding sequence of
 zero-curvature pairs by using a Lenard
 recursion formalism (see Refs.\cite{38} for more details).   Throughout this section let us make 
 the following assumption.

\newtheorem{hyp1}{Hypothesis}[section]
\begin{hyp1}\label{hpysis1}
      In the stationary case we assume that $u:\mathbb{C}\rightarrow\mathbb{C}$ satisfies
  \begin{equation}
    u\in C^\infty( \mathbb{C}),~\partial_x^ku\in L^\infty
    ( \mathbb{C}),~k\in \mathbb{N}_0.
  \end{equation}
    In the time-dependent case we suppose  $u:\mathbb{C}^2\rightarrow\mathbb{C}$ satisfies
  \begin{equation}
    \begin{split}
    & u(\cdot,t)\in C^\infty ( \mathbb{C}),~\partial_x^ku(\cdot,t)\in L^\infty
    ( \mathbb{C}),~k\in \mathbb{N}_0, t\in  \mathbb{C},\\
    & u(x,\cdot),u_{xx}(x,\cdot) \in C^1( \mathbb{C}),~x\in  \mathbb{C}.\\
  \end{split}
  \end{equation}
\end{hyp1}

   We start by the following $3\times 3$ matrix isospectral problem
 \begin{equation}\label{1}
    \psi_x=U\psi,\quad \psi=\left(\begin{array}{c}
                              \psi_1 \\
                              \psi_2\\
                              \psi_3
                            \end{array}\right),\quad
    U=\left(
        \begin{array}{ccc}
          0 & 1 & 0\\
          0 & 0 & 1 \\
          -mz^{-1} & 1 & 0 \\
        \end{array}
      \right),
 \end{equation}
where $m=u-u_{xx}$, the function $u$ is a potential, and $z$
is a constant spectral parameter independent of variable $x$. Next,
we introduce two Lenard operators
  \begin{eqnarray}
      K&=&4\partial-5\partial^3+\partial^5,\\
      J&=&3(2m\partial+\partial
      m)(\partial-\partial^3)^{-1}(m\partial+2\partial m).
  \end{eqnarray}
Obviously, $K$ and $J$ are two skew-symmetric operators. A direct
calculation shows that
 \begin{eqnarray*}
    K^{-1}&=&(\partial-\partial^3)^{-1}(4-\partial^2)^{-1},\\
    J^{-1}&=&\frac{1}{27}m^{-2/3}\partial^{-1}m^{-1/3}
             (\partial-\partial^3)m^{-1/3}\partial^{-1}m^{-2/3},
\end{eqnarray*}
and  we further define an operator
    \begin{equation*}
        \mathscr{L}=K^{-1}J=3(\partial-\partial^3)^{-1}(4-\partial^2)^{-1}
        (2m\partial+\partial
      m)(\partial-\partial^3)^{-1}(m\partial+2\partial m).
    \end{equation*}

   Choose $G_{0}=\frac{1}{6} \in \mathrm{ker}K$, the Lenard's
recursive sequence are defined as follows
  \begin{equation}\label{2.1}
     G_{j-1}=\mathscr{L}^{-1} G_j, \quad j=1,2,\ldots
  \end{equation}
Hence $G_j$ are uniquely determined, for example, the first two
elements read as
    $$G_0=\frac{1}{6}, \quad
    G_1=(\partial-\partial^3)^{-1}uu_x. $$

In order to obtain the DP hierarchy associated with the spectral
problem (\ref{1}), we first solve the stationary zero-curvature equation
 \begin{equation}\label{2}
    V_x-[U,V]=0,\quad V=(V_{ij})_{3\times 3}
 \end{equation}
with
 \begin{equation}\label{3.0}
   V=\left(
       \begin{array}{ccc}
         V_{11} & V_{12}& V_{13} \\
         V_{21} & V_{22} & V_{23} \\
         V_{31} & V_{32} & V_{33} \\
       \end{array}
     \right),
 \end{equation}
where each entry $V_{ij}$ is a Laurent expansion in $z$
  \begin{equation}\label{3}
       V_{ij}=\sum_{\ell=0}^{n}V_{ij}^{(\ell)}(G_\ell)z^{2(n-\ell+1)}
        \quad i,j=1,\ldots,3, \quad \ell=0,\ldots,n.
  \end{equation}
Equation (\ref{2}) can be rewritten as
     \begin{eqnarray}\label{4}
       V_{11,x} &=& V_{21}+z^{-1}mV_{13}, \nonumber\\
       V_{12,x} &=& V_{22}-V_{11}-V_{13}, \nonumber\\
       V_{13,x} &=& V_{23}-V_{12}, \nonumber\\
       V_{21,x} &=& V_{31}+z^{-1}mV_{23}, \nonumber\\
       V_{22,x} &=& V_{32}-V_{21}-V_{23}, \\
       V_{23,x} &=& V_{33}-V_{22}, \nonumber\\
       V_{31,x} &=& z^{-1}m(V_{33}-V_{11})+V_{21}, \nonumber\\
       V_{32,x} &=& -z^{-1}mV_{12}+V_{22}-V_{31}-V_{33}, \nonumber\\
       V_{33,x} &=& -z^{-1}mV_{13}+V_{23}-V_{32}.\nonumber
     \end{eqnarray}
Inserting (\ref{3}) into (\ref{4}) yields
   \begin{eqnarray}\label{5}
     V_{11}^{(\ell)} &=& z^{-1}(4-\partial^2)G_\ell
     +3z^{-2}\partial(\partial-\partial^3)^{-1}(m\partial+2\partial m)G_\ell, \nonumber\\
     V_{12}^{(\ell)} &=& 3z^{-1}G_{\ell,x}
     -3z^{-2}(\partial-\partial^3)^{-1}(m\partial+2\partial m)G_\ell, \nonumber\\
     V_{13}^{(\ell)} &=& -6z^{-1}G_\ell, \nonumber\\
     V_{21}^{(\ell)}&=& z^{-1}(4-\partial^2)G_{\ell,x}
     +3z^{-2}(\partial^2(\partial-\partial^3)^{-1}(m\partial+2\partial m)G_\ell
     +2mG_\ell), \nonumber\\
     V_{22}^{(\ell)} &=& -2z^{-1}(G_\ell-G_{\ell,xx}),
   \end{eqnarray}
   \begin{eqnarray*}
     V_{23}^{(\ell)} &=& -3z^{-1}G_{\ell,x}
     -3z^{-2}(\partial-\partial^3)^{-1}(m\partial+2\partial m)G_\ell, \nonumber\\
     V_{31}^{(\ell)} &=& z^{-1}(4-\partial^2)G_{\ell,xx}+
      3z^{-2}(\partial+z^{-1}m)(\partial-\partial^3)^{-1}(m\partial+2\partial m)G_\ell,\nonumber\\
     V_{32}^{(\ell)} &=& -z^{-1}(\partial-\partial^3)G_\ell-
     3z^{-2}(\partial^{-1}(m\partial+2\partial m)G_\ell
     -2mG_\ell), \nonumber\\
     V_{33}^{(\ell)} &=&z^{-1} (-2 G_\ell- G_{\ell,xx})
     -3z^{-2}(\partial(\partial-\partial^3)^{-1}(m\partial+2\partial
     m)G_\ell).\nonumber
   \end{eqnarray*}
Substituting  (\ref{4}) and (\ref{5}) into (2.7), we can show that  Lenard sequence $G_{\ell}$ satisfy the
Lenard equation
   \begin{equation}
        KG_{\ell}=z^{-2}JG_{\ell},\ \ell=0, 1, \cdots.
   \end{equation}
For our use in Theorem \ref{them6.2}, we introduce the following notations
\begin{eqnarray*}
  V_{11}^{(\ell,0)} &=& (4-\partial^2)G_{\ell},\qquad V_{11}^{(\ell,1)}=3\partial(\partial-\partial^3)^{-1}(m\partial+2\partial m)G_\ell, \\
   V_{12}^{(\ell,0)}  &=& G_{\ell,x},\qquad~~~~~~~~~  V_{12}^{(\ell,1)} =-3(\partial-\partial^3)^{-1}(m\partial+2\partial m)G_\ell, \\
   V_{13}^{(\ell,0)}&=& -6G_\ell,\qquad ~~~~~~~~  V_{13}^{(\ell,1)}=0,\\
    V_{21}^{(\ell,0)}&=& (4-\partial^2)G_{\ell,x}, \\ V_{21}^{(\ell,1)}& = & 3(\partial^2(\partial-\partial^3)^{-1}(m\partial+2\partial m)G_\ell
     +2mG_\ell),\\
     V_{22}^{(\ell,0)}&=& -2(G_\ell-G_{\ell,xx}), \qquad V_{22}^{(\ell,1)}=0,\\
    V_{23}^{(\ell,0)}&=& -3G_{\ell,x},\qquad  V_{23}^{(\ell,1)}=-3 (\partial-\partial^3)^{-1}(m\partial+2\partial m)G_\ell,\\
    V_{31}^{(\ell,0)}&=& (4-\partial^2)G_{\ell,xx},\\
    V_{31}^{(\ell,1)}&=& 3(\partial+z^{-1}m)(\partial-\partial^3)^{-1}(m\partial+2\partial m)G_\ell,\\
    V_{32}^{(\ell,0)}&=& -(\partial-\partial^3)G_\ell, \quad V_{32}^{(\ell,1)}=-3\partial^{-1} (m\partial+2\partial m)G_\ell
     -2mG_\ell),\\
      V_{33}^{(\ell,0)}&=& -2 G_\ell- G_{\ell,xx},\qquad  V_{33}^{(\ell,1)}=-3(\partial(\partial-\partial^3)^{-1}(m\partial+2\partial
     m)G_\ell).
    \end{eqnarray*}

Let $\psi$ satisfy the spectral problem ({\ref{1}) and an auxiliary
problem
    \begin{equation} \label{6}
       \psi_{t_n}=V\psi.
    \end{equation}
where $V$ is defined by (\ref{3.0}) and (\ref{3}).  The compatibility condition between  (\ref{1}) and (\ref{6}) yields the
zero-curvature equation
$$U_{t_n}-V_x+[U,V]=0,$$
which is equivalent to the DP  hierarchy
   \begin{equation}\label{7}
    \mathrm{DP}_n(u)= m_{t_n}-X_n=0, \qquad n\geq 0,
   \end{equation}
where the vector fields are given by
$$X_n =JG_n=J\mathscr{L}^nG_0, \quad n\geq 0.$$

  By definition, the set of solutions of (\ref{7}), with $n$ ranging
in $\mathbb{N}_0$, represents the class of algebro-geometric DP
solutions. At times it convenient to abbreviate algebro-geometric
stationary DP solutions $u$ simply as DP potentials.

  The system of equations $\mathrm{DP}_0(u)=0$ represents the
  $\mathrm{DP}$ equation.

  In order to derive the corresponding plane algebraic curve, we
  consider the
  stationary zero-curvature equation
  \begin{equation}\label{8}
           z^{1/2}V_x=[U,z^{1/2}V],
  \end{equation}
  which   is equivalent to
  (\ref{2}), but the term $z^{1/2}V$ can ensure that the following algebraic curve
   is in positive  powers of  $z$.

   A direct calculation
   shows that the matrix $yI-z^{1/2}V$ also satisfies the stationary
   zero-curvature equation, then we conclude  that
        $$\frac{d}{dx}(\mathrm{det}(yI-z^{1/2}V))=0,$$
   which implies that the characteristic polynomial $\mathrm{det}(yI-z^{1/2}V)$ of
   Lax matrix $z^{1/2}V$ is independent of the
   variable $x$. Therefore we define the algebraic curve
        \begin{equation}\label{a0}
          \mathcal {F}_r(z,y)=\mathrm{det}(yI-z^{1/2}V)=
          y^3+yS_r(z)-T_r(z),
        \end{equation}
   where $S_r(z)$ and $T_r(z)$ are polynomials with constant
   coefficients of $z,$
        \begin{eqnarray}
          S_r(z)&=&z\left(\left|\begin{array}{cc}
                    V_{11} & V_{12} \\
                    V_{21} & V_{22}
                  \end{array}\right|
          +
          \left|\begin{array}{cc}
                  V_{22} & V_{23} \\
                  V_{32} & V_{33}
                \end{array}\right|
           +
           \left|\begin{array}{cc}
                  V_{11} & V_{13} \\
                  V_{31} & V_{33}
                \end{array}\right|\right), \\
           T_r(z)&=&z^{3/2}\left|\begin{array}{ccc}
                          V_{11} & V_{12} & V_{13} \\
                          V_{21} & V_{22} & V_{23} \\
                          V_{31} & V_{32} & V_{33}
                        \end{array}\right|.
        \end{eqnarray}
   In order to ensure the polynomials with integer powers,
   we introduce the $z=\tilde{z}^2$, the algebraic curve becomes,
         \begin{equation}\label{2.19000}
          \mathcal {F}_r(\tilde{z},y)=
          y^3+yS_r(\tilde{z})-T_r(\tilde{z}),
         \end{equation}
   where $S_r(\tilde{z})$ and $T_r(\tilde{z})$ are polynomials with constant
   coefficients of $\tilde{z},$
        \begin{eqnarray}\label{2.20}
          S_r(\tilde{z})&=&\tilde{z}^2\left(\left|\begin{array}{cc}
                    V_{11} & V_{12} \\
                    V_{21} & V_{22}
                  \end{array}\right|
          +
          \left|\begin{array}{cc}
                  V_{22} & V_{23} \\
                  V_{32} & V_{33}
                \end{array}\right|
           +
           \left|\begin{array}{cc}
                  V_{11} & V_{13} \\
                  V_{31} & V_{33}
                \end{array}\right|\right) \nonumber\\
            &=&\sum_{j=0}^{4n+2}S_{r,j}\tilde{z}^{8n+6-2j},\\
          T_r(\tilde{z})&=&\tilde{z}^3\left|\begin{array}{ccc}
                          V_{11} & V_{12} & V_{13} \\
                          V_{21} & V_{22} & V_{23} \\
                          V_{31} & V_{32} & V_{33}
                         \end{array}\right|\nonumber
                        \\
          & = &\sum_{j=0}^{6n+4} T_{r,j}\tilde{z}^{12n+9-2j}.
        \end{eqnarray}
  We note that $T_r(\tilde{z})$ is a polynomial of degree $r$~$(r=3(4n+3))$ with respect to
  $\tilde{z}$,
  then $\mathcal{F}_r(\tilde{z},y)=0$ naturally leads to the plane third order
  algebraic
  curve $\mathcal{K}_{r-2}$ of genus $r-2\in\mathbb{N}$ (see Remark 2.2 and Remark 2.3),
        \begin{equation}\label{9}
          \mathcal{K}_{r-2}:\mathcal{F}_r(\tilde{z},y)=
          y^3+yS_r(\tilde{z})-T_r(\tilde{z})=0, \quad r=12n+9.
        \end{equation}
   The algebraic curve $\mathcal{K}_{r-2}$ in (\ref{9}) is compactified by
   joining three points at infinity
   $$P_{\infty_i}, \ i=1,2,3,$$
   but for notational
   simplicity the compactification is also denoted by
   $\mathcal{K}_{r-2}$. Points on
   $$\mathcal{K}_{r-2} \setminus \{P_{\infty_i}\},\
   i=1,2,3$$
    are
   represented as pairs $P=(\tilde{z},y(P))$, where $y(\cdot)$ is the meromorphic
   function on $\mathcal{K}_{r-2}$ satisfying
   $$\mathcal{F}_r(\tilde{z},y(P))=0.$$
   The complex structure on $\mathcal{K}_{r-2}$ is defined in the usual
   way by introducing local coordinates
   $$\zeta_{Q_0}:P \rightarrow
   \zeta=\tilde{z}-\tilde{z}_0$$
  near points
   $$Q_0=(\tilde{z}_0,y(Q_0))\in \mathcal{K}_{r-2} \backslash\{P_0=(0,0)\},$$
    which
   are neither branch nor singular points of $\mathcal{K}_{r-2}$;
   near $P_0=(0,0)$,  the local coordinate is
   \begin{equation}\label{r0021}
     \zeta_{P_0}: P \rightarrow  \zeta=\tilde{z}^{\frac{1}{3}},
   \end{equation}
   and similarly at branch and singular points of $\mathcal{K}_{r-2};$
   near the points $P_{\infty_i} \in \mathcal{K}_{r-2}$,
    the local coordinates are
  \begin{equation}\label{n001}
  \zeta_{P_{\infty_i}}:P  \rightarrow \zeta=\tilde{z}^{-1},\quad i=1,2,3.
  \end{equation}

   The holomorphic map
   $\ast,$ changing sheets, is defined by
       \begin{eqnarray}\label{3.3}
       &&\ast: \begin{cases}
                        \mathcal{K}_{r-2}\rightarrow\mathcal{K}_{r-2},
                       \\
                       P=(\tilde{z},y_j(\tilde{z}))\rightarrow
                       P^\ast=(\tilde{z},y_{j+1(\mathrm{mod}~
                       3)}(\tilde{z})), \quad j=0,1,2,
                      \end {cases}
                     \nonumber \\
      && P^{\ast \ast}:=(P^\ast)^\ast, \quad \mathrm{etc}.,
       \end{eqnarray}
   where $y_j(\tilde{z}),\, j=0,1,2$ denote the three branches of $y(P)$
   satisfying $\mathcal{F}_{r}(\tilde{z},y)=0$.

   Finally, positive divisors on $\mathcal{K}_{r-2}$ of degree $r-2$
   are denoted by
        \begin{equation}\label{3.4}
          \mathcal{D}_{P_1,\ldots,P_{r-2}}:
             \begin{cases}
              \mathcal{K}_{r-2}\rightarrow \mathbb{N}_0,\\
              P\rightarrow \mathcal{D}_{P_1,\ldots,P_{r-2}}(P)=
                \begin{cases}
                  \textrm{ $k$ if $P$ occurs $k$
                      times in $\{P_1,\ldots,P_{r-2}\},$}\\
                   \textrm{ $0$ if $P \notin
                     $$ \{P_1,\ldots,P_{r-2}\}.$}
                \end{cases}
             \end{cases}
        \end{equation}
        In particular, the divisor $\left(\phi(\cdot)\right)$ of a meromorphic function
        $\phi(\cdot)$ on $\mathcal{K}_{r-2}$ is defined by
        \begin{eqnarray}\label{3.4a0}
          \left(\phi(\cdot)\right):
           \mathcal{K}_{r-2} \rightarrow \mathbb{Z},\quad
          P\mapsto \omega_{\phi}(P),
          \end{eqnarray}
        where $\omega_{\phi}(P)=m_0\in\mathbb{Z}$ if
        $(\phi\circ\zeta_{P}^{-1})(\zeta)=\sum_{n=m_0}^{\infty}c_n(P)\zeta^n$
        for some $m_0\in\mathbb{Z}$ by using a chart $(U_{P}, \zeta_P)$ near $P\in\mathcal{K}_{r-2}.$
        \newtheorem{rem254}[hyp1]{Remark}
        \begin{rem254}\label{rem254}
        In this paper, we make the following two assumptions about the curve $\mathcal{K}_{r-2}$:\\
        $(i)$ The affine plane algebraic curve $\mathcal{K}_{r-2}$ is nonsingular.\\
        $(ii)$ The leading coefficients of
        $S_r(z)$, $T_r(z)$ satisfy
        \begin{equation}\label{2500}
        \pm\frac{2\sqrt{-3}}{9}S_{r,0}^{3/2}-T_{r,0}\neq 0.
        \end{equation}
        Multiplying the
        polynomial $T_r(\tilde{z})$
        by a constant $\hbar\in\mathbb{R}$ \emph{(}or $\mathbb{C}$\emph{)},
        one easily finds the curve $(\ref{9})$ changes into
        \begin{equation*}
         \widetilde{\mathcal{F}}_r(\tilde{z},y)=
          y^3+yS_r(\tilde{z})-\hbar T_r(\tilde{z})=0.
         \end{equation*}
         Since there exists a constant $\hbar$ such that
         $$\frac{2\sqrt{-3}}{9}S_{r,0}^{3/2}-\hbar T_{r,0}\neq 0,$$
         we assume $(\ref{2500})$ is always true for the curve
         $\mathcal{K}_{r-2}$ $(\ref{9})$ without loss of generality.
        \end{rem254}

        Next, we give a few words about computing the genus of the curve
        $(\ref{9})$ under the two assumptions in Remark $\ref{rem254}$.
        \newtheorem{rem252}[hyp1]{Remark}
        \begin{rem252}\label{rem252}
        In this paper, we denote by $\overline{\mathcal{K}}$ the associated
        projective curve of a affine curve $\mathcal{K}.$
        There are two approaches to compute the genus $g$ of $\overline{\mathcal{K}}_{r-2}$.
        One of them is to use the formula
         \begin{equation}\label{genus}
        g=(n-1)(n-2)/2,
        \end{equation}
        where $n$ is the degree of corresponding homogeneous polynomial of
        $\overline{\mathcal{K}}_{r-2},$ if the curve $\mathcal{K}_{r-2}$ is nonsingular
        $($smooth$).$
        The Fermat curve is a celebrated example of smooth projective curves.
        In general, the projective curve $\overline{\mathcal{K}}_{r-2}$ may be singular
        even the associated affine curve $\mathcal{K}_{r-2}$ is nonsingular.
        In this case one has to account for the singularities at infinity
        and properly amend the genus formula $(\ref{genus})$ according to the results
        of Clebsch, M. Noether, and Pl\emph{$\ddot{u}$}cker.
        Alternative and more efficient way is to use
        a special case of Riemann-Hurwitz formula. The $g$-number $g$ \emph{\cite{58}} of
        $\mathcal{K}_{r-2}$ and hence the genus of $\mathcal{K}_{r-2}$ if
        $\mathcal{K}_{r-2}$ is nonsingular $($smooth$),$  is
        \begin{equation}\label{genus1}
        g=1-N+B/2, \quad \text{with}\quad B=\sum_{P\in\mathcal{K}_{r-2}}(k(P)-1),
        \end{equation}
        where N is the number of sheets of $\mathcal{K}_{r-2},$
        $B$ is the total branching number of
        sheets of $\mathcal{K}_{r-2},$
        and $k(P)-1$ is the branching order of $P\in\mathcal{K}_{r-2}.$
        In the current DP case,
        one easily finds $N=3$. Next, one accounts for the computation
        of $B$. The discriminant $\Delta(\tilde{z})$ of the curve $(\ref{9})$ defined by
         $\Delta(\tilde{z})=27T_r^2(\tilde{z})+4S_r^3(\tilde{z})=\tilde{z}^2
         \Delta_1(\tilde{z}),$
         where $\Delta_1(\tilde{z})$
         a polynomial of degree $2r-2$ with $\Delta_1(0)\neq 0.$ Hence the Riemann
         surface defined by the
         compactification of $(\ref{9})$ can have at most $2r$ double points.
         However, since $\tilde{z}=0$ is a triple root of equation $(\ref{9})$,
         there are at most $2r-2$ double points on $\mathcal{K}_{r-2}$.
         Then if all
         branch points except $P_0$ are distinct double points, one obtains
         $($taking into account the triple point at $P_0$$)$
         \begin{eqnarray*}
          B&=&\sum_{P\in\mathcal{K}_{r-2}}(k(P)-1)
           =\!\sum_{P\in\mathcal{K}_{r-2}\backslash\{P_0\}}(k(P)-1)+(k(P_0)-1).\\
           &=&(2r-2)+2=24n+18
         \end{eqnarray*}
         Substituting the value of $N, B$ into the Riemann-Hurwitz formula
         $(\ref{genus1})$, we \newpage
         \noindent derive $g=12n+7=r-2.$

        Obviously, the DP-type curve $\mathcal{K}_{r-2}$ differs from other kinds of
        algebraic curves \emph{(}such as KdV-type, AKNS-type, Boussinesq-type, etc.\emph{)}
        in the sense that it is compactified by three distinct points
        $P_{\infty_i}$ \emph{(}$i=1,2,3$\emph{)} at infinity. Moreover,
        the genus of $\mathcal{K}_{r-2}$ is not $r-2$ if we remove the assumption $(ii)$ in
        Remark $2.2$.
        In the KdV \emph{(}or AKNS, Boussinesq\emph{)} case,
        the topological genus is uniquely determined as long as
        the given affine curve is nonsingular. However, in the DP case, the only assumption
        that the affine
        curve is nonsingular can not ensure its topological genus is of one type.
        Thus we add a condition $(\ref{2500})$
        to the curve $\mathcal{K}_{r-2}$.
        \end{rem252}
       \newtheorem{rem251}[hyp1]{Remark}
        \begin{rem251}
        We investigate what happens at the point infinity on our DP-type curve $\mathcal{K}_{r-2}$.
        Following the treatment in $\emph{\cite{11a}}$ we substitute the variable
        $v=\tilde{z}^{-1}$ into $(\ref{9})$ yields
        \begin{eqnarray}\label{r001}
        (v^{4n+3}y)^3&+&(S_{r,0}+S_{r,1}v^2+\ldots+S_{r,4n+2}v^{8n+4})v^{4n+3}y\nonumber \\
        &-&(T_{r,0}+\ldots+T_{r,6n+4}v^{12n+8}) =0.
        \end{eqnarray}
        Let $v_1=v^{4n+3}y$,  $(\ref{r001})$ becomes
        \begin{equation}\label{r002}
        v^3_1+S_{r,0}v_1-T_{r,0}=0
        \end{equation}
        as $v\rightarrow 0$ \emph{(}corresponding to $\tilde{z}\rightarrow \infty$\emph{)}. This
        corresponds to three distinct points $P_{\infty_j},$
        $j=1,2,3$ at infinity \emph{(}each with multiplicity one\emph{)}, given by the three points
        $(0,\aleph_j)$ for $j=1,2,3,$ where $\aleph_j$ $(j=1,2,3)$ are the three distinct roots
        of equation $(\ref{r002})$. As each point at infinity has multiplicity
        one , none are branch points, and consequently each admits the local coordinate
        $(\ref{n001})$ for $|\tilde{z}|$ sufficiently large.

        Similarly, near point $P_0=(0,0)\in\mathcal{K}_{r-2},$ one finds
        $y^3=0$ by taking $\tilde{z}\rightarrow 0$ in $(\ref{9})$. This corresponds to one
        point of multiplicity three at $\tilde{z}=0$. We therefore use the coordinate
        $(\ref{r0021})$
        at the branch point $P_0$.
       \end{rem251}
    \section{The stationary  DP formalism}
    In this section, we are devoted to a detailed study of the
    stationary DP hierarchy.
    Our principle tools are derived from a fundamental meromorphic
    function $\phi$ on the algebraic curve $\mathcal{K}_{r-2}$.
    With the help of $\phi$ we study the Baker-Akhiezer vector $\psi,$ and Dubrovin-type
    equations.

    First, we give a brief description about the Baker-Akhiezer functions.
    The exponential $e^z$ is analytic in $\mathbb{C}$ and has an essential singularity at the \nopagebreak[4]
    \hfill point $z=\infty.$ If $q(z)$ is a rational function, then $f(z)=e^{q(z)}$ is analytic in
    \newpage
   \noindent $\bar{\mathbb{C}}=\mathbb{CP}^1$
   everywhere except at the poles of $q(z)$, where $f(z)$ has essential singular points.
  In the last century Clebsh and Gordan considered a generalizing
  functions of exponential type to Riemann surfaces of higher genus.
  Baker noted that such functions of exponential type can be expressed in terms of theta functions
  of Riemann surfaces. Akhiezer first directed attention to the fact that under certain
  conditions functions of exponential type on hyperelliptic Riemann surfaces are eigenfunctions
  of second-order linear diferential operators. Following the established tradition, we call
  functions of exponential type on Riemann surfaces Baker-Akhiezer functions.

 Next, we introduce the stationary vector Baker-Akhiezer function $\psi=(\psi_1,\psi_2,\psi_3)^t$
       \begin{equation}\label{3.5}
         \begin{split}
          & \psi_x(P,x,x_0)=U(u(x),\tilde{z}(P))\psi(P,x,x_0),\\
           & \tilde{z}V(u(x),\tilde{z}(P))\psi(P,x,x_0)=y(P)\psi(P,x,x_0),\\
          & \psi_2(P,x_0,x_0)=1; \quad P=(\tilde{z},y)\in \mathcal{K}_{r-2}
           \setminus \{P_{\infty_i},P_0\},~i=1,2,3,~x\in \mathbb{C}.
         \end{split}
       \end{equation}
  Closely related to $\psi(P,x,x_0)$ is the following meromorphic
  function $\phi(P,x)$ on $\mathcal{K}_{r-2}$ defined by
       \begin{equation}\label{3.6}
         \phi(P,x)=\tilde{z}\frac{ \psi_{2,x}(P,x,x_0)}{\psi_2(P,x,x_0)},
         \quad P\in \mathcal{K}_{r-2},~ x\in \mathbb{C}
       \end{equation}
  such that
       \begin{equation}\label{3.7}
         \psi_2(P,x,x_0)=\mathrm{exp}\left(\tilde{z}^{-1}\int_{x_0}^x
         \phi(P,x^\prime) dx^\prime \right),
         \quad P\in \mathcal{K}_{r-2}\setminus \{P_{\infty_i},P_0\},
         \quad i=1,2,3.
       \end{equation}
  Since $\phi$
  is the fundamental ingredient for the construction of algebro-geometric solutions
  of the stationary DP hierarchy, we next seek its connection with the recursion formalism
  of Section 2.
  By using (\ref{3.5}),  a direct calculation gives
       \begin{equation}\label{3.8}
         \phi=\tilde{z}\frac{yV_{31}+C_r}{yV_{21}+A_r}
         =\frac{\tilde{z}F_r}{y^2V_{31}-yC_{r}+D_r}
         =\tilde{z}\frac{y^2V_{21}-yA_r+B_r}{E_r},
       \end{equation}
where
       \begin{eqnarray}\label{3.9}
          \begin{split}
            A_r&=\tilde{z}(V_{23}V_{31}-V_{33}V_{21})\\
             &  =\tilde{z}[V_{23}V_{31}+V_{21}(V_{22}+V_{11})],\\
            B_r&=\tilde{z}^2[V_{22}(V_{11}V_{21}+V_{23}V_{31})
                  -V_{21}(V_{12}V_{21}+V_{23}V_{32})],\\
           C_r&= \tilde{z}(V_{21}V_{32}-V_{22}V_{31})\\
           &    = \tilde{z}[V_{21}V_{32}+V_{31}(V_{11}+V_{33})],\\
           D_r&=\tilde{z}^2[V_{31}(V_{11}V_{33}-V_{13}V_{31})+
               V_{32}(V_{21}V_{33}-V_{23}V_{31})],
          \end{split}
       \end{eqnarray}
       \begin{eqnarray}\label{3.10}
         \begin{split}
          E_r&=\tilde{z}^2[V_{23}(V_{21}V_{33}-V_{11}V_{21}-V_{23}V_{31})+V_{13}V_{21}^2],\\
            F_r&=\tilde{z}^2[V_{31}(V_{22}V_{32}-V_{11}V_{32}+V_{12}V_{31})-V_{21}V_{32}^2].
         \end{split}
       \end{eqnarray}
   The quantities $A_r,\ldots,F_r$ in (\ref{3.9}) and (\ref{3.10})
   are of course not independent each other. There exist various
   interrelationships between them and $S_r,$ $T_r,$ some of which
   are summarized below.
     \newtheorem{lem3.1}{Lemma}[section]
     \begin{lem3.1}
       Let $(\tilde{z},x)\in \mathbb{C}^2.$ Then
       \begin{eqnarray}\label{3.11}
         \begin{split}
            & V_{21}F_r=V_{31}D_r-C_r^2-V_{31}^2S_r,\\
           & A_rF_r=T_rV_{31}^2+C_rD_r,
         \end{split}
       \end{eqnarray}
       \begin{eqnarray}\label{3.12}
          \begin{split}
           &V_{31}E_r=V_{21}B_r-A_r^2-V_{21}^2S_r,\\
           &E_rC_r=T_rV_{21}^2+A_rB_r,
          \end{split}
       \end{eqnarray}
       \begin{eqnarray}\label{3.13}
          \begin{split}
           &V_{21}D_r+V_{31}B_r-V_{21}V_{31}S_r+A_rC_r=0,\\
           &T_rV_{21}V_{31}+S_rC_rV_{21}+S_rA_rV_{31}-A_rD_r-B_rC_r=0,\\
           &E_rF_r=-T_rC_rV_{21}-T_rA_rV_{31}+B_rD_r,\\
          \end{split}
       \end{eqnarray}
       \begin{eqnarray}\label{3.14}
         \begin{split}
           &E_{r,x}=-2S_rV_{21}+3B_r,\\
           &
           V_{31}F_{r,x}=-2V_{31}^2S_r+3V_{31}D_r+2\tilde{z}^{-2}mV_{33}F_r+V_{31}mJ_r,
           \end{split}
           \end{eqnarray}
       where
         $$ J_r=V_{22}^2V_{32}-V_{11}V_{22}V_{32}-V_{13}V_{31}V_{32}
          -V_{23}V_{32}^2+2V_{21}V_{32}V_{12}+V_{31}V_{33}V_{12}.$$
     \end{lem3.1}
   \textbf{Proof.}~~Using (\ref{9}) and (\ref{3.8}), we have
       \begin{eqnarray*}
          F_rV_{21}y+F_rA_r&=&V_{31}^2y^3+(V_{31}D_r-C_r^2)y+C_rD_r \\
           &=& (V_{31}D_r-C_r^2-V_{31}^2S_r)y+T_rV_{31}^2+C_rD_r,
       \end{eqnarray*}
       \begin{eqnarray*}
          E_rV_{31}y+E_rC_r&=&V_{21}^2y^3+(V_{21}B_r-A_r^2)y+A_rB_r\\
              &=& (V_{21}B_r-A_r^2-V_{21}^2S_r)y+T_rV_{21}^2+A_rB_r,
       \end{eqnarray*}
       \begin{eqnarray*}
         E_rF_r&=& (y^2V_{21}-yA_r+B_r)(y^2V_{31}-yC_{r}+D_r)\\
               &=& (V_{21}D_r+V_{31}B_r-V_{21}V_{31}S_r+A_rC_r)y^2
               \\
               &&+(T_rV_{21}V_{31}+S_rC_rV_{21}+S_rA_rV_{31}-A_rD_r-B_rC_r)y.\\
               &&  ~~~~-T_rC_rV_{21}-T_rA_rV_{31}+B_rD_r.
       \end{eqnarray*}
   By comparing the same powers of $y$, we arrive at
   (\ref{3.11})-(\ref{3.13}). With the help of (\ref{3.10}) and the stationary
    zero-curvature equation (\ref{4}),  we have
        \begin{eqnarray*}
          E_{r,x}&=&\tilde{z}^2[V_{21}(V_{33}^2-V_{11}V_{33}-V_{22}V_{33}+V_{22}V_{11})
           -V_{33}V_{23}V_{31}\\
           &&+2V_{22}V_{23}V_{31}-V_{31}V_{23}V_{11}+2V_{31}V_{13}V_{21}
             -V_{23}V_{21}V_{32}-V_{12}V_{21}^2]\\
           &=& -2S_rV_{21}+3B_r,
        \end{eqnarray*}
        \begin{eqnarray*}
           V_{31}F_{r,x}&=&
           \tilde{z}^2[2V_{31}^2(V_{21}V_{12}-V_{11}V_{22}+V_{13}V_{31}-V_{11}V_{33}+V_{32}V_{23}
           -V_{22}V_{33})\\
           &&+V_{31}^2V_{11}V_{22}-3V_{31}^3V_{13}+3V_{31}^2V_{11}V_{33}
           -3V_{31}^2V_{23}V_{32}+V_{31}^2V_{22}V_{33}]\\
          & &~~~~+2\tilde{z}^{-2}mV_{33}F_r+mV_{31}J_r\\
          &=&
          -2V_{31}^2S_r+3V_{31}D_r+2\tilde{z}^{-2}mV_{33}F_r+mV_{31}J_r,
        \end{eqnarray*}
        which is just (\ref{3.14}). \quad $\square$ \\

   By inspection of (\ref{5}) and (\ref{3.10}), one infers that
   $E_r$ and $\tilde{z}^2 F_r$ are polynomials with respect to $\tilde{z}$ of
   degree $r-5$ and $r-3$, respectively.
   Let $\{\mu_j(x)\}_{j=1,\ldots,r-5}$ and $\{\nu_j(x)\}_{j=1,\ldots,r-3}$
   denote the zeros of $E_r(x)$ and $\tilde{z}^2 F_r(x)$, respectively.
   Hence we may write
        \begin{equation}\label{3.15}
           E_r=u \prod_{j=1}^{r-5}(\tilde{z}-\mu_j(x)),
        \end{equation}
        \begin{equation}\label{3.16}
         F_r=-uu_x^2 \tilde{z}^{-2} \prod_{j=1}^{r-3}
         (\tilde{z}-\nu_j(x)).
         \end{equation}
         Defining
         \begin{eqnarray}\label{3.17}
            \hat{\mu}_j(x)
            =\left(\mu_j(x),-\frac{A_r(\mu_j(x),x)}{V_{21}(\mu_j(x),x)}\right)
            \in \mathcal{K}_{r-2}, ~
            j=1,\ldots,r-5,~x\in\mathbb{C},
         \end{eqnarray}
         \begin{eqnarray}\label{3.18}
           \hat{\nu}_j(x)
            =\left(\nu_j(x),-\frac{C_r(\nu_j(x),x)}{V_{31}(\nu_j(x),x)}\right)
            \in \mathcal{K}_{r-2},~
            j=1,\ldots,r-3,~x\in\mathbb{C}.
         \end{eqnarray}
   One infers from (\ref{3.8}) that the divisor
   $(\phi(P,x))$ of $\phi(P,x)$ is given by
   \begin{equation}\label{3.19}
           (\phi(P,x))=\mathcal{D}_{P_0,\underline{\hat{\nu}}(x)}(P)
           -\mathcal{D}_{P_{\infty_1},\underline{\hat{\mu}}(x)}(P),
         \end{equation}
   where
   \begin{equation*}
     \underline{\hat{\nu}}(x)=\{\hat{\nu}_1(x),\ldots,\hat{\nu}_{r-3}(x)\},\quad \underline{\hat{\mu}}(x)=\{P_{\infty_2},P_{\infty_3},\hat{\mu}_1(x),
           \ldots,\hat{\mu}_{r-5}(x)\}.
     \end{equation*}
         That is, $P_0,\hat{\nu}_1(x),\ldots,\hat{\nu}_{r-3}(x)$ are the
    $r-2$ zeros of $\phi(P,x)$ and
    $P_{\infty_1},P_{\infty_2},$
    $ P_{\infty_3},\hat{\mu}_1(x), \ldots,\hat{\mu}_{r-5}(x)$ its $r-2$ poles.

   Since from (\ref{3.3}), $y_j(\tilde{z}),\, j=0,1,2$
   satisfy $\mathcal{F}_{r}(\tilde{z},y)=0$, that is
        \begin{equation}\label{3.21}
          (y-y_0(\tilde{z}))(y-y_1(\tilde{z}))(y-y_2(\tilde{z}))
          =y^3+yS_r(\tilde{z})-T_r(\tilde{z})=0,
        \end{equation}
    then we can easily get
        \begin{equation}\label{3.22}
           \begin{split}
             & y_0+y_1+y_2=0,\\
             & y_0y_1+y_0y_2+y_1y_2=S_r(\tilde{z}),\\
             & y_0y_1y_2=T_r(\tilde{z}),\\
             & y_0^2+y_1^2+y_2^2=-2S_r(\tilde{z}),\\
             & y_0^3+y_1^3+y_2^3=3T_r(\tilde{z}),\\
             & y_0^2y_1^2+y_0^2y_2^2+y_1^2y_2^2=S_r^2(\tilde{z}).
           \end{split}
        \end{equation}
   Further properties of $\phi(P,x)$ and $\psi_2(P,x,x_0)$ are
   summarized as follows.
   \newtheorem{the3.2}[lem3.1]{Theorem}
    \begin{the3.2}\label{them3.20}
      Assume $(\ref{3.5})$ and $(\ref{3.6})$, $P=(\tilde{z},y)\in
      \mathcal{K}_{r-2}\setminus \{P_{\infty_i},P_0\},$ $i=1,2,3,$ and let
      $(\tilde{z},x,x_0)\in \mathbb{C}^3$. Then
         \begin{eqnarray}\label{3.23}
            &&\phi_{xx}(P,x)+3\tilde{z}^{-1}\phi(P,x)\phi_x(P,x)
             +\tilde{z}^{-2}\phi^3(P,x)
              -\frac{m_x(x)}{m(x)}\phi_x(P,x)\nonumber\\
              &&~~~-\phi(P,x)-\tilde{z}^{-1}\frac{m_x(x)}{m(x)}\phi^2(P,x)
                +m(x)\tilde{z}^{-1}+\frac{m_x(x)}{m(x)}\tilde{z}=0,
         \end{eqnarray}
         \begin{eqnarray}\label{3.24}
          \phi(P,x)\phi(P^\ast,x)\phi(P^{\ast\ast},x)=
            -\tilde{z}^3\frac{F_r(\tilde{z},x)}{E_r(\tilde{z},x)},
         \end{eqnarray}
         \begin{eqnarray}\label{3.25}
          \phi(P,x)+\phi(P^\ast,x)+\phi(P^{\ast\ast},x)=
            \tilde{z}\frac{E_{r,x}(\tilde{z},x)}{E_r(\tilde{z},x)},
         \end{eqnarray}
         \begin{eqnarray}\label{3.26}
         \frac{1}{\phi(P,x)}+\frac{1}{\phi(P^\ast,x)}
                    +\frac{1}{\phi(P^{\ast\ast},x)}
          &=&\frac{F_{r,x}(\tilde{z},x)}{\tilde{z}F_r(\tilde{z},x)}-
             \frac{m(x)J_r(\tilde{z},x)}{\tilde{z}F_r(\tilde{z},x)}
             \nonumber \\
                &-&\frac{2m(x)V_{33}(\tilde{z},x)}
                   {\tilde{z}^3V_{31}(\tilde{z},x)},
         \end{eqnarray}
         \begin{eqnarray}\label{3.27}
          y(P)\phi(P,x)+y(P^\ast)\phi(P^\ast,x)
                   +y(P^{\ast\ast})\phi(P^{\ast\ast},x)= \nonumber\\
            \tilde{z}\frac{3T_r(\tilde{z})V_{21}(\tilde{z},x)
              +2S_r(\tilde{z})A_r(\tilde{z},x)} {E_r(\tilde{z},x)},
         \end{eqnarray}
         \begin{eqnarray}\label{3.28}
          \psi_2(P,x,x_0)\psi_2(P^\ast,x,x_0)\psi_2(P^{\ast\ast},x,x_0)
          =
              \frac{E_r(\tilde{z},x)}{E_r(\tilde{z},x_0)},
         \end{eqnarray}
         \begin{eqnarray}\label{3.29}
          \psi_{2,x}(P,x,x_0)\psi_{2,x}(P^\ast,x,x_0)
                  \psi_{2,x}(P^{\ast\ast},x,x_0)=
            -\frac{F_r(\tilde{z},x)}{E_r(\tilde{z},x_0)},
         \end{eqnarray}
         \begin{eqnarray}\label{3.30}
          &&\psi_2(P,x,x_0)=
          \left[
         \frac{E_r(\tilde{z},x)}{E_r(\tilde{z},x_0)}\right]^{1/3} \nonumber
         \\
         && \times ~
         \mathrm{exp} \left(
              \int_{x_0}^x  \frac{y(P)^2V_{21}(\tilde{z},x^\prime)
                -y(P)A_r(\tilde{z},x^\prime)
                   +\frac{2}{3}S_r(\tilde{z})V_{21}(\tilde{z},x^\prime)}
                   {E_r(\tilde{z},x^\prime)}
                   dx^\prime \right).
         \nonumber \\
         \end{eqnarray}
    \end{the3.2}
   \textbf{Proof.}~~A straightforward calculation shows that (\ref{3.23})
   holds. Next, we prove (\ref{3.24})-(\ref{3.30}).
   From (\ref{3.6}), (\ref{3.8}), (\ref{3.11})-(\ref{3.14}) and
   (\ref{3.22}), we have
       \begin{eqnarray*}
     && \phi(P,x)\phi(P^\ast,x)\phi(P^{\ast\ast},x)=
          \tilde{z}\frac{y_0V_{31}+C_r}{y_0V_{21}+A_r} \times
          \tilde{z}\frac{y_1V_{31}+C_r}{y_1V_{21}+A_r} \times
          \tilde{z}\frac{y_2V_{31}+C_r}{y_2V_{21}+A_r}\\
     && =\tilde{z}^3\frac
      {y_0y_1y_2(V_{31})^3+C_r(V_{31})^2(y_0y_1+y_0y_2+y_1y_2)
              +C_r^2V_{31}(y_0+y_1+y_2)+C_r^3}
      {y_0y_1y_2(V_{21})^3+A_r(V_{21})^2(y_0y_1+y_0y_2+y_1y_2)
              +A_r^2V_{21}(y_0+y_1+y_2)+A_r^3}\\
      &&=\tilde{z}^3\frac
        {T_r(V_{31})^3+C_r(V_{31})^2S_r+C_r^3}
        {T_r(V_{21})^3+A_r(V_{21})^2S_r+A_r^3}\\
      &&=\tilde{z}^3\frac
        {T_r(V_{31})^3+C_r(V_{31}D_r-V_{21}F_r-C_r^2)+C_r^3}
        {T_r(V_{21})^3+A_r(V_{21}B_r-V_{31}E_r-A_r^2)+A_r^3}\\
      &&=-\tilde{z}^3\frac{F_r(\tilde{z},x)}{E_r(\tilde{z},x)},
       \end{eqnarray*}
      \begin{eqnarray*}
        \phi(P,x)+\phi(P^\ast,x)+\phi(P^{\ast\ast},x)&=&\tilde{z}
        \frac{V_{21}(y_0^2+y_1^2+y_2^2)-A_r(y_0+y_1+y_2)+3B_r}{E_r}\\
        &=&\tilde{z}\frac{-2S_rV_{21}+3B_r}{E_r}\\
        &=&\tilde{z}\frac{E_{r,x}(\tilde{z},x)}{E_r(\tilde{z},x)},
      \end{eqnarray*}
      \begin{eqnarray*}
        \frac{1}{\phi(P,x)}+\frac{1}{\phi(P^\ast,x)}+\frac{1}{\phi(P^{\ast\ast},x)}
        &=&\frac{V_{31}(y_0^2+y_1^2+y_2^2)-C_r(y_0+y_1+y_2)+3D_r}{\tilde{z}F_r}\\
        &=&\frac{-2S_rV_{31}+3D_r}{\tilde{z}F_r}\\
        &=&\frac{F_{r,x}(\tilde{z},x)}{\tilde{z}F_r(\tilde{z},x)}-
             \frac{mJ_r(\tilde{z},x)}{\tilde{z}F_r(\tilde{z},x)}
                -2\frac{mV_{33}}{\tilde{z}^3V_{31}},
      \end{eqnarray*}
      \begin{eqnarray*}
       &&y(P)\phi(P,x)+y(P^\ast)\phi(P^\ast,x)+y(P^{\ast\ast})\phi(P^{\ast\ast},x)\\
       &&\qquad \qquad =\tilde{z}\frac{V_{21}(y_0^3+y_1^3+y_2^3)-A_r(y_0^2+y_1^2+y_2^2)
            +B_r(y_0+y_1+y_2)}{E_r}\\
       && \qquad \qquad =\tilde{z}\frac{3T_rV_{21}+2S_rA_r}{E_r},
      \end{eqnarray*}
      \begin{eqnarray*}
       \psi_2(P,x,x_0)\psi_2(P^\ast,x,x_0)\psi_2(P^{\ast\ast},x,x_0)&=&
         \mathrm{exp}\left( \tilde{z}^{-1}\int_{x_0}^x
         [\phi(P,x^\prime)+\phi(P^\ast,x^\prime)+\phi(P^{\ast\ast},x^\prime)]
          dx^\prime\right)\\
        &=& \mathrm{exp} \left( \int_{x_0}^x
             \frac{E_{r,x^\prime}}{E_r} dx^\prime \right)\\
        &=&
        \frac{E_r(\tilde{z},x)}{E_r(\tilde{z},x_0)},
      \end{eqnarray*}
      \begin{eqnarray*}
       \psi_{2,x}(P,x,x_0)\psi_{2,x}(P^\ast,x,x_0)\psi_{2,x}(P^{\ast\ast},x,x_0)
         &=&\tilde{z}^{-1} \psi_2(P,x,x_0)\phi(P,x)  \times
         \tilde{z}^{-1} \psi_2(P^\ast,x,x_0)\phi(P^\ast,x)\\
         &\times &
         \tilde{z}^{-1} \psi_2(P^{\ast\ast},x,x_0)\phi(P^{\ast\ast},x)\\
           &=&-\frac{F_r(\tilde{z},x)}{E_r(\tilde{z},x_0)}.
      \end{eqnarray*}
   Using (\ref{3.7}), (\ref{3.8}) and (\ref{3.14}), we obtain
      \begin{eqnarray*}
         \psi_2(P,x,x_0)&=&\mathrm{exp}
            \left(\tilde{z}^{-1}\int_{x_0}^x \phi(P,x^\prime) dx^\prime \right)\\
         &=& \mathrm{exp} \left(\tilde{z}^{-1}
           \int_{x_0}^x \tilde{z} \frac{y^2V_{21}-yA_r
                   +\frac{2S_rV_{21}+E_{r,x}}{3}}{E_r} dx^\prime
                   \right)  \\
         &=& \mathrm{exp} \left(
              \int_{x_0}^x  \frac{y^2V_{21}-yA_r
                   +\frac{2}{3}S_rV_{21}}{E_r} dx^\prime
              +\frac{1}{3} \int_{x_0}^x \frac{E_{r,x^\prime}}{E_r}
              dx^\prime \right)\\
         &=&
         \left[\frac{E_r(\tilde{z},x)}{E_r(\tilde{z},x_0)}\right]^{1/3}
         \mathrm{exp} \left(
              \int_{x_0}^x  \frac{y^2V_{21}-yA_r
                   +\frac{2}{3}S_rV_{21}}{E_r} dx^\prime \right),
      \end{eqnarray*}
      which implies (\ref{3.30}). \quad $\square$ \\

   Next, we derive Dubrovin-type equations which is first-order
   coupled systems of differential equations, and govern the dynamics
   of the zeros $\mu_j(x)$ and $\nu_j(x)$ of $E_r(\tilde{z},x)$
   and $F_r(\tilde{z},x)$
   with respect to $x$.

   \newtheorem{lem3.3}[lem3.1]{Lemma}
     \begin{lem3.3}\label{lemma3.3}
         Assume $(\ref{7})$ to hold in the stationary case.\\

        $(\mathrm{i})$
     Suppose the zeros $\{\mu_j(x)\}_{j=1,\ldots,r-5}$
     of $E_r(\tilde{z},x)$ remain distinct for $ x \in
     \Omega_\mu,$ where $\Omega_\mu \subseteq \mathbb{C}$ is open
     and connected. Then
     $\{\mu_j(x)\}_{j=1,\ldots,r-5}$ satisfy the system of
     differential equations,
    \begin{equation}\label{3.31}
       \mu_{j,x}(x)=-
         \frac{[S_r(\mu_j(x))+3y(\hat{\mu}_j(x))^2]V_{21}(\mu_j(x),x)}
            {u
            \prod_{\scriptstyle k=1 \atop \scriptstyle k \neq j }^{r-5}
            (\mu_j(x)-\mu_k(x))}, \quad j=1,\ldots,r-5,
        \end{equation}
   with initial conditions
       \begin{equation}\label{3.31.1}
         \{\hat{\mu}_j(x_0)\}_{j=1,\ldots,r-5}
         \in \mathcal{K}_{r-2},
       \end{equation}
   for some fixed $x_0 \in \Omega_\mu$. The initial value
   problem $(\ref{3.31})$, $(\ref{3.31.1})$ has a unique solution
   satisfying
        \begin{equation}\label{3.31.2}
         \hat{\mu}_j \in C^\infty(\Omega_\mu,\mathcal{K}_{r-2}),
         \quad j=1,\ldots,r-5.
        \end{equation}

   $(\mathrm{ii})$
     Suppose the zeros $\{\nu_j(x)\}_{j=1,\ldots,r-3}$
     of $F_r(\tilde{z},x)$ remain distinct for $ x \in
     \Omega_\nu,$ where $\Omega_\nu \subseteq \mathbb{C}$ is open
     and connected. Then
     $\{\nu_j(x)\}_{j=1,\ldots,r-3}$ satisfy the system of
     differential equations,
    \begin{eqnarray}\label{3.32}
      \nu_{j,x}(x)&=&
       \nu_j(x)^{2}\frac{[S_r(\nu_j(x))+3y(\hat{\nu}_j(x))^2]V_{31}(\nu_j(x),x)
            +m(x)J_r(\nu_j(x),x)}
       {uu_x^2
            \prod_{\scriptstyle k=1 \atop \scriptstyle k \neq j }^{r-3}
            (\nu_j(x)-\nu_k(x))}, \nonumber \\
       && ~~~~~~~~~~~~~~~~~~~~~~~~~~~~~~~~~~~
                    j=1,\ldots,r-3
    \end{eqnarray}
    with initial conditions
       \begin{equation}\label{3.32.1}
         \{\hat{\nu}_j(x_0)\}_{j=1,\ldots,r-3}
         \in \mathcal{K}_{r-2},
       \end{equation}
   for some fixed $x_0 \in \Omega_\nu$. The initial value
   problem $(\ref{3.32})$, $(\ref{3.32.1})$ has a unique solution
   satisfying
        \begin{equation}\label{3.32.2}
         \hat{\nu}_j \in C^\infty(\Omega_\nu,\mathcal{K}_{r-2}),
         \quad j=1,\ldots,r-3.
        \end{equation}
     \end{lem3.3}
 \textbf{Proof.}~~From (\ref{3.11}) and (\ref{3.12}), substituting
       $\tilde{z}=\mu_j(x)$ and $\nu_j(x)$ respectively, we
       have
       \begin{equation}\label{3.33}
           V_{21}^2(\mu_j(x),x)S_r(\mu_j(x))
             -V_{21}(\mu_j(x),x)B_r(\mu_j(x),x)+A_r^2(\mu_j(x),x)=0,
       \end{equation}
       \begin{equation}\label{3.34}
        V_{31}^2(\nu_j(x),x)S_r(\nu_j(x))
             -V_{31}(\nu_j(x),x)D_r(\nu_j(x),x)+C_r^2(\nu_j(x),x)=0.
       \end{equation}
  Then it is easy to get
        \begin{eqnarray}\label{3.35}
          B_r(\mu_j(x),x)&=&V_{21}(\mu_j(x),x)S_r(\mu_j(x))
             +\frac{A_r^2(\mu_j(x),x)}{V_{21}(\mu_j(x),x)}\nonumber \\
          &=& [S_r(\mu_j(x))+y(\hat{\mu}_j(x))^2]V_{21}(\mu_j(x),x),
        \end{eqnarray}
         \begin{eqnarray}\label{3.36}
          D_r(\nu_j(x),x)&=&V_{31}(\nu_j(x),x)S_r(\nu_j(x))
             +\frac{C_r^2(\nu_j(x),x)}{V_{31}(\nu_j(x),x)}\nonumber \\
          &=& [S_r(\nu_j(x))+y(\hat{\nu}_j(x))^2]V_{31}(\nu_j(x),x).
         \end{eqnarray}
  Inserting (\ref{3.35}) and (\ref{3.36}) into (\ref{3.14})
  respectively, we obtain
         \begin{equation}\label{3.37}
           E_{r,x}(\mu_j(x),x)=[S_r(\mu_j(x))+3y(\hat{\mu}_j(x))^2]V_{21}(\mu_j(x),x),
         \end{equation}
         \begin{equation}\label{3.38}
            F_{r,x}(\nu_j(x),x)=[S_r(\nu_j(x))+3y(\hat{\nu}_j(x))^2]V_{31}(\nu_j(x),x)
            +m(x)J_r(\nu_j(x),x).
         \end{equation}
  On the other hand, derivatives of  (\ref{3.15}) and (\ref{3.16})
  with respect to $x$ are
         \begin{equation}\label{3.39}
            E_{r,x}|_{\tilde{z}=\mu_j(x)}=-u\mu_{j,x}(x)
            \prod_{\scriptstyle k=1 \atop \scriptstyle k \neq j }^{r-5}
            (\mu_j(x)-\mu_k(x)),
         \end{equation}
         \begin{equation}\label{3.40}
            F_{r,x}|_{\tilde{z}=\nu_j(x)}=uu_x^2 \nu_j(x)^{-2}\nu_{j,x}(x)
            \prod_{\scriptstyle k=1 \atop \scriptstyle k \neq j }^{r-3}
            (\nu_j(x)-\nu_k(x)).
         \end{equation}
  Comparing (\ref{3.37})-(\ref{3.40}) leads to  ({\ref{3.31}) and
  (\ref{3.32}).
   \quad $\square$

\newtheorem{them3.001}[lem3.1]{Remark}
\begin{them3.001}
In Lemma $\ref{lemma3.3}$, we assume that $\{\mu_j(x)\}_{
j=1,\ldots,r-5} $ are pairwise distinct. However, if two or more of
$\{\mu_j(x)\}_{j=1,\ldots,r-5}$ coincide at $x=x_0,$ the
Dubrovin-type equation $(\ref{3.31})$ is ill-defined and the
stationary algorithm breaks down at such value of $x.$ Moreover,
$\theta(\underline{\tilde{z}}(P,\underline{\hat{\mu}}(x)))
 =\theta(\underline{\tilde{z}}(P,\underline{\hat{\nu}}(x)))\equiv 0. $
Therefore,
when attempting to solve the Dubrovin-type equation $(\ref{3.31})$,
they must be augmented with appropriate divisor
$\mathcal{D}_{\underline{\hat{\mu}}(x_0)}\in\sigma^{r-2}\mathcal{K}_{r-2}$
as initial conditions. The similar analysis holds for
$\{\nu_j(x)\}_{j=1,\ldots,r-3}$.
\end{them3.001}

\section{Stationary algebro-geometric solutions}

     In this section we continue our study of the stationary DP
     hierarchy, and will obtain explicit Riemann theta function
     representations for the meromorphic function $\phi$, the
     Baker-Akhiezer function $\psi_2$, and the
     algebro-geometric solutions $u$ for the stationary DP hierarchy.

  \newtheorem{lem4.1}{Lemma}[section]
   \begin{lem4.1}\label{lem4.1}
     Let $x \in \mathbb{C}$.

     $\mathrm{(i)}$
     Near $P_{\infty_1} \in \mathcal{K}_{r-2}$,
     in terms of the local coordinate $\zeta=\tilde{z}^{-1}, $ we
     have
        \begin{equation}\label{4.1}
          \phi(P,x) \underset{\zeta\rightarrow 0}{=}
          \frac{1}{\zeta} \sum_{j=0}^\infty \kappa_j(x)
          \zeta^j \quad \textrm{as $P\rightarrow P_{\infty_1},$}
        \end{equation}
    where
    \begin{eqnarray}
       &&\kappa_0= \frac{u_{x}(x)}{u(x)}, \qquad \kappa_1=0,\label{4.2}\\
      && \kappa_{2,xx}+3\left(\kappa_{0,x}\kappa_2+\kappa_0\kappa_{2,x}\right)+3\kappa_0^2\kappa_2-\kappa_2+m
         =\frac{m_x}{m}\left(\kappa_{2,x}+2\kappa_0\kappa_2\right),\nonumber\\\label{4.2001}\\
       &&\kappa_3=0, \quad\ldots\ldots\nonumber\\
     &&  \kappa_{2\varsigma,xx}+3\sum_{i=0}^{\varsigma}
     \kappa_{2i}\kappa_{2\varsigma-2i,x}+\sum_{i=0}^{\varsigma}
     \sum_{\ell=0}^{\varsigma-i}\kappa_{2i}
          \kappa_{2\ell}\kappa_{2\varsigma-2i-2\ell}
          -\kappa_{2\varsigma}\nonumber\\
       &&~~~~~~~~~    =\frac{m_x}{m}\left(\sum_{i=0}^{\varsigma}\kappa_{2i}\kappa_{2\varsigma-2i}
          \right),\label{4.2002}\\
             &&   \kappa_{2\varsigma+1}=0,
             \quad \varsigma \geq 2,\quad \varsigma \in \mathbb{N}.\label{4.2003}
       \end{eqnarray}

   $\mathrm{(ii)}$
        Near $P_0\in\mathcal{K}_{r-2},$
        in terms of the local coordinate $\zeta=\tilde{z}^{\frac{1}{3}},$ we have
        \begin{equation}\label{4.2a}
          \phi(P,x) \underset{\zeta\rightarrow 0}{=}
           \sum_{j=0}^\infty \iota_j(x)
          \zeta^{j+1} \quad \textrm{as $P\rightarrow P_{0},$}
        \end{equation}
        where
        \begin{eqnarray}\label{4.2b}
        &&\iota_0=-m^{\frac{1}{3}},\qquad \iota_1=0,\nonumber\\
        &&\iota_2=\frac{(m_x/m)\iota_0^2-3\iota_0\iota_{0,x}}{3\iota_0^2}=0,\qquad \iota_3=0,\nonumber\\
        &&\iota_4=\frac{\frac{m_x}{m}\iota_{0,x}+\iota_0-\iota_{0,xx}}{3\iota_0^2}, \qquad \iota_5=0,\nonumber\\
        &&\iota_6=\frac{\frac{m_x}{m}(2\iota_0\iota_4-1)-3(\iota_{0,x}\iota_4+\iota_0\iota_{4,x})}{3\iota_0^2},\qquad \iota_7=0,\\
        &&\ldots\ldots\nonumber\\
        &&\iota_{2\varsigma}=\frac{\frac{m_x}{m}(2\iota_0\iota_{2\varsigma-2}+\iota_{2\varsigma-4,x})
        +\iota_{2\varsigma-4}-\iota_{2\varsigma-4,xx}-3(\iota_0\iota_{2\varsigma-2,x}+\iota_{0,x}
        \iota_{2\varsigma-2})}{3\iota_0^2},\nonumber\\
        &&\iota_{2\varsigma+1}=0,\qquad \varsigma\geq4,\qquad \varsigma\in\mathbb{N}.\label{4.2b0}
        \end{eqnarray}

        \end{lem4.1}
    \textbf{Proof.}~~The existence of these asymptotic expansions
     (\ref{4.1}) and (\ref{4.2a}) in terms of local coordinates
     $\zeta=\tilde{z}^{-1}$ near $P_{\infty_1}$ and
     $\zeta=\tilde{z}^{\frac{1}{3}}$ near $P_0$ is clear from the
     explicit form of $\phi$ in (\ref{3.8}).
     Insertion of the polynomials
     $V_{ij}~(i,j=1,2,3)$ then, in principle, yields the explicit expansion
     coefficients in (\ref{4.1}) and (\ref{4.2a}). For example, $\kappa_0=u_x(x)/u(x)$
     and $\kappa_1=0$ in (\ref{4.2}).
     However, this is a cumbersome procedure, especially with regard to
     the next to leading coefficients in (\ref{4.1}). Much more efficient
     is the actual computation of these coefficients utilizing the Riccati-type
     equation (\ref{3.23}). Indeed, in
     terms of the local coordinate $\zeta=\tilde{z}^{\frac{1}{3}},$
     then (\ref{3.23}) can be written as
          \begin{equation}\label{4.3c}
          \begin{split}
         &\phi_{xx}(P,x)+3\zeta^{-3}\phi(P,x)\phi_x(P,x)+\zeta^{-6}\phi^3(P,x)-\frac{m_x}{m}\phi_x(P,x)\\
         &-\phi(P,x)-\zeta^{-3}\frac{m_x}{m}\phi^2(P,x)+m\zeta^{-3}+\frac{m_x}{m}\zeta^3=0\\
         \end{split}
         \end{equation}
     near the point $P_0$. Substituting a power series ansatz
          \begin{equation*}
          \phi \underset{\zeta\rightarrow 0}{=}
            \sum_{j=0}^\infty \iota_j(x)
           \zeta^{j+1} \quad \textrm{as $P\rightarrow P_{0},$}
          \end{equation*}
     into (\ref{4.3c}) and comparing the same powers of $\zeta$,
     then yields (\ref{4.2b}).

    Similarly, in terms of the local coordinate $\zeta=\tilde{z}^{-1},$
    equation (\ref{3.23}) can be written as
          \begin{eqnarray}\label{4.3}
            &&\phi_{xx}(P,x)+3\zeta\phi(P,x)\phi_x(P,x)
             +\zeta^{2}\phi^3(P,x)
              -\phi(P,x)-\frac{m_x(x)}{m(x)}\phi_x(P,x)\nonumber\\
              &&~~~~-\zeta\frac{m_x(x)}{m(x)}\phi^2(P,x)
                +m(x)\zeta+\frac{m_x(x)}{m(x)}\zeta^{-1}=0
          \end{eqnarray}
    near the point $P_{\infty_1}.$ Substituting a power series ansatz
         \begin{equation*}
          \phi \underset{\zeta\rightarrow 0}{=}
          \frac{1}{\zeta} \sum_{j=0}^\infty \kappa_j(x)
          \zeta^j \quad \textrm{as $P\rightarrow P_{\infty_1},$}
        \end{equation*}
    into (\ref{4.3}) and comparing the same powers of $\zeta$,
    then yields the indicated Laurent series relations (\ref{4.2001})
    and (\ref{4.2002}).  Finally, (\ref{4.2003}) and
    (\ref{4.2b0}) arise from the technical treatment in section 2
    ($z=\tilde{z}^2,$ see (\ref{2.19000})).
        \quad $\square$

        \newtheorem{rem1987}[lem4.1]{Remark}
        \begin{rem1987}
        We have derived the explicit expressions for $\kappa_0,
        \kappa_{2\varsigma+1}, \varsigma\in\mathbb{N}_0$ in Lemma $\ref{lem4.1}~
        $. However, the coefficients $\kappa_{2\varsigma},
        \varsigma\in\mathbb{N}$
        in the high-energy expansion of $\phi$ are still implicit,
        since \emph{(\ref{4.2001})} and $(\ref{4.2002})$ involve the
        $x$-derivatives of $\kappa_{2\varsigma},\varsigma\in\mathbb{N}$ and
        hence yields a series of second order ODEs \emph{(}or PDEs
        in time-dependent case\emph{)} with variable coefficients.
        In the process of solving other
        integrable evolution equations such as classical Thirring
        system\emph{(}near the points $P_{0,\pm},$ see \emph{\cite{14}}\emph{)},
        Camassa-Holm hierarchy \emph{(}near the points
        $P_{\infty\pm}$, see \emph{\cite{13,14}}\emph{)}, if we directly insert a ansatz into a Riccati-type equation,
        analogous problem will arise.
        The DP hierarchy shares some similarities
        with the CH hierarchy at this point. Since the
        concrete expressions $\kappa_j, j\geq2, j\in\mathbb{N}$
        are useless in the process of finding the algebro-geometric
        solutions of DP hierarchy,
        we are not intend to take effort to write out their explicit
        forms from \emph{(\ref{3.8})}.
        \end{rem1987}

     We assume
     $\mathcal{K}_{r-2}$ to be nonsingular for the remainder of this
     section. We now introduce the holomorphic differentials $\eta_l(P)$ on
   $\mathcal{K}_{r-2}$ defined by
     \begin{equation}\label{4.23}
        \eta_l(P)=\frac{1}{3y(P)^2+S_r(\tilde{z})}
          \begin{cases}
             \tilde{z}^{l-1} d \tilde{z}, \quad 1 \leq l \leq
             8n+5,\\
              y(P)\tilde{z}^{l-8n-6}d \tilde{z}, \quad 8n+6 \leq l
               \leq 12n+7.
          \end{cases}
     \end{equation}
      and choose an
     appropriate fixed homology basis $\{a_j, b_j\}_{j=1}^{r-2}$ on $\mathcal{K}_{r-2}$
     in such a way that the intersection matrix of cycles satisfies
     $$a_j\circ b_k=\delta_{j,k},\quad a_j\circ a_k=0,\quad b_j\circ b_k=0,\quad j,k=1,\ldots, r-2. $$
    Define an invertible matrix $E \in GL(r-2, \mathbb{C})$ as
    follows
       \begin{equation}\label{4.24}
          \begin{split}
        & E=(E_{j,k})_{(r-2) \times (r-2)}, \quad E_{j,k}=
           \int_{a_k} \eta_j, \\
        &  \underline{e}(k)=(e_1(k),\ldots, e_{r-2}(k)), \quad
           e_j(k)=(E^{-1})_{j,k},
           \end{split}
       \end{equation}
    and the normalized holomorphic differentials
        \begin{equation}\label{4.25}
          \omega_j= \sum_{l=1}^{r-2} e_j(l)\eta_l, \quad
          \int_{a_k} \omega_j = \delta_{j,k}, \quad
          \int_{b_k} \omega_j= \Gamma_{j,k}, \quad
          j,k=1, \ldots ,r-2.
        \end{equation}
    One can see that the matrix $\Gamma=(\Gamma_{i,j})_{(r-2)\times(r-2)}$ is symmetric, and it has a
    positive-definite imaginary part.

     Next, choosing a convenient base point $Q_0 \in
     \mathcal{K}_{r-2} \setminus \{P_{\infty_1},P_0\}$, the vector of Riemann
     constants $\underline{\Xi}_{Q_0}$ is given by (A.45) \cite{14}, and the Abel maps
      $\underline{A}_{Q_0}(\cdot) $ and
      $\underline{\alpha}_{Q_0}(\cdot)$ are defined by
         \begin{eqnarray*}
           \underline{A}_{Q_0}:\mathcal{K}_{r-2} \rightarrow
           J(\mathcal{K}_{r-2})&=&\mathbb{C}^{r-2}/L_{r-2},
         \end{eqnarray*}
         \begin{eqnarray*}
          P \mapsto \underline{A}_{Q_0} (P)&=& (A_{Q_0,1}(P),\ldots,
           A_{Q_0,r-2} (P)) \\
           &=&\left(\int_{Q_0}^P\omega_1,\ldots,\int_{Q_0}^P\omega_{r-2}\right)
           (\mathrm{mod}~L_{r-2}),
         \end{eqnarray*}
     and
         \begin{eqnarray*}
          && \underline{\alpha}_{Q_0}:
          \mathrm{Div}(\mathcal{K}_{r-2}) \rightarrow
          J(\mathcal{K}_{r-2}),\\
          && \mathcal{D} \mapsto \underline{\alpha}_{Q_0}
          (\mathcal{D})= \sum_{P\in \mathcal{K}_{r-2}}
           \mathcal{D}(P)\underline{A}_{Q_0} (P),
         \end{eqnarray*}
    where $L_{r-2}=\{\underline{z}\in \mathbb{C}^{r-2}|
           ~\underline{z}=\underline{N}+\Gamma\underline{M},
           ~\underline{N},~\underline{M}\in \mathbb{Z}^{r-2}\}.$

    For brevity, define the function
      $\underline{z}:\mathcal{K}_{r-2} \times
      \sigma^{r-2}\mathcal{K}_{r-2} \rightarrow \mathbb{C}^{r-2}$ by\footnote{$\sigma^{r-2}\mathcal{K}_{r-2}$=
      $\underbrace{\mathcal{K}_{r-2}\times\ldots\times\mathcal{K}_{r-2}}_{r-2}.$}
     \begin{eqnarray}\label{4.4}
           \underline{z}(P,\underline{Q})&=& \underline{\Xi}_{Q_0}
           -\underline{A}_{Q_0}(P)+\underline{\alpha}_{Q_0}
             (\mathcal{D}_{\underline{Q}}), \nonumber \\
           P\in \mathcal{K}_{r-2},\,
           \underline{Q}&=&(Q_1,\ldots,Q_{r-2})\in
           \sigma^{r-2}\mathcal{K}_{r-2},
         \end{eqnarray}
     here $\underline{z}(\cdot,\underline{Q}) $ is
     independent of the choice of base point $Q_0$.
     The Riemann theta
     function $\theta(\underline{z})$ associated with $\mathcal{K}_{r-2}$ and the homology is
      defined by
     $$\theta(\underline{z})=\sum_{\underline{n}\in\mathbb{Z}}\exp\left(2\pi i<\underline{n},\underline{z}>+\pi i<\underline{n},\underline{n}\Gamma>\right),\quad \underline{z}\in\mathbb{C}^{r-2},$$
     where $<\underline{B},\underline{C}>=\overline{\underline{B}}\cdot\underline{C}^t=\sum_{j=1}^{r-2}\overline{B}_jC_j$
     denotes the scalar product in $\mathbb{C}^{r-2}$.

       The normalized differential $\omega_{P_{\infty_1} P_0}^{(3)}(P)$
       of the third kind is the unique differential holomorphic on
       $\mathcal{K}_{r-2} \setminus \{P_{\infty_1},P_0\}$ with simple
       poles at $P_{\infty_1}$ and $P_0$ with residues $\pm 1$,
       respectively, that is,
         \begin{eqnarray}\label{4.5}
            \begin{split}
              & \omega_{P_{\infty_1} P_0}^{(3)}(P) \underset
              {\zeta \rightarrow 0}{=} (\zeta^{-1}+O(1))d \zeta,
              \quad \textrm{as $P \rightarrow P_{\infty_1},$}\\
              & \omega_{P_{\infty_1} P_0}^{(3)}(P) \underset
              {\zeta \rightarrow 0}{=} (-\zeta^{-1}+O(1))d \zeta,
              \quad \textrm{as $P \rightarrow P_0.$}
            \end{split}
         \end{eqnarray}
       In particular,
         $$\int_{a_j}\omega_{P_{\infty_1} P_0}^{(3)}(P)=0,\quad j=1,\ldots,r-2.$$
        Then
            \begin{eqnarray} \label{4.6}
              \begin{split}
               & \int_{Q_0}^P \omega_{P_{\infty_1} P_0}^{(3)}(P) \underset
                  {\zeta \rightarrow 0}{=} \mathrm{ln} \zeta +
                  e^{(3)}(Q_0)+O(\zeta),
               \quad \textrm{as $P \rightarrow P_{\infty_1},$}\\
              & \int_{Q_0}^P \omega_{P_{\infty_1} P_0}^{(3)}(P) \underset
                  {\zeta \rightarrow 0}{=} -\mathrm{ln} \zeta +
                  e^{(3)}(Q_0)+O(\zeta),
               \quad \textrm{as $P \rightarrow P_0,$}
              \end{split}
            \end{eqnarray}
     where $e^{(3)}(Q_0)$ is an integration constant.

     The theta function representation of $\phi(P,x)$ then reads as
     follows.

     \newtheorem{the4.2}[lem4.1]{Theorem}
     \begin{the4.2}\label{them4.30}
        Assume that the curve $\mathcal{K}_{r-2}$ is nonsingular.
        Let $P=(\tilde{z},y) \in \mathcal{K}_{r-2} \setminus
        \{P_{\infty_1},  P_0\}$ and let $x,x_0 \in \Omega_\mu$, where
        $\Omega_\mu \subseteq \mathbb{C}$ is open and connected.
        Suppose that $\mathcal{D}_{\underline{\hat{\mu}}(x)}$, or
        equivalently, $\mathcal{D}_{\underline{\hat{\nu}}(x)}$ is
        nonspecial\footnote{The definition of a nonspecial divisor see \cite{11b}.} for $x \in \Omega_\mu$. Then
        \begin{equation}\label{4.7}
            \phi(P,x)=-m^{\frac{1}{3}}(x)\frac{\theta(\underline{\tilde{z}}(P,\underline{\hat{\nu}}(x)))
            \theta(\underline{\tilde{z}}(P_{0},\underline{\hat{\mu}}(x)))}
            {\theta(\underline{\tilde{z}}(P_{0},\underline{\hat{\nu}}(x)))
            \theta(\underline{\tilde{z}}(P,\underline{\hat{\mu}}(x)))}
            \mathrm{exp}\left(e^{(3)}(Q_0)
            -\int_{Q_0}^P \omega_{P_{\infty_1} P_0}^{(3)}\right).
         \end{equation}
     \end{the4.2}
    \textbf{Proof.}~~Let $\Phi$ be defined by the right-hand side of
    (\ref{4.7}) with the aim to prove that $\phi=\Phi$. From
    (\ref{4.6}) it follows that
         \begin{equation}\label{4.8}
           \begin{split}
             &\mathrm{exp}\left(e^{(3)}(Q_0)
                -\int_{Q_0}^P \omega_{P_{\infty_1} P_0}^{(3)}\right)
                 \underset{\zeta \rightarrow 0}{=}\zeta^{-1}+O(1),
                \quad \textrm{as $P \rightarrow P_{\infty_1},$}\\
             & \mathrm{exp}\left(e^{(3)}(Q_0)
                -\int_{Q_0}^P \omega_{P_{\infty_1} P_0}^{(3)}\right)
                 \underset{\zeta \rightarrow 0}{=}\zeta+O(\zeta^2),
                \quad \textrm{as $P \rightarrow P_0.$}\\
           \end{split}
           \end{equation}
     Using (\ref{3.19}) we immediately know that $\phi$ has simple
     poles at $\underline{\hat{\mu}}(x)$ and $P_{\infty_1}$, and simple
     zeros at $P_0$ and $\underline{\hat{\nu}}(x)$. By (\ref{4.7})
     and a special case of Riemann's vanishing theorem \cite{11b, 14, 14.0}, we see that
     $\Phi$ shares the same properties. Hence, using the Riemann-Roch
     theorem (\cite{11b, 14, 14.0})
     yields that the holomorphic function $\Phi/\phi=\gamma$,
     where $\gamma$ is a constant with respect to $P$.
     Finally, considering the asymptotic expansion of $\Phi$ and $\phi$ near
     $P_0$, we obtain
       \begin{equation}\label{4.9}
             \frac{\Phi}{\phi}\underset{\zeta \rightarrow 0}{=}
              \frac{-m^{1/3}(1+O(\zeta))(\zeta+O(\zeta^2))}{-m^{1/3}\zeta+O(\zeta^2)}
               \underset{\zeta \rightarrow 0}{=}1+O(\zeta),
              \quad \textrm{as $P \rightarrow  P_{0},$}
           \end{equation}
     from which we conclude that $\gamma=1$, where we used
     (\ref{4.8}) and (\ref{4.2a}). Hence, we prove (\ref{4.7}).
      \quad $\square$ \\

    Furthermore, let $\omega_{P_0,3}^{(2)}(P)$ denote
    the normalized differential
    of the second kind  which is holomorphic on
    $\mathcal{K}_{r-2}\setminus \{P_0\}$
    with a pole of order $3$ at $P_0$,
         \begin{equation*}
         \omega_{P_0,3}^{(2)}(P)=\frac{\tilde{z}^{-1}d\tilde{z}}
    {3(3y(P)^2+S_r(\tilde{z}))}+\sum_{j=1}^{r-2}\lambda_j\eta_j(P)
      \underset{\zeta \rightarrow 0}{=}(\zeta^{-3}+O(1))d \zeta,
      \quad \textrm{as $P \rightarrow P_{0}$},
       \end{equation*}
    where the constants $\{\lambda_j\}_{j=1,\ldots,r-2} \in \mathbb{C}$
    are determined by the normalization condition
    $$\int_{a_j}\omega_{P_0,3}^{(2)}(P)=0,\quad j=1,\ldots,r-2,$$
    and the differentials
     $\{\eta_j(P)\}_{j=1,\ldots,r-2}$ (defined in (\ref{4.23}))
    form a basis for the space of holomorphic
    differentials.
    Moreover, we define the vector of $b$-periods of
      $\omega_{P_0,3}^{(2)}$,
     \begin{equation}\label{6.32a0}
        \hat{\underline{U}}_{3}^{(2)}=(\hat{U}_{3,1}^{(2)},
        \ldots,\hat{U}_{3,r-2}^{(2)}) ,
        \quad \hat{U}_{3,j}^{(2)}=
         \frac{1}{2\pi i}\int_{b_j}\omega_{P_0,3}^{(2)}, \quad j=1,\ldots,r-2.
      \end{equation}
       Then
         \begin{equation*}
          \int_{Q_0}^P \omega_{P_0,3}^{(2)}(P)
         \underset{\zeta \rightarrow 0}{=}-\frac{1}{2} \zeta^{-2}
         +e_3^{(2)}(Q_0)   + O(\zeta),  \quad \textrm{as $P \rightarrow P_{0}$},
         \end{equation*}
          \begin{equation*}
          \int_{Q_0}^P \omega_{P_0,3}^{(2)}(P)
         \underset{\zeta \rightarrow 0}{=}
         e_3^{(2)}(Q_0)   +  f_3^{(2)}(Q_0)\zeta^2+ O(\zeta^4) ,\quad \textrm{as $P \rightarrow P_{\infty_1}$},
         \end{equation*}
      where $e_3^{(2)}(Q_0)$,  $f_3^{(2)}(Q_0)$ are integration constants.

     Similarly, the theta function representation of the
     Baker-Akhiezer function $\psi_2(P,x,x_0)$ is summarized in the
     following theorem.

     \newtheorem{the4.3}[lem4.1]{Theorem}
       \begin{the4.3}\label{them4.40}
         Assume that the curve $\mathcal{K}_{r-2}$ is nonsingular.
        Let $P=(\tilde{z},y) \in \mathcal{K}_{r-2} \setminus
        \{P_{\infty_1}, P_0\}$ and let $x,x_0 \in \Omega_\mu$, where
        $\Omega_\mu \subseteq \mathbb{C}$ is open and connected.
        Suppose that $\mathcal{D}_{\underline{\hat{\mu}}(x)}$, or
        equivalently, $\mathcal{D}_{\underline{\hat{\nu}}(x)}$ is
        nonspecial for $x \in \Omega_\mu$. Then
           \begin{eqnarray}\label{4.10}
             \psi_2(P,x,x_0)&=&\frac
        {\theta(\underline{\tilde{z}}(P,\underline{\hat{\mu}}(x)))
        \theta(\underline{\tilde{z}}(P_{0},\underline{\hat{\mu}}(x_0)))}
        {\theta(\underline{\tilde{z}}(P_{0},\underline{\hat{\mu}}(x)))
        \theta(\underline{\tilde{z}}(P,\underline{\hat{\mu}}(x_0)))} \\
         &&
        \times ~
        \mathrm{exp} \Bigg( \int_{x_0}^x 2m^{\frac{1}{3}}(x^\prime) dx^\prime
       \Big( \int_{Q_0}^P
       \omega_{P_0,3}^{(2)}-e_3^{(2)}(Q_0)\Big)\Bigg).
         \nonumber
          \end{eqnarray}
        \end{the4.3}
    \textbf{Proof.}~~Assume temporarily that
          \begin{equation}\label{4.11}
             \mu_j(x) \neq \mu_k(x),
             \quad \textrm{ for $j \neq k$ and $x \in \widetilde{\Omega}_\mu
             \subseteq \Omega_\mu $},
          \end{equation}
    where $\widetilde{\Omega}_\mu$ is open and connected. For the
    Baker-Akhiezer function $\psi_2$ we will use the same strategy
    as was used in the previous proof. Let $\Psi$ denote the
    right-hand side of (\ref{4.10}). We intend to prove $\psi_2 =
    \Psi$. For that purpose we first investigate the
    local zeros and poles of $\psi_2$. Since
          \begin{equation}\label{4.12}
         \psi_2(P,x,x_0)=\mathrm{exp}\left(\tilde{z}^{-1}\int_{x_0}^x
         \phi(P,x^\prime) dx^\prime \right),
         \end{equation}
    we can see that the zeros and poles of $\psi_2$ can come only
    from simple poles in the integrand (with positive and negative
    residues respectively).
     By using the definition
    (\ref{3.8}) of $\phi$, (\ref{3.14}) and the Dubrovin equations
    (\ref{3.31}), we obtain
        \begin{eqnarray} \label{4.13}
           \phi(P,x)&=& \tilde{z}\frac{y^2V_{21}-yA_r+B_r}{E_r}
           \nonumber \\
           &=&\tilde{z}\frac{y^2V_{21}-yA_r+\frac{2}{3}V_{21}S_r+\frac{1}{3}E_{r,x}}
             {E_r} \nonumber \\
           &=& \tilde{z}\left(\frac{1}{3}V_{21}\frac{3y^2+S_r}{E_r}
               + \frac{1}{3}\frac{E_{r,x}}{E_r}
              +\frac{1}{3}\frac{-3yA_r+V_{21}S_r}{E_r}\right)
              \nonumber \\
           &=& \tilde{z}\left(\frac{2}{3}V_{21}\frac{3y^2+S_r}{E_r}
               +\frac{1}{3}\frac{E_{r,x}}{E_r}
               -\frac{V_{21}y(y+\frac{A_r}{V_{21}})}{E_r}\right).
        \end{eqnarray}
   Hence
         \begin{eqnarray}\label{4.14}
   \phi(P,x)&=& \mu_j\left(-\frac{2}{3}\frac{\mu_{j,x}}{\tilde{z}-\mu_j}
               -\frac{1}{3}\frac{\mu_{j,x}}{\tilde{z}-\mu_j}
               +O(1)\right) \nonumber \\
           &=&
              -\mu_j\frac{\mu_{j,x}}{\tilde{z}-\mu_j}+O(1)
              \quad \textrm{as $\tilde{z} \rightarrow \mu_j(x)$},
        \end{eqnarray}
   where
    $$y \rightarrow
    y(\hat{\mu}_j(x))=-\frac{A_r(\mu_j(x))}{V_{21}(\mu_j(x))}, \quad
     \textrm{as $\tilde{z} \rightarrow \mu_j(x)$}.$$
   More concisely,
         \begin{equation}\label{4.15}
          \phi(P,x)=\mu_j(x)\frac{\partial}{\partial x}\mathrm{ln}(\tilde{z}-\mu_j(x))
           +O(1),\quad \textrm{for $P$ near $\hat{\mu}_j(x)$},
         \end{equation}
  which together with (\ref{4.12}) yields
       \begin{eqnarray}\label{4.16}
         \psi_2(P,x,x_0)&=& \mathrm{exp} \left( \int_{x_0}^x dx^\prime
          \left(\frac{\partial}{\partial x^\prime}\mathrm{ln}(\tilde{z}-\mu_j(x^\prime))
           +O(1)\right)\right) \nonumber \\
          &=& \frac{\tilde{z}-\mu_j(x)}{\tilde{z}-\mu_j(x_0)}O(1)
          \nonumber \\
          &=&
            \begin{cases}
            (\tilde{z}-\mu_j(x))O(1) \quad \textrm{for $P$ near
                     $\hat{\mu}_j(x) \neq \hat{\mu}_j(x_0)$},\\
            O(1) \quad \textrm{for $P$ near
                     $\hat{\mu}_j(x)=\hat{\mu}_j(x_0)$},\\
            (\tilde{z}-\mu_j(x_0))^{-1}O(1) \quad \textrm{for $P$ near
                     $\hat{\mu}_j(x_0) \neq \hat{\mu}_j(x)$},
            \end{cases}
       \end{eqnarray}
   where $O(1) \neq 0$ in (\ref{4.16}). Consequently, all zeros and
   poles of $\psi_2$ and $\Psi$ on $\mathcal{K}_{r-2} \setminus
   \{P_{\infty_1}, P_0\}$ are simple and coincident. It remains to identify
   the behavior of $\psi_2$ and $\Psi$ near $P_{\infty_1}$  and $P_0$.

   (i) Near $P_{\infty_1}$: from (\ref{4.1}), we infer
          \begin{equation}\label{4.17}
           \exp\left(\tilde{z}^{-1}\int_{x_0}^x dx^\prime \phi(P,x^\prime)\right)
             \underset{\zeta \rightarrow 0}{=}
             1+\int_{x_0}^x\kappa_0(x^\prime)dx^\prime+O(\zeta^2)\quad
               \textrm{as $P \rightarrow P_{\infty_1}$}.
          \end{equation}
   Taking into account the expression (\ref{4.10}) for $\Psi$, then
   shows that  $\psi_2$ and $\Psi$ have identical exponential behavior
   near $P_{\infty_1}$.

   (ii) Near $P_0$: from (\ref{4.2a}), we arrive at
          \begin{equation*}
           \tilde{z}^{-1}\int_{x_0}^x dx^\prime \phi(P,x^\prime)
             \underset{\zeta \rightarrow 0}{=}
              -\int_{x_0}^{x}m^{\frac{1}{3}}(x^\prime)dx^\prime\left(\zeta^{-2}+O(\zeta^2)\right)\quad
               \textrm{as $P \rightarrow P_{0}$},
          \end{equation*}
   Taking into account the expression (\ref{4.10}) for $\Psi$, then
   shows that  $\psi_2$ and $\Psi$ have identical exponential behavior
   up to order $O(\zeta^2)$ near $P_0$.

   The uniqueness result
   for Baker-Akhiezer functions (\cite{12,14,14.0,15.0}) then completes the proof
   $\psi_2=\Psi$ as both functions share the same singularities and
   zeros. The extension of this result from $x \in
   \widetilde{\Omega}_\mu$ to $x \in   \Omega_\mu$ then simply
   follows from the continuity of $\underline{\alpha}_{Q_0}$ and the
   hypothesis of $\mathcal{D}_{\underline{\hat{\mu}}(x)}$ being
   nonspecial for $x \in \Omega_\mu$. \quad $\square$ \\

   The asymptotic behavior of $y(P)$ and $S_r$ near $P_{\infty_1}$
   are summarized as follows.

   \newtheorem{lem4.4}[lem4.1]{Lemma}
   \begin{lem4.4}
     \begin{equation}\label{4.18}
     y(P) \underset{\zeta \rightarrow 0}{=} -\frac{1}{3}\varrho
     \zeta^{-4n-3}(1+\alpha_0 \zeta^2+ \alpha_1 \zeta^4+
      O(\zeta^6)), \quad
     \textrm{as $P \rightarrow P_{\infty_1}$,}
     \end{equation}
     \begin{equation}\label{4.19}
       S_r \underset{\zeta \rightarrow 0}{=} -\frac{1}{3}
       \zeta^{-8n-6}(1+\beta_0 \zeta^2 +\beta_1 \zeta^4 +
       O(\zeta^6)), \quad \textrm{as $P \rightarrow P_{\infty_1}$,}
     \end{equation}
   where $\varrho= -3\aleph_1$ and $\aleph_1$ is the root of
   of algebraic equation $(\ref{r002})$ corresponding to the point $P_{\infty_1}\in\mathcal{K}_{r-2}.$
   \end{lem4.4}
   \textbf{Proof.}~~From (\ref{3.5}) and (\ref{3.6}), we arrive at
       \begin{equation}\label{4.20}
         y(P)=V_{21}\frac{(\tilde{z}^3-\phi\tilde{z}^2)}{m}+V_{22}\tilde{z}
           +V_{23}\phi.
       \end{equation}
    Then, in terms of the local coordinate $\zeta=\tilde{z}^{-1}$,
    insertion of (\ref{5}) and (\ref{4.1}) into (\ref{4.20}) yields
       \begin{eqnarray*}\label{4.21}
         &&y(P)=\frac{1}{m}\sum_{\ell=0}^n
         V_{21}^{(\ell)}(G_\ell)\zeta^{-4(n+1-\ell)}
         (\zeta^{-3}-\zeta^{-2}\sum_{j=0}^\infty \kappa_j
         \zeta^{j-1})  \nonumber \\
        && \qquad +\sum_{\ell=0}^n
         V_{22}^{(\ell)}(G_\ell)\zeta^{-4(n+1-\ell)-1}\nonumber
      \end{eqnarray*}
      \begin{eqnarray}
          && \qquad \qquad  +\sum_{\ell=0}^n
         V_{23}^{(\ell)}(G_\ell)\zeta^{-4(n+1-\ell)}
            \sum_{j=0}^\infty \kappa_j
         \zeta^{j-1} \nonumber \\
         &&\underset{\zeta \rightarrow 0}{=} -\frac{1}{3}\varrho
           \zeta^{-4n-3}(1+\alpha_0 \zeta^2+\alpha_1 \zeta^4+
            O(\zeta^6)), \quad
             \textrm{as $P \rightarrow P_{\infty_1}$.}
       \end{eqnarray}
   Similarly, we recall the definition of $S_r$,
       \begin{equation}\label{4.22}
         S_r=\tilde{z}^2(V_{11}V_{22}+V_{11}V_{33}+V_{22}V_{33}
         -V_{12}V_{21}-V_{13}V_{31}-V_{23}V_{32}),
       \end{equation}
  Insertion of (\ref{4.1}) into (\ref{4.22}) leads to (\ref{4.19}).
  \quad $\square$\\

    A straightforward Laurent expansion of (\ref{4.23}),
    (\ref{4.24}) and (\ref{4.25}) near $P_{\infty_1}$ yields the
    following results.

    \newtheorem{lem4.5}[lem4.1]{Lemma}
      \begin{lem4.5}
        Assume the curve $\mathcal{K}_{r-2}$ to be nonsingular. Then
        the vector of normalized holomorphic differentials $\underline{\omega}$ have the Laurent
        series
         \begin{equation}\label{4.26}
           \underline{\omega}=(\omega_1,\ldots,\omega_{r-2})
             \underset{\zeta \rightarrow
             0}{=}(\underline{\rho}_0+\underline{\rho}_1 \zeta
             +O(\zeta^2))d\zeta
         \end{equation}
        near $P_{\infty_1}$
        with
           \begin{eqnarray*}
             \underline{\rho}_0 &=& \frac{-3}{\varrho^2-1}\underline{e}(8n+5)
             +\frac{\varrho}{\varrho^2-1}\underline{e}(r-2),\\
            \underline{\rho}_1 &=& \frac{-3}{\varrho^2-1}\underline{e}(8n+4)
             +\frac{\varrho}{\varrho^2-1}\underline{e}(r-3),
           \end{eqnarray*}
      where $\varrho=-3\aleph_1$, $\aleph_1$ is given in Lemma
      $4.5$.
      \end{lem4.5}
   \textbf{Proof.}~~Using (\ref{4.18}) and (\ref{4.19}), the local coordinate
   $\zeta=\tilde{z}^{-1}$ near $P_{\infty_1}$, we obtain
        \begin{equation}\label{4.27}
          3y^2+S_r\underset{\zeta \rightarrow
          0}{=}\frac{1}{3}\zeta^{-8n-6}[\varrho^2-1+(2\varrho^2\alpha_0
          -\beta_0)\zeta^2+(2\varrho^2\alpha_1+\varrho^2\alpha_0^2-\beta_1)
          \zeta^4+O(\zeta^6)].
        \end{equation}
   Then
     \begin{eqnarray}\label{4.28}
      &&\frac{1}{3y^2+S_r}\underset{\zeta \rightarrow 0}{=}3\zeta^{8n+6}
      \Big[\frac{1}{\varrho^2-1}-\frac{2\varrho^2\alpha_0-\beta_0}{(\varrho^2-1)^2}
      \zeta^2+\Big(\frac{2\varrho^2\alpha_1+\varrho^2\alpha_0^2-\beta_1}
        {-(\varrho^2-1)^2}+ \nonumber \\
       && \qquad \qquad
       \frac{(2\varrho^2\alpha_0-\beta_0)^2}{(\varrho^2-1)^3}\Big)\zeta^4
       +O(\zeta^6) \Big].
      \end{eqnarray}
  From (\ref{4.23}), (\ref{4.25}) and (\ref{4.28}), we have
       \begin{eqnarray}\label{4.29}
          \omega_j&=& \sum_{l=1}^{r-2} e_j(l)\eta_l
             =\sum_{l=1}^{8n+5}e_j(l)
             \frac{\tilde{z}^{l-1}d\tilde{z}}{3y^2+S_r}
             +\sum_{l=8n+6}^{r-2} e_j(l)
             \frac{y\tilde{z}^{l-8n-6}d\tilde{z}}{3y^2+S_r}
              \nonumber  \\
          &=&-\sum_{l=1}^{8n+5}e_j(l)
             \frac{\zeta^{-l-1}d\zeta}{3y^2+S_r}
             -\sum_{l=8n+6}^{r-2} e_j(l)
             \frac{y\zeta^{-l+8n+4}d\zeta}{3y^2+S_r}
             \nonumber \\
          &   \underset{\zeta \rightarrow 0}{=}&
           -\sum_{l=1}^{8n+5}3e_j(l)\zeta^{-l+8n+5}
       \Big[\frac{1}{\varrho^2-1}-\frac{2\varrho^2\alpha_0-\beta_0}{(\varrho^2-1)^2}
        \zeta^2+\Big(\frac{2\varrho^2\alpha_1+\varrho^2\alpha_0^2-\beta_1}
           {-(\varrho^2-1)^2} \nonumber \\
        && \quad
         +\frac{(2\varrho^2\alpha_0-\beta_0)^2}{(\varrho^2-1)^3}\Big)\zeta^4
           +O(\zeta^6) \Big] d\zeta + \sum_{l=8n+6}^{r-2} \varrho e_j(l)
           \zeta^{-l+r-2}
           \Big[\frac{1}{\varrho^2-1}- \nonumber \\
          &&  \quad
           \frac{2\varrho^2\alpha_0-\beta_0}{(\varrho^2-1)^2}
            \zeta^2+\Big(\frac{2\varrho^2\alpha_1+\varrho^2\alpha_0^2-\beta_1}
             {-(\varrho^2-1)^2}+
             \frac{(2\varrho^2\alpha_0-\beta_0)^2}{(\varrho^2-1)^3}\Big)\zeta^4
              +O(\zeta^6) \Big] \nonumber \\
          && \quad
          \times ~  [1+\alpha_0 \zeta^2+ \alpha_1 \zeta^4+
      O(\zeta^6)] d \zeta \nonumber \\
         &  \underset{\zeta \rightarrow 0}{=}&
         \Big(\frac{-3}{\varrho^2-1}e_j(8n+5)+\frac{\varrho}{\varrho^2-1}e_j(r-2)
          +[\frac{-3}{\varrho^2-1}e_j(8n+4)+\frac{\varrho}{\varrho^2-1}\nonumber \\
          && \quad \times~
          e_j(r-3)]\zeta +O(\zeta^2)\Big) d \zeta,
       \end{eqnarray}
    which yields (\ref{4.26}). \quad $\square$

    \newtheorem{the4.6}[lem4.1]{Theorem}
      \begin{the4.6}\label{them4.70}
        Assume that the curve $\mathcal{K}_{r-2}$ is nonsingular and
        let $x,x_0 \in \mathbb{C}.$ Then
          \begin{equation}\label{4.30}
       \underline{\alpha}_{Q_0}(\mathcal{D}_{\underline{\hat{\mu}}(x)})
      =\underline{\alpha}_{Q_0}(\mathcal{D}_{\underline{\hat{\mu}}(x_0)})
      +\frac{1}{3}\underline{e}(r-2)(x-x_0)
      + \underline{e}(8n+5) \int_{x_0}^x dx^\prime
            (\Psi_1(\underline{\mu})),
          \end{equation}
          \begin{equation}\label{4.31}
      \underline{\alpha}_{Q_0}(\mathcal{D}_{\underline{\hat{\nu}}(x)})
      =\underline{\alpha}_{Q_0}(\mathcal{D}_{\underline{\hat{\nu}}(x_0)})
      +\frac{1}{3}\underline{e}(r-2)(x-x_0)
      + \underline{e}(8n+5) \int_{x_0}^x dx^\prime
            (\Psi_1(\underline{\mu})),
          \end{equation}
      where $\Psi_1(\underline{\mu})=- \sum_{j=1}^{12n+4} \mu_j(x)$.
      In particular, the Abel map does not linearize the divisor
       $\mathcal{D}_{\underline{\hat{\mu}}(\cdot)}$ and
       $\mathcal{D}_{\underline{\hat{\nu}}(\cdot)}$.
     \end{the4.6}
     \textbf{Proof.}~~We prove only (\ref{4.30}) as (\ref{4.31}) can be
     obtained from (\ref{4.30}) and Abel's theorem. Assume
     temporarily that
          \begin{equation}\label{4.32}
            \mu_j(x) \neq \mu_{j^\prime}(x) \quad
             \textrm{for $ j \neq j^\prime$ and $x \in
             \widetilde{\Omega}_\mu \subseteq \mathbb{C},$}
          \end{equation}
          where $\widetilde{\Omega}_\mu$ is open and connected. Then
          using (\ref{3.31}), (\ref{4.23}) and (\ref{4.25}), one computes
          \begin{eqnarray*}
           &&\frac{d}{dx}
           \alpha_{Q_0,l}(\mathcal{D}_{\underline{\hat{\mu}}(x)})
           =
            \frac{d}{dx} \sum_{j=1}^{12n+4}
           \int_{Q_0}^{\hat{\mu}_j}\omega_l
           =
             \sum_{j=1}^{12n+4} \mu_{j,x}(x) \omega_l(\hat{\mu}_j(x)) \nonumber \\
           &&=
            \sum_{j=1}^{12n+4} \mu_{j,x}(x) \sum_{k=1}^{r-2}e_l(k)
              \eta_k \nonumber \\
           &&=
             \sum_{j=1}^{12n+4}
             \frac{-[S_r(\mu_j)+3y(\hat{\mu}_j)^2]V_{21}(\mu_j(x))}
             {u
            \prod_{\scriptstyle p=1 \atop \scriptstyle p \neq j }^{12n+4}
            (\mu_j-\mu_p)}
            \Bigg( \sum_{k=1}^{8n+5}e_l(k)
            \frac{\mu_j^{k-1}}{S_r(\mu_j)+3y(\hat{\mu}_j)^2} \nonumber \\
            && \qquad
            +\sum_{k=8n+6}^{r-2}e_l(k)
            \frac{y(\hat{\mu}_j)\mu_j^{k-8n-6}}{S_r(\mu_j)+3y(\hat{\mu}_j)^2}
            \Bigg)
               \nonumber \\
            &&=
             \sum_{j=1}^{12n+4} \frac{-V_{21}(\mu_j(x))}
             {u \prod_{\scriptstyle p=1 \atop \scriptstyle p \neq j }^{12n+4}
             (\mu_j-\mu_p)}
             \sum_{k=1}^{8n+5}e_l(k)\mu_j^{k-1}
             +\sum_{j=1}^{12n+4} \frac{-V_{21}(\mu_j(x))y(\hat{\mu}_j)}
             {u \prod_{\scriptstyle p=1 \atop \scriptstyle p \neq j }^{12n+4}
             (\mu_j-\mu_p)} \nonumber \\
            && \qquad
               \times \sum_{k=8n+6}^{r-2}e_l(k)\mu_j^{k-8n-6} \nonumber \\
            &&=
              \sum_{k=1}^{8n+5}e_l(k)
              \sum_{j=1}^{12n+4} \frac{-V_{21}(\mu_j(x))\mu_j^{k-1}}
             {u \prod_{\scriptstyle p=1 \atop \scriptstyle p \neq j }^{12n+4}
             (\mu_j-\mu_p)}
             +\sum_{k=8n+6}^{r-2}e_l(k) \\
            && \qquad
             \times
             \sum_{j=1}^{12n+4}
             \frac{-V_{21}(\mu_j(x))y(\hat{\mu}_j)\mu_j^{k-8n-6}}
             {u \prod_{\scriptstyle p=1 \atop \scriptstyle p \neq j }^{12n+4}
             (\mu_j-\mu_p)} \\
            &&=
            \sum_{k=1}^{8n+5}e_l(k)
             \sum_{j=1}^{12n+4} \frac{-(u\mu_j^{4n}+ a_0\mu_j^{4n-2}
             + \cdots)\mu_j^{k-1}}
             {u \prod_{\scriptstyle p=1 \atop \scriptstyle p \neq j }^{12n+4}
             (\mu_j-\mu_p)}
             +\sum_{k=8n+6}^{r-2}e_l(k) \\
            && \qquad
             \times
             \sum_{j=1}^{12n+4} \frac{(\frac{1}{3}u\mu_j^{8n+2}+ b_0\mu_j^{8n}
             + \cdots)\mu_j^{k-8n-6}}
             {u \prod_{\scriptstyle p=1 \atop \scriptstyle p \neq j }^{12n+4}
             (\mu_j-\mu_p)}.
          \end{eqnarray*}

          Using the standard Largange interpolation
           argument then yields
          \begin{equation}\label{4.33}
            \frac{d}{dx}
           \alpha_{Q_0,l}(\mathcal{D}_{\underline{\hat{\mu}}(x)})
           =\Psi_1(\underline{\mu})e_l(8n+5)+\frac{1}{3}e_l(r-2).
          \end{equation}
     Then we have
           \begin{equation}\label{4.34}
      \underline{\alpha}_{Q_0}(\mathcal{D}_{\underline{\hat{\mu}}(x)})
      =\underline{\alpha}_{Q_0}(\mathcal{D}_{\underline{\hat{\mu}}(x_0)})
      +\frac{1}{3}\underline{e}(r-2)(x-x_0)
      + \underline{e}(8n+5) \int_{x_0}^x dx^\prime
            (\Psi_1(\underline{\mu})).
           \end{equation}
    The equality (\ref{4.31}) follows from the linear equivalence
    $$\mathcal{D}_{P_{\infty_1}\underline{\hat{\mu}}(x)}
     \sim \mathcal{D}_{P_0\underline{\hat{\nu}}(x)},$$
     that is,
      $$\underline{A}_{Q_0}(P_{\infty_1})+
      \underline{\alpha}_{Q_0}(\mathcal{D}_{\underline{\hat{\mu}}(x)})
      =\underline{A}_{Q_0}(P_0)
      +\underline{\alpha}_{Q_0}(\mathcal{D}_{\underline{\hat{\nu}}(x)}),$$
     and (\ref{4.34}). The extension of all these results from $x
     \in \widetilde{\Omega}_\mu$ to $x \in \mathbb{C}$ then simply
     follows from the continuity of $\underline{\alpha}_{Q_0}$ and
     the hypothesis of $\mathcal{D}_{\underline{\hat{\mu}}(x)}$
     being nonspecial on $\Omega_\mu$. \quad   $\square$ \\

     Next, we provide an explicit representation for the stationary
     DP solutions $u$ in terms of the Riemann theta function
     associated with $\mathcal{K}_{r-2}$, assuming the affine part
     of $\mathcal{K}_{r-2}$ to be nonsingular.

     \newtheorem{the4.7}[lem4.1]{Theorem}
       \begin{the4.7}\label{them4.80}
       Assume that $u$ satisfies the $n$-th stationary DP equation \emph{(\ref{7})},
       that is, $X_n(u)=0,$
       and the curve $\mathcal{K}_{r-2}$ is nonsingular.
        Let $x \in \Omega_\mu$, where $\Omega_\mu \subseteq
         \mathbb{C}$ is open and connected. Suppose that
         $\mathcal{D}_{\underline{\hat{\mu}}(x)}$, or equivalently
         $\mathcal{D}_{\underline{\hat{\nu}}(x)}$ is nonspecial for
         $x\in \Omega_\mu$. Then

          \begin{equation}\label{4.35}
          u(x)=u(x_0)\frac{\theta(\underline{\tilde{z}}(P_0,\underline{\hat{\mu}}(x_0)))
          \theta(\underline{\tilde{z}}
          (P_{\infty_1},\underline{\hat{\mu}}(x)))}
           {\theta(\underline{\tilde{z}}
          (P_{\infty_1},\underline{\hat{\mu}}(x_0)))
          \theta(\underline{\tilde{z}}(P_0,\underline{\hat{\mu}}(x)))}.
       \end{equation}

       \end{the4.7}
  \textbf{Proof.}~~Using Theorem \ref{them4.40}, one can write $\psi_2$ near
  $P_{\infty_1}$ in the coordinate $\zeta =\tilde{z}^{-1}$, as
    \begin{eqnarray}\label{4.36}
        &&
          \psi_2(P,x,x_0) \underset{\zeta \rightarrow 0}{=}
            \left(\sigma_0(x)+\sigma_1(x)\zeta + \sigma_2(x)\zeta^2+ O(\zeta^3)\right)
             \\
        &&  \qquad \times ~
          \mathrm{exp}\left(\left(\int_{x_0}^{x}2m^{\frac{1}{3}}(x^\prime)dx^\prime\right)\left(f_3^{(2)}(Q_0)
          \zeta^2+O(\zeta^4)\right)\right),
           \quad \textrm{as $P \rightarrow P_{\infty_1}$}, \nonumber
          \end{eqnarray}
  where the terms $ \sigma_0(x), \sigma_1(x)$ and $\sigma_2(x)$ in (\ref{4.36})
  come from the Taylor expansion about  $P_{\infty_1}$ of the ratios
  of the theta functions in (\ref{4.10}). That is
      \begin{eqnarray*}
      \frac{\theta\left(\underline{\tilde{z}}(P,\underline{\hat{\mu}}(x))\right)}
           {\theta\left(\underline{\tilde{z}}(P_{0},\underline{\hat{\mu}}(x))\right)}
      &=& \frac
  {\theta\left(\underline{\Xi}_{Q_0}-\underline{A}_{Q_0}(P)+\underline{\alpha}_{Q_0}
  (\mathcal{D}_{\underline{\hat{\mu}}(x)})\right)}
  {\theta\left(\underline{\Xi}_{Q_0}-\underline{A}_{Q_0}(P_{0})
  +\underline{\alpha}_{Q_0}(\mathcal{D}_{\underline{\hat{\mu}}(x)})\right)}
       \\
      &=&  \frac
  {\theta \Big(\underline{\Xi}_{Q_0}-\underline{A}_{Q_0}(P_{\infty_1})
  +\underline{\alpha}_{Q_0} (\mathcal{D}_{\underline{\hat{\mu}}(x)})
  +\int_P^{P_{\infty_1}}\underline{\omega} \Big)}
  {\theta\left(\underline{\Xi}_{Q_0}-\underline{A}_{Q_0}(P_{0})
  +\underline{\alpha}_{Q_0}(\mathcal{D}_{\underline{\hat{\mu}}(x)})\right)}
       \\
  \end{eqnarray*}
  \begin{eqnarray}
  && \underset{\zeta \rightarrow 0}{=}
    \frac
  {\theta \Big(\underline{\Xi}_{Q_0}-\underline{A}_{Q_0}(P_{\infty_1})
  +\underline{\alpha}_{Q_0} (\mathcal{D}_{\underline{\hat{\mu}}(x)})
  -\rho_{0,j}\zeta-\frac{1}{2}\rho_{1,j}\zeta^2+O(\zeta^3) \Big)}
  {\theta\left(\underline{\Xi}_{Q_0}-\underline{A}_{Q_0}(P_{0})
  +\underline{\alpha}_{Q_0}(\mathcal{D}_{\underline{\hat{\mu}}(x)})\right)}
      \nonumber \\
      && \underset{\zeta \rightarrow 0}{=}
    \frac{1}{\theta_1} \Bigg[ \theta_0- \sum_{j=1}^{r-2}
    \frac{\partial \theta_0}{\partial \tilde{z}_j}\rho_{0,j}\zeta
    -\frac{1}{2} \sum_{j=1}^{r-2} \Bigg(
    \frac{\partial \theta_0}{\partial \tilde{z}_j}\rho_{1,j}
       -\sum_{k=1}^{r-2}
    \frac{\partial^2 \theta_0}{\partial \tilde{z}_j \partial \tilde{z}_k}
    \rho_{0,j}\rho_{0,k} \Bigg)\zeta^2+O(\zeta^3) \Bigg]
     \nonumber \\
   && \underset{\zeta \rightarrow 0}{=}
    \frac{\theta_0-\partial_x \theta_0 \zeta
       +(\frac{1}{2}\partial_x^2 \theta_0 -
        \partial_{\underline{U}_3^{(2)}}\theta_0)\zeta^2
          +O(\zeta^3)}{\theta_1}\nonumber
       \\
   && \underset{\zeta \rightarrow 0}{=}
    \frac{\theta_0}{\theta_1}- \frac{\partial_x  \, \theta_0 }{\theta_1} \, \zeta +
      \frac{\frac{1}{2}\partial_x^2 \theta_0 -
        \partial_{\underline{U}_3^{(2)}}\theta_0}{\theta_1}
    \zeta^2 + O(\zeta^3),
     \quad \textrm{as $P \rightarrow P_{\infty_1}$},\label{aoteman01}
  \end{eqnarray}
  where
    \begin{equation*}\label{4.37}
  \theta_0=\theta_0(x)=
  \theta(\underline{\tilde{z}}(P_{\infty_1},\underline{\hat{\mu}}(x)))=
  \theta\left(\underline{\Xi}_{Q_0}-\underline{A}_{Q_0}(P_{\infty_1})
  +\underline{\alpha}_{Q_0}(\mathcal{D}_{\underline{\hat{\mu}}(x)})\right),
    \end{equation*}
    \begin{equation*}\label{4.37a}
  \theta_1=\theta_1(x)=\theta(\underline{\tilde{z}}(P_{0},\underline{\hat{\mu}}(x)))
  =\theta\left(\underline{\Xi}_{Q_0}-\underline{A}_{Q_0}(P_{0})
  +\underline{\alpha}_{Q_0}(\mathcal{D}_{\underline{\hat{\mu}}(x)})\right),
    \end{equation*}
    and \begin{equation*}\label{4.38}
   \partial_{\underline{U}_3^{(2)}}= \sum_{j=1}^{r-2}
    U_{3,j}^{(2)}\frac{\partial}{\partial \tilde{z}_j}
   \end{equation*}
    denotes the directional derivative in the direction of the
    vector of $b$-periods $\underline{U}_3^{(2)}$, defined by
   \begin{equation}\label{4.39}
    \underline{U}_3^{(2)}=(U_{3,1}^{(2)},\ldots,U_{3,r-2}^{(2)}),
    \quad U_{3,j}^{(2)}=\frac{1}{2 \pi i} \int_{b_j}
     \omega_{P_{\infty_1,3}}^{(2)}, \quad j=1,\ldots,r-2,
   \end{equation}
   with $\omega_{P_{\infty_1,3}}^{(2)}$ holomorphic on
   $\mathcal{K}_{r-2} \setminus \{P_{\infty_1}\}$
   with a pole of order $3$ at $P_{\infty_1}$,
   \begin{equation}\label{4.40}
     \omega_{P_{\infty_1,3}}^{(2)}(P)
      \underset{\zeta \rightarrow 0}{=}(\zeta^{-3}+O(1))d \zeta
      \quad \textrm{as $P \rightarrow P_{\infty_1}$}.
   \end{equation}
  Similarly, we have
      \begin{eqnarray*}
       \frac{\theta\left(\underline{\tilde{z}}(P_0,\underline{\hat{\mu}}(x_0))\right)}
      {\theta\left(\underline{\tilde{z}}(P,\underline{\hat{\mu}}(x_0))\right)}
      &=& \left(\frac{\theta\left(\underline{\tilde{z}}(P,\underline{\hat{\mu}}(x))\right)}
      {\theta\left(\underline{\tilde{z}}(P_0,\underline{\hat{\mu}}(x))\right)} \right)^{-1}
      \Big |_{x=x_0}     \nonumber \\
      &\underset{\zeta\rightarrow 0}{=}&
      \left(\frac{\theta_0}{\theta_1}
      \left(  1- \frac{\partial_x\theta_0}{\theta_0}\zeta + O(\zeta^2)  \right)\right)^{-1}
      \Big |_{x=x_0}   \nonumber \\
      &\underset{\zeta \rightarrow 0}{=}&
      \frac{\theta_1}{\theta_0}
      \left(1+ \partial_x\ln\theta_0\,\zeta+ O(\zeta^2)\right)
      \Big |_{x=x_0}  \nonumber
     \end{eqnarray*}
     \begin{eqnarray}\label{4.41}
      \underset{\zeta \rightarrow 0}{=}
      \frac{\theta_1(x_0)}{\theta_0(x_0)}
      \left(1+ \partial_x\ln\theta_0(x)\Big |_{x=x_0}\,\zeta+ O(\zeta^2)\right),
               \textrm{ as $P \rightarrow P_{\infty_1}$}.
      \end{eqnarray}
      Then the Taylor expansion about $\psi_2$ is as follows
      \begin{eqnarray}\label{4.42}
            \psi_2(P,x,x_0)
        &\underset{\zeta \rightarrow 0}{=}&
            \frac
        {\theta(\underline{\tilde{z}}(P,\underline{\hat{\mu}}(x)))
        \theta(\underline{\tilde{z}}(P_{0},\underline{\hat{\mu}}(x_0)))}
        {\theta(\underline{\tilde{z}}(P_{0},\underline{\hat{\mu}}(x)))
        \theta(\underline{\tilde{z}}(P,\underline{\hat{\mu}}(x_0)))}
         \nonumber  \\
        &&  \times ~
          \mathrm{exp}\left(\left(\int_{x_0}^{x}2m^{\frac{1}{3}}(x^\prime)dx^\prime\right)\left(f_3^{(2)}(Q_0)
          \zeta^2+O(\zeta^4)\right)\right)
              \nonumber  \\
        & \underset{\zeta \rightarrow 0}{=} &
          \Big[  \frac{\theta_1(x_0)}{\theta_0(x_0)} \,\frac{\theta_0(x)}{\theta_1(x)} \, + \,
          \left(  \frac{\theta_1(x_0)}{\theta_0(x_0)} \,\frac{\theta_0(x)}{\theta_1(x)} \, \partial_x\ln \theta_0(x)\Big |_{x=x_0}
          -\frac{\theta_1(x_0)}{\theta_0(x_0)} \frac{\partial_x\theta_0(x)}{\theta_1(x)}\right) \zeta  \nonumber \\
         &&  + O(\zeta^2)\Big] ~\times ~
         \mathrm{exp}\left(\left(\int_{x_0}^{x}2m^{\frac{1}{3}}(x^\prime)dx^\prime\right)\left(f_3^{(2)}(Q_0)
          \zeta^2+O(\zeta^4)\right)\right)
           \nonumber \\
         & \underset{\zeta \rightarrow 0}{=} &
          \Big[  \frac{\theta_1(x_0)}{\theta_0(x_0)} \,\frac{\theta_0(x)}{\theta_1(x)} \, + \,
          \frac{\theta_1(x_0)}{\theta_0(x_0)} \,\frac{\theta_0(x)}{\theta_1(x)} \,\left( \partial_x\ln \theta_0(x)\Big |_{x=x_0}
          - \partial_x\ln \theta_0(x) \right) \zeta  \nonumber \\
         &&  + O(\zeta^2)\Big] ~\times ~
         \left( 1+ \left(f_3^{(2)}(Q_0)\int_{x_0}^{x}2m^{\frac{1}{3}}(x^\prime)dx^\prime \right)\zeta^2 + O(\zeta^4)  \right), \nonumber\\
         &&~~~~~~~~~~~~~~~~~~~~~\qquad\qquad ~~~~~~~~~~~~~~~~~
         \textrm{as $P\rightarrow P_{\infty_1}.$}
         \end{eqnarray}
         Hence, comparing the same powers of $\zeta$ in (\ref{4.36}) and
         (\ref{4.42}) gives
          \begin{equation}\label{4.43}
            \begin{split}
            &  \sigma_0(x)= \frac{\theta_1(x_0)}{\theta_0(x_0)} \,\frac{\theta_0(x)}{\theta_1(x)},\\
            &  \sigma_1(x)= \frac{\theta_1(x_0)}{\theta_0(x_0)} \,\frac{\theta_0(x)}{\theta_1(x)} \,\left( \partial_x\ln \theta_0(x)\Big |_{x=x_0}-\partial_x\ln \theta_0(x)\right).       \\
            \end{split}
          \end{equation}
          If we set
           \begin{equation}\label{4.44}
             \psi_2 \underset{\zeta \rightarrow 0}{=}
              (\sigma_0(x) + \sigma_1(x)\zeta + \sigma_2(x)\zeta^2+ O(\zeta^3))
              ~ \mathrm{exp}(\Delta), \quad
              \textrm{as $P \rightarrow P_{\infty_1}$}
           \end{equation}
            with
   \begin{eqnarray*}
   \exp\left(\Delta\right)&=&\mathrm{exp}
   \left(\left(\int_{x_0}^{x}2m^{\frac{1}{3}}(x^\prime)
   dx^\prime\right)\left(f_3^{(2)}(Q_0)
          \zeta^2+O(\zeta^4)\right)\right),\\
   &=&\left( 1+ \left(f_3^{(2)}(Q_0)\int_{x_0}^{x}2m^{\frac{1}{3}}(x^\prime)
   dx^\prime \right)\zeta^2 + O(\zeta^4)  \right),\\
  \end{eqnarray*}
  then we  compute its $x$-derivatives as ($P \rightarrow
  P_{\infty_1}$)
          \begin{eqnarray}\label{4.45}
            \begin{split}
            &\psi_{2,x}\underset{\zeta \rightarrow 0}{=}
            \left(\sigma_{0,x}+\sigma_{1,x}\zeta+O(\zeta^2)\right)\exp\left(\Delta\right)
             + \left(\left(2f_3^{(2)}(Q_0)m^{\frac{1}{3}}\right)\zeta^2+ O(\zeta^4)\right)\psi_2\\
            &~~\quad\underset{\zeta \rightarrow 0}{=}\sigma_{0,x}+\sigma_{1,x}\zeta+ O(\zeta^2),\\
            &\psi_{2,xx}\underset{\zeta \rightarrow 0}{=}
            \sigma_{0,xx}+\sigma_{1,xx}\zeta+ O(\zeta^2),\\
            &\psi_{2,xxx}\underset{\zeta \rightarrow 0}{=}
            \sigma_{0,xxx}+\sigma_{1,xxx}\zeta+ O(\zeta^2).\\
            \end{split}
            \end{eqnarray}
    By eliminating $\psi_1$ and $\psi_3$ in (\ref{1}), we arrive at
       \begin{equation}\label{4.46}
         \psi_{2,xxx}=-m\tilde{z}^{-2}+\frac{m_x}{m}\psi_{2,xx}
         -\frac{m_x}{m}\psi_2+\psi_{2,x}.
       \end{equation}
  Substituting (\ref{4.45}) into (\ref{4.46}) and comparing the
  coefficients of  $\zeta^0$, we obtain
  \begin{equation*}
    \sigma_{0,xxx}=\frac{m_x}{m}(\sigma_{0,xx}-\sigma_0)+\sigma_{0,x},
  \end{equation*}
  namely,
  \begin{equation*}
   \frac{( \sigma_{0,xx}-\sigma_{0})_x}{\sigma_{0,xx}-\sigma_{0}}=\frac{m_x}{m}
   =\frac{(u(x)-u_{xx}(x))_x}{u(x)-u_{xx}(x)},
  \end{equation*}
 which together with the first line of (\ref{4.43}) leads to (\ref{4.35}).
  \quad  $\square$

  \newtheorem{rem4.9}[lem4.1]{Remark}
    \begin{rem4.9}
        We note the unusual fact that $P_0$, as opposed to
        $P_{\infty_i},$ $i=1,2,3$, is the essential singularity of
        $\psi_2$. What makes matters worse is the intricate
        $x$-dependence of the leading-order exponential term in $
        \psi_2$, near $P_0$, as displayed in $(\ref{4.10})$. This is
        in sharp contrast to standard Baker-Akhiezer functions that
        typically feature a linear behavior with respect to $x$ in
        connection with their essential singularities of the type
        $\mathrm{exp}((x-x_0)\zeta^{-2})$ near $\zeta=0$.
        Therefore,
        in Theorem $\ref{them4.70}$, the Abel map does not provide the proper
        change of variables to linearize the divisor
        $\mathcal{D}_{\underline{\hat{\mu}}(x)}$ in the DP context
        is in sharp contrast to standard integrable soliton
        equations such as the KdV and AKNS hierarchies.
    \end{rem4.9}

\section{The time-dependent DP formalism}
   In this section we extend the results of Section 3 to the
   time-dependent DP hierarchy.  We
   employ the notations $\widetilde{G}_j,$ $\widetilde{V}$,
   $\widetilde{V}_{ij}$, etc., in order to distinguish them from $G_j$,
   $V$, $V_{ij}$, etc. In addition, we indicate that the individual $p$th DP flow by a
   separate time variable $t_p \in \mathbb{C}$. In analogy to
   (\ref{3.5}), we introduce the time-dependent vector Baker-Akhiezer
   function $\psi=(\psi_1,\psi_2,\psi_3)^t$ by
      \begin{equation}\label{5.1}
        \begin{split}
          & \psi_x(P,x,x_0,t_p,t_{0,p})=U(u(x,t_p),\tilde{z}(P))
               \psi(P,x,x_0,t_p,t_{0,p}),\\
          &  \psi_{t_p}(P,x,x_0,t_p,t_{0,p})=\widetilde{V}(u(x,t_p),\tilde{z}(P))
               \psi(P,x,x_0,t_p,t_{0,p}), \\
          &   \tilde{z}V(u(x,t_p),\tilde{z}(P))\psi(P,x,x_0,t_p,t_{0,p})
                  =y(P)\psi(P,x,x_0,t_p,t_{0,p}), \\
          &   \psi_2(P,x_0,x_0,t_{0,p},t_{0,p})=1,
               \quad x,t_p \in \mathbb{C},
        \end{split}
      \end{equation}
   where $\widetilde{V}=(\widetilde{V}_{ij})_{3 \times 3}$, and
         \begin{equation}\label{5.2}
       \widetilde{V}_{ij}=\sum_{l=0}^{p}
       \widetilde{V}_{ij}^{(l)}(\widetilde{G}_l)\tilde{z}^{4(p-l+1)}
        \quad i,j=1,\ldots,3, \quad l=0,\ldots,p
       \end{equation}
   with $\widetilde{V}_{ij}^{(l)}(\widetilde{G}_l)$ determined by
   $\widetilde{G}_l$, which is defined in (\ref{2.1}) by substituting $\widetilde{G}_l$ for $G_l$.

   The compatibility conditions of the first three expressions in
   (\ref{5.1}) yield that
       \begin{equation}\label{5.3}
          \begin{split}
         &
          U_{t_p}(\tilde{z})-\widetilde{V}_{x}(\tilde{z})
          +[U(\tilde{z}),\widetilde{V}(\tilde{z})]=0, \\
         &
         -V_x(\tilde{z})+[U(\tilde{z}),V(\tilde{z})]=0,\\
         &
        -V_{t_p}(\tilde{z})+[\widetilde{V}(\tilde{z}),V(\tilde{z})]=0.
           \end{split}
       \end{equation}
   A direct calculation shows that $yI-\tilde{z}V(\tilde{z})$
   satisfies the last two equations in (\ref{5.3}). Then the
   characteristic polynomial of Lax matrix $\tilde{z}V(\tilde{z})$
   for the DP hierarchy is an independent constant of variables $x$
   and $t_p$ with the expansion
         \begin{equation}\label{5.4}
           \mathrm{det}(yI-\tilde{z}V)=y^3+yS_r(\tilde{z})-T_r(\tilde{z}),
         \end{equation}
   where $S_r(\tilde{z})$ and $T_r(\tilde{z})$ are defined as in
   (\ref{2.20}).
   Then the time-dependent DP curve $\mathcal{K}_{r-2}$ is defined
   by
        \begin{equation}\label{5.5}
          \mathcal{K}_{r-2}:\mathcal{F}_r(\tilde{z},y)=
            y^3+yS_r(\tilde{z})-T_r(\tilde{z})=0.
        \end{equation}
  In analogy to (\ref{3.6}), we can define
  the following meromorphic
  function $\phi(P,x,t_p)$ on $\mathcal{K}_{r-2}$
  the fundamental ingredient for the construction of algebro-geometric solutions
  of the time-dependent DP hierarchy,
         \begin{equation}\label{5.6}
         \phi(P,x,t_p)=\tilde{z}\frac{\partial_x \psi_2(P,x,x_0,t_p,t_{0,p})}
          {\psi_2(P,x,x_0,t_p,t_{0,p})}, \quad P \in
          \mathcal{K}_{r-2}, ~ x\in \mathbb{C}.
        \end{equation}
  Using (\ref{5.1}), a direct calculation shows that
        \begin{eqnarray}\label{5.7}
          \phi(P,x,t_p) &=& \tilde{z}
          \frac{yV_{31}(\tilde{z},x,t_p)+C_r(\tilde{z},x,t_p)}
             {yV_{21}(\tilde{z},x,t_p)+A_r(\tilde{z},x,t_p)}
          \nonumber \\
            &=& \tilde{z} \frac{F_r(\tilde{z},x,t_p)}
            {y^2V_{31}(\tilde{z},x,t_p)-yC_r(\tilde{z},x,t_p)+D_r(\tilde{z},x,t_p)}
            \nonumber \\
          &=& \tilde{z}
  \frac{y^2V_{21}(\tilde{z},x,t_p)-yA_r(\tilde{z},x,t_p)+B_r(\tilde{z},x,t_p)}
             {E_r(\tilde{z},x,t_p)},
        \end{eqnarray}
  where $P=(\tilde{z},y) \in \mathcal{K}_{r-2}$, $(x,t_p)\in
  \mathbb{C}^2$, and $A_r(\tilde{z},x,t_p),$ $B_r(\tilde{z},x,t_p),$
  $C_r(\tilde{z},x,t_p),$  $D_r(\tilde{z},x,t_p),$  $E_r(\tilde{z},x,t_p),$
   $F_r(\tilde{z},x,t_p)$ and $J_r(\tilde{z},x,t_p)$ are defined as
  in  (\ref{3.9}) and (\ref{3.10}). Hence the interrelationships
  among them (\ref{3.11})-(\ref{3.14}) also hold in the time-dependent case.

  Similarly, we denote by $\{\mu_j(x,t_p)\}_{j=1,\ldots,r-5}$
  and $\{\nu_j(x,t_p)\}_{j=1,\ldots,r-3}$ the zeros of
  $E_r(\tilde{z},x,t_p)$ and $\tilde{z}^2F_r(\tilde{z},x,t_p)$,
  respectively. Thus, we may write
      \begin{equation}\label{5.8}
           E_r(\tilde{z},x,t_p)=u(x,t_p) \prod_{j=1}^{r-5}
            (\tilde{z}-\mu_j(x,t_p)),
        \end{equation}
        \begin{equation}\label{5.9}
         F_r(\tilde{z},x,t_p)=-u(x,t_p)u_x^2(x,t_p)
          \tilde{z}^{-2} \prod_{j=1}^{r-3}(\tilde{z}-\nu_j(x,t_p)).
         \end{equation}
  Defining
       \begin{eqnarray}\label{5.10}
            \hat{\mu}_j(x,t_p)
            &=&
            \Big(\mu_j(x,t_p),y(\hat{\mu}_j(x,t_p))\Big)
             \nonumber \\
            &=&
            \left(\mu_j(x,t_p),-\frac{A_r(\mu_j(x,t_p),x,t_p)}
            {V_{21}(\mu_j(x,t_p),x,t_p)}\right)
            \in \mathcal{K}_{r-2}, \\
            &&  \qquad \qquad
             j=1,\ldots,r-5,~(x,t_p)\in\mathbb{C}^2, \nonumber
         \end{eqnarray}
         \begin{eqnarray}\label{5.11}
           \hat{\nu}_j(x,t_p)
            &=&
            \Big(\nu_j(x,t_p),y(\hat{\nu}_j(x,t_p))\Big)
             \nonumber \\
            &=&
            \left(\nu_j(x,t_p),
            -\frac{C_r(\nu_j(x,t_p),x,t_p)}{V_{31}(\nu_j(x,t_p),x,t_p)}\right)
            \in \mathcal{K}_{r-2}, \\
            && \qquad \qquad
            j=1,\ldots,r-3,~(x,t_p)\in\mathbb{C}^2. \nonumber
         \end{eqnarray}
    One infers from (\ref{5.7}) that the divisor $(\phi(P,x,t_p))$
    of $\phi(P,x,t_p)$ is given by
          \begin{equation}\label{5.12}
           (\phi(P,x,t_p))=
           \mathcal{D}_{P_0,\underline{\hat{\nu}}(x,t_p)}(P)
           -\mathcal{D}_{P_{\infty_1},\underline{\hat{\mu}}(x,t_p)}(P),
         \end{equation}
   where
      $$ \underline{\hat{\nu}}(x,t_p)
     =\{\hat{\nu}_1(x,t_p),\ldots,\hat{\nu}_{r-3}(x,t_p)\},$$
     $$\underline{\hat{\mu}}(x,t_p)
     =\{P_{\infty_2},P_{\infty_3},\hat{\mu}_1(x,t_p),\ldots,\hat{\mu}_{r-5}(x,t_p)\}.$$
    That is,
    $P_0,\hat{\nu}_1(x,t_p),\ldots,\hat{\nu}_{r-3}(x,t_p)$ are the
    $r-2$ zeros of $\phi(P,x,t_p)$ and
    $P_{\infty_1},P_{\infty_2},$
    $ P_{\infty_3},\hat{\mu}_1(x,t_p), \ldots,\hat{\mu}_{r-5}(x,t_p)$ its $r-2$
    poles. \\

    Further properties of $\phi(P,x,t_p)$ are summarized as follows.
    \newtheorem{the5.1}{Theorem}[section]
      \begin{the5.1}
        Assume $(\ref{5.1})$ and $(\ref{5.6})$, $P=(\tilde{z},y)\in
      \mathcal{K}_{r-2}\setminus \{P_{\infty_ i}, P_0\},$ $i=1,2,3,$ and let
      $(\tilde{z},x,t_p)\in \mathbb{C}^3$. Then
     \begin{eqnarray}\label{5.13}
        &&
         \phi_{xx}(P,x,t_p)+3\tilde{z}^{-1}\phi(P,x,t_p)\phi_x(P,x,t_p)
           +\tilde{z}^{-2}\phi^3(P,x,t_p) \nonumber \\
       && \qquad
        -\frac{m_x(x,t_p)}{m(x,t_p)}\phi_x(P,x,t_p)
          -\tilde{z}^{-1}\frac{m_x(x,t_p)}{m(x,t_p)}\phi^2(P,x,t_p)
            \nonumber \\
        &&  \qquad \qquad
          -\phi(P,x,t_p)
           +m(x,t_p)\tilde{z}^{-1}+\frac{m_x(x,t_p)}{m(x,t_p)}\tilde{z}=0,
       \end{eqnarray}
       \begin{eqnarray}\label{5.14}
         \phi_{t_p}(P,x,t_p)&=&\tilde{z} \partial_x
          \Big( \frac{\widetilde{V}_{21}(\tilde{z},x,t_p)}{m(x,t_p)}
             (\tilde{z}^2-\tilde{z}\phi_x(P,x,t_p)-\phi^2(P,x,t_p))
          \nonumber \\
         &&
          + \widetilde{V}_{22}(\tilde{z},x,t_p)
          +\widetilde{V}_{23}(\tilde{z},x,t_p)\tilde{z}^{-1}\phi(P,x,t_p)
          \Big),
       \end{eqnarray}
        \begin{eqnarray}\label{5.15}
          \phi(P,x,t_p)\phi(P^\ast,x,t_p)\phi(P^{\ast\ast},x,t_p)=
            -\tilde{z}^3\frac{F_r(\tilde{z},x,t_p)}{E_r(\tilde{z},x,t_p)},
        \end{eqnarray}
      \begin{eqnarray}\label{5.16}
          \phi(P,x,t_p)+\phi(P^\ast,x,t_p)+\phi(P^{\ast\ast},x,t_p)=
            \tilde{z}\frac{E_{r,x}(\tilde{z},x,t_p)}{E_r(\tilde{z},x,t_p)},
      \end{eqnarray}
       \begin{eqnarray}\label{5.17}
         && \frac{1}{\phi(P,x,t_p)}+\frac{1}{\phi(P^\ast,x,t_p)}
                    +\frac{1}{\phi(P^{\ast\ast},x,t_p)}
          =\frac{F_{r,x}(\tilde{z},x,t_p)}{\tilde{z}F_r(\tilde{z},x,t_p)}
          \nonumber \\
          &&  \qquad \qquad
          - \frac{m(x,t_p)J_r(\tilde{z},x,t_p)}{\tilde{z}F_r(\tilde{z},x,t_p)}
                -\frac{2m(x,t_p)V_{33}(\tilde{z},x,t_p)}
                   {\tilde{z}^3V_{31}(\tilde{z},x,t_p)},
         \end{eqnarray}
       \begin{eqnarray}\label{5.18}
          y(P)\phi(P,x,t_p)+y(P^\ast)\phi(P^\ast,x,t_p)
                   +y(P^{\ast\ast})\phi(P^{\ast\ast},x,t_p)= \nonumber\\
            \tilde{z}\frac{3T_r(\tilde{z})V_{21}(\tilde{z},x,t_p)
              +2S_r(\tilde{z})A_r(\tilde{z},x,t_p)}
              {E_r(\tilde{z},x,t_p)}.
       \end{eqnarray}
    \end{the5.1}
  \textbf{Proof.}~~Equation (\ref{5.13}) follows from (\ref{5.1}) and
  (\ref{5.7}). Relation (\ref{5.14}) can be proved as follows.
  Differentiating (\ref{5.6}) with respect to $t_p$ and using
  (\ref{5.1}), we have
      \begin{eqnarray}\label{5.19}
        (\phi)_{t_p}&=&\tilde{z} \partial_x
         \frac{\widetilde{V}_{21}\psi_1+\widetilde{V}_{22}\psi_2
               +\widetilde{V}_{23}\psi_3}{\psi_2}
         \nonumber \\
        &=& \tilde{z} \partial_x  \Big[\widetilde{V}_{21}
        \frac{(-\psi_{2,xx}+\psi_2)\tilde{z}^2}{m\psi_2}
          +\widetilde{V}_{22}+\widetilde{V}_{23}
          \frac{\psi_3}{\psi_2} \Big]
         \nonumber \\
        &=& \tilde{z} \partial_x  \Big[
          \frac{\widetilde{V}_{21}}{m}(-\tilde{z}\phi_x-\phi^2+\tilde{z}^2)
           +\widetilde{V}_{22}+\widetilde{V}_{23}\tilde{z}^{-1}\phi
           \Big].
      \end{eqnarray}
  Moreover, (\ref{5.15})-(\ref{5.18}) can be derived as in Theorem \ref{them3.20}.
  \quad $\square$ \\

  Next, we consider the $t_p$-dependence of $E_r$ and $F_r$.

  \newtheorem{lem5.2}[the5.1]{Lemma}
    \begin{lem5.2}
      Assume $(\ref{5.1}),$ $(\ref{5.3})$ and let
      $(\tilde{z},x,t_p)\in \mathbb{C}^3$. Then
     \begin{equation} \label{5.20}
         E_{r,t_p}(\tilde{z},x,t_p)=E_{r,x}(\tilde{z},x,t_p)
          \Big(\widetilde{V}_{23}
          -\frac{\widetilde{V}_{21}}{V_{21}}
          V_{23} \Big)
        +E_r(\tilde{z},x,t_p)3 \Big(\widetilde{V}_{22}
         -\frac{\widetilde{V}_{21}}{V_{21}}
          V_{22}\Big),
     \end{equation}
      \begin{eqnarray}\label{5.21}
       F_{r,t_p}(\tilde{z},x,t_p)&=&F_{r,x}(\tilde{z},x,t_p)
         \widetilde{V}_{32}
         -J_r(\tilde{z},x,t_p)
        (\tilde{z}^2\widetilde{V}_{31}+m\widetilde{V}_{32})
           \nonumber \\
         &&
        +F_r(\tilde{z},x,t_p)
        \Big(3\widetilde{V}_{22}+3\widetilde{V}_{23,x}
         -\frac{2mV_{33}}{\tilde{z}^2V_{31}}
          (\frac{\tilde{z}^2}{m}\widetilde{V}_{31}+\widetilde{V}_{32})
         \Big). \nonumber \\
        \end{eqnarray}
    \end{lem5.2}
  \textbf{Proof.}~~From (\ref{5.1}) and (\ref{5.6}), we obtain
     \begin{equation}\label{5.22}
       \tilde{z}\phi_x +\phi^2=
        \frac{m}{V_{21}}(-\tilde{z}^{-1}y+V_{22}
        +\tilde{z}^{-1}V_{23}\phi)+\tilde{z}^2.
     \end{equation}
  Hence, one can compute
      \begin{eqnarray*}
      &&
      \tilde{z}\phi_x(P,x,t_p)+\tilde{z}\phi_x(P^\ast,x,t_p)
       +\tilde{z}\phi_x(P^{\ast\ast},x,t_p)
         \nonumber \\
      &&  ~~~~~~~~~~~~~~~~~~
      +\phi^2(P,x,t_p)+\phi^2(P^\ast,x,t_p)+\phi^2(P^{\ast\ast},x,t_p)
        \nonumber \\
       &&
          =\frac{m}{V_{21}}(-\tilde{z}^{-1}y_0+V_{22}
          +\tilde{z}^{-1}V_{23}\phi(P))+\tilde{z}^2
         \nonumber \\
       && ~~~~~~~~~~~~~~~~~~
          +
          \frac{m}{V_{21}}(-\tilde{z}^{-1}y_1+V_{22}
          +\tilde{z}^{-1}V_{23}\phi(P^\ast))+\tilde{z}^2
             \nonumber \\
        && ~~~~~~~~~~~~~~~~~~
        +\frac{m}{V_{21}}(-\tilde{z}^{-1}y_2+V_{22}
           +\tilde{z}^{-1}V_{23}\phi(P^{\ast\ast}))+\tilde{z}^2
        \nonumber \\
        &&
         =-\frac{m\tilde{z}^{-1}(y_0+y_1+y_2)}{V_{21}}
         +3\frac{mV_{22}}{V_{21}}+3\tilde{z}^2
        \nonumber \\
        && ~~~~~~~~~~~~~~~~~~~
         + \frac{m\tilde{z}^{-1}V_{23}}{V_{21}}
           (\phi(P)+\phi(P^\ast)+\phi(P^{\ast\ast}))
         \nonumber
       \end{eqnarray*}
       \begin{eqnarray}\label{5.23}
        &&
        =3\frac{mV_{22}}{V_{21}}+\frac{m\tilde{z}^{-1}V_{23}}{V_{21}}
         (\phi(P)+\phi(P^\ast)+\phi(P^{\ast\ast}))+3\tilde{z}^2,
      \end{eqnarray}
  and
      \begin{eqnarray}\label{5.24}
       &&
        \partial_{t_p} (\phi(P,x,t_p)+\phi(P^\ast,x,t_p)+\phi(P^{\ast\ast},x,t_p))
        \nonumber \\
        && \qquad
        =\partial_{t_p} \Big(\tilde{z}\frac{E_{r,x}(\tilde{z},x,t_p)}
             {E_r(\tilde{z},x,t_p)} \Big)
        \nonumber \\
         && \qquad
         =\tilde{z} \partial_{t_p} \partial_{x}
          (\mathrm{ln}E_r(\tilde{z},x,t_p))
        \nonumber \\
         && \qquad
         =\tilde{z} \partial_{x} \partial_{t_p}
          (\mathrm{ln}E_r(\tilde{z},x,t_p)).
      \end{eqnarray}
   On the other hand, from (\ref{5.14}), we can see that
       \begin{eqnarray}\label{5.25}
       &&
        \partial_{t_p} (\phi(P,x,t_p)+\phi(P^\ast,x,t_p)+\phi(P^{\ast\ast},x,t_p))
        \nonumber \\
        && \quad
          = \tilde{z} \partial_x
          \Big( \frac{\widetilde{V}_{21}}{m}
             (\tilde{z}^2-\tilde{z}\phi_x(P,x,t_p)-\phi^2(P,x,t_p))
          \nonumber \\
         && \qquad
          + \widetilde{V}_{22}+\widetilde{V}_{23}\tilde{z}^{-1}\phi(P,x,t_p)
          \Big)
          \nonumber \\
         && \quad
           +\tilde{z} \partial_x
           \Big( \frac{\widetilde{V}_{21}}{m}
             (\tilde{z}^2-\tilde{z}\phi_x(P^\ast,x,t_p)-\phi^2(P^\ast,x,t_p))
          \nonumber \\
         && \qquad
          + \widetilde{V}_{22}+\widetilde{V}_{23}\tilde{z}^{-1}\phi(P^\ast,x,t_p)
          \Big)
          \nonumber \\
         && \quad
         +\tilde{z} \partial_x
          \Big( \frac{\widetilde{V}_{21}}{m}
         (\tilde{z}^2-\tilde{z}\phi_x(P^{\ast\ast},x,t_p)-\phi^2(P^{\ast\ast},x,t_p))
          \nonumber \\
         && \qquad
          + \widetilde{V}_{22}+\widetilde{V}_{23}\tilde{z}^{-1}\phi(P^{\ast\ast},x,t_p)
          \Big).
       \end{eqnarray}
  Without loss of generality, we take the integration constants as
  zero and then obtain
       \begin{eqnarray*}
       \partial_{t_p}(\mathrm{ln}E_r(\tilde{z},x,t_p))&=&
        -\frac{\widetilde{V}_{21}}{m}\tilde{z}
          (\phi_x(P,x,t_p)+\phi_x(P^\ast,x,t_p)+\phi_x(P^{\ast\ast},x,t_p)
         \nonumber \\
        &&
        -\frac{\widetilde{V}_{21}}{m}(\phi^2(P,x,t_p)+\phi^2(P^\ast,x,t_p)
            +\phi^2(P^{\ast\ast},x,t_p)
         \nonumber \\
        &&
         + \tilde{z}^{-1}\widetilde{V}_{23}
         (\phi(P,x,t_p)+\phi(P^\ast,x,t_p)+\phi(P^{\ast\ast},x,t_p))
         \nonumber \\
        &&
        +3\widetilde{V}_{22}+3\frac{\widetilde{V}_{21}}{m}\tilde{z}^2
        \nonumber \\
        &=&
        \tilde{z}^{-1}\Big(\widetilde{V}_{23}
        -\frac{\widetilde{V}_{21}}{V_{21}}V_{23}\Big)
        (\phi(P)+\phi(P^\ast)+\phi(P^{\ast\ast}))
        \nonumber \\
        &&
        +3\widetilde{V}_{22}-3\frac{\widetilde{V}_{21}}{V_{21}}V_{22}
        \nonumber
        \end{eqnarray*}
   \begin{eqnarray}\label{5.26}
        &=&
          \tilde{z}^{-1}\Big(\widetilde{V}_{23}
        -\frac{\widetilde{V}_{21}}{V_{21}}V_{23}\Big)
        \Big(\tilde{z}\frac{E_{r,x}}{E_r}\Big)
         +3\widetilde{V}_{22}-3\frac{\widetilde{V}_{21}}{V_{21}}V_{22}
        \nonumber \\
           &=&
        \Big(\widetilde{V}_{23}
        -\frac{\widetilde{V}_{21}}{V_{21}}V_{23}\Big)
        \Big(\frac{E_{r,x}}{E_r}\Big)
         +3\widetilde{V}_{22}-3\frac{\widetilde{V}_{21}}{V_{21}}V_{22},
        \end{eqnarray}
  which implies
     \begin{equation}\label{5.27}
      E_{r,t_p}(\tilde{z},x,t_p)=E_{r,x}\Big(\widetilde{V}_{23}
            -\frac{\widetilde{V}_{21}}{V_{21}}V_{23}\Big)
      +E_r   3\Big(
      \widetilde{V}_{22}-\frac{\widetilde{V}_{21}}{V_{21}}V_{22}
      \Big).
     \end{equation}
  Relation ({\ref{5.21}) can be proved as follows. Using
  (\ref{5.3}), (\ref{5.13}), (\ref{5.15}), (\ref{5.17}) and
  (\ref{5.23}), we have
      \begin{eqnarray*}
      &&
        \partial_{t_p} \Big(-\tilde{z}^3 \frac{F_r(\tilde{z},x,t_p)}
           {E_r(\tilde{z},x,t_p)} \Big) =
        \partial_{t_p}
        [\phi(P,x,t_p)\phi(P^\ast,x,t_p)\phi(P^{\ast\ast},x,t_p)]
          \\
      &&
       = \phi_{t_p}(P)\phi(P^\ast)\phi(P^{\ast\ast})  +
        \phi(P)\phi_{t_p}(P^\ast)\phi(P^{\ast\ast})  +
         \phi(P)\phi(P^\ast)\phi_{t_p}(P^{\ast\ast})
          \\
      &&
      =\phi(P^\ast)\phi(P^{\ast\ast})
       \left(\tilde{z} \partial_x  \Big[
          \frac{\widetilde{V}_{21}}{m}(-\tilde{z}\phi_x(P)-\phi^2(P)+\tilde{z}^2)
           +\widetilde{V}_{22}+\widetilde{V}_{23}\tilde{z}^{-1}\phi(P)
           \Big]\right)
          \\
       && \qquad
       +\phi(P)\phi(P^{\ast\ast})
         \left(\tilde{z} \partial_x  \Big[
          \frac{\widetilde{V}_{21}}{m}(-\tilde{z}\phi_x(P^\ast)
           -\phi^2(P^\ast)+\tilde{z}^2)
           +\widetilde{V}_{22}+\widetilde{V}_{23}\tilde{z}^{-1}\phi(P^\ast)
           \Big]\right)
          \\
      &&  \qquad
         +\phi(P)\phi(P^{\ast})
         \left(\tilde{z} \partial_x  \Big[
          \frac{\widetilde{V}_{21}}{m}(-\tilde{z}\phi_x(P^{\ast\ast})
           -\phi^2(P^{\ast\ast})+\tilde{z}^2)
           +\widetilde{V}_{22}+\widetilde{V}_{23}\tilde{z}^{-1}\phi(P^{\ast\ast})
           \Big]\right)
          \\
      &&
         =\phi(P)\phi(P^\ast)\phi(P^{\ast\ast}) \Bigg[-\frac{\tilde{z}^2}{m}
   \widetilde{V}_{31} \partial_x \mathrm{ln}\phi(P)\phi(P^\ast)\phi(P^{\ast\ast})
   -\frac{\tilde{z}\widetilde{V}_{21,x}}{m}(\phi(P)+\phi(P^\ast)+\phi(P^{\ast\ast}))
          \\
      && \qquad
        +\tilde{z}(\frac{\tilde{z}^2\widetilde{V}_{21,x}}{m}+\widetilde{V}_{21}
        +\widetilde{V}_{22,x})\Big(\frac{1}{\phi(P)}+\frac{1}{\phi(P^\ast)}
         +\frac{1}{\phi(P^{\ast\ast})}\Big)-3\frac{\tilde{z}^2\widetilde{V}_{21}}{m}
         +3\widetilde{V}_{23,x} \\
      &&  \qquad
        + \frac{\widetilde{V}_{21}}{m}\Big(\tilde{z}\phi_x(P)+\phi^2(P)
        + \tilde{z}\phi_x(P^\ast)+\phi^2(P^\ast)
        +
        \tilde{z}\phi_x(P^{\ast\ast})+\phi^2(P^{\ast\ast})\Big)\Bigg]
      \\
        &=&
         \phi(P)\phi(P^\ast)\phi(P^{\ast\ast}) \Bigg[-\frac{\tilde{z}^2}{m}
   \widetilde{V}_{31} \partial_x \mathrm{ln}\phi(P)\phi(P^\ast)\phi(P^{\ast\ast})
   -\frac{\tilde{z}\widetilde{V}_{21,x}}{m}(\phi(P)+\phi(P^\ast)+\phi(P^{\ast\ast}))
         \nonumber \\
      && \qquad
        +\tilde{z}(\frac{\tilde{z}^2\widetilde{V}_{21,x}}{m}+\widetilde{V}_{21}
        +\widetilde{V}_{22,x})\Big(\frac{1}{\phi(P)}+\frac{1}{\phi(P^\ast)}
         +\frac{1}{\phi(P^{\ast\ast})}\Big)-3\frac{\tilde{z}^2\widetilde{V}_{21}}{m}
         +3\widetilde{V}_{23,x}
         \nonumber \\
      &&  \qquad
        + \frac{\widetilde{V}_{21}}{m}\Big(\frac{3mV_{22}}{V_{21}}
     + \frac{\tilde{z}^{-1}V_{23}}{V_{21}}(\phi(P)+\phi(P^\ast)+\phi(P^{\ast\ast}))
     + 3\tilde{z}^2 \Big) \Bigg]
     \nonumber
     \end{eqnarray*}
      \begin{eqnarray}\label{5.28}
     &=&
       -\tilde{z}^3\frac{F_r}{E_r} \Bigg[-\frac{\tilde{z}^2}{m}\widetilde{V}_{31}
           \Big(\frac{F_{r,x}}{F_r}-\frac{E_{r,x}}{E_r}\Big)
       -\frac{\tilde{z}^2\widetilde{V}_{21,x}}{m}\frac{E_{r,x}}{E_r}
       + 3\frac{\widetilde{V}_{21}}{V_{21}}V_{22}
       + \frac{\widetilde{V}_{21}}{V_{21}}V_{23}\frac{E_{r,x}}{E_r}
       \nonumber  \\
     && \qquad
        +\Big(\frac{\tilde{z}^2}{m}\widetilde{V}_{21,x}+\widetilde{V}_{21}
           +\widetilde{V}_{22,x}\Big)
         \Big(\frac{F_{r,x}}{F_r}-\frac{mJ_r}{F_r}
              -\frac{2mV_{33}}{\tilde{z}^2V_{31}}\Big)
         + 3\widetilde{V}_{23,x} \Bigg],
      \end{eqnarray}
   which implies that
      \begin{eqnarray}\label{5.29}
        &&
        \frac{F_{r,t_p}}{E_r}-\frac{F_rE_{r,t_p}}{E_r^2}
        \nonumber \\
       &=&
        \frac{F_r}{E_r}
         \Bigg[-\frac{\tilde{z}^2}{m}\widetilde{V}_{31}
           \Big(\frac{F_{r,x}}{F_r}-\frac{E_{r,x}}{E_r}\Big)
          -\frac{\tilde{z}^2\widetilde{V}_{21,x}}{m}\frac{E_{r,x}}{E_r}
          + \frac{\widetilde{V}_{21}}{V_{21}}V_{23}\frac{E_{r,x}}{E_r}
          + 3\frac{\widetilde{V}_{21}}{V_{21}}V_{22}
       \nonumber \\
       &&         \qquad
        +\Big(\frac{\tilde{z}^2}{m}\widetilde{V}_{21,x}+\widetilde{V}_{21}
           +\widetilde{V}_{22,x}\Big)
         \Big(\frac{F_{r,x}}{F_r}-\frac{mJ_r}{F_r}
              -\frac{2mV_{33}}{\tilde{z}^2V_{31}}\Big)
          + 3\widetilde{V}_{23,x} \Bigg]
       \nonumber \\
       &=&
       \frac{F_{r,x}}{E_r} \Big(-\frac{\tilde{z}^2}{m}\widetilde{V}_{31}
         +\Big(\frac{\tilde{z}^2}{m}\widetilde{V}_{21,x}+\widetilde{V}_{21}
           +\widetilde{V}_{22,x}\Big)\Big)
           \nonumber \\
       &&  \qquad
         +\frac{E_{r,x}F_r}{E_r^2} \Big(\frac{\tilde{z}^2}{m}\widetilde{V}_{31}
         -\frac{\tilde{z}^2\widetilde{V}_{21,x}}{m}
         + \frac{\widetilde{V}_{21}}{V_{21}}V_{23}\Big)
          \nonumber \\
       &&  \qquad
        +\frac{F_r}{E_r}\Big(3\frac{V_{22}}{V_{21}}\widetilde{V}_{21}
             +3\widetilde{V}_{23,x}
           -\frac{2mV_{33}}{\tilde{z}^2V_{31}}
          (\frac{\tilde{z}^2}{m}\widetilde{V}_{21,x}+\widetilde{V}_{21}+
          \widetilde{V}_{22,x})\Big)
          \nonumber \\
       &&  \qquad
        -\frac{mJ_r}{E_r}
        \Big(\frac{\tilde{z}^2}{m}\widetilde{V}_{21,x}+\widetilde{V}_{21}+
          \widetilde{V}_{22,x}\Big).
        \end{eqnarray}
  Then substituting  (\ref{5.27}) and the following formulas
  $$\widetilde{V}_{21,x}=\widetilde{V}_{31}+\tilde{z}^{-2}m\widetilde{V}_{23},$$
  $$ \widetilde{V}_{22,x}=\widetilde{V}_{32}-\widetilde{V}_{21}-
    \widetilde{V}_{23}$$
  into (\ref{5.29}), we obtain (\ref{5.21}).
  \quad $\square$ \\

  The properties of $\psi_2(P,x,x_0,t_p,t_{0,p})$ are summarized as
  follows.
   \newtheorem{the5.3}[the5.1]{Theorem}
    \begin{the5.3}
     Assume $(\ref{5.1})$ and $(\ref{5.6})$, $P=(\tilde{z},y)\in
      \mathcal{K}_{r-2}\setminus \{P_{\infty_ i},P_0\},$ $i=1,2,3,$ and let
      $(\tilde{z},x,x_0,t_p,t_{0,p})\in \mathbb{C}^5$. Then
    \begin{eqnarray}\label{5.30}
      \psi_{2,{t_p}}(P,x,x_0,t_p,t_{0,p})&=& \Big(
       \frac{\widetilde{V}_{21}(\tilde{z},x,t_p)}{m(x,t_p)}
        (\tilde{z}^2-\tilde{z}\phi_x(P,x,t_p)-\phi^2(P,x,t_p))
         +\widetilde{V}_{22}(\tilde{z},x,t_p)
       \nonumber \\
     &&
       +\widetilde{V}_{23}(\tilde{z},x,t_p)
        \tilde{z}^{-1}\phi(P,x,t_p) \Big)
        \psi_2(P,x,x_0,t_p,t_{0,p}),
    \end{eqnarray}
    \begin{eqnarray}\label{5.31}
           &&
           \psi_2(P,x,x_0,t_p,t_{0,p})= \mathrm{exp} \Bigg(\tilde{z}^{-1}
       \int_{x_0}^x \phi(P,x^\prime,t_p) dx^\prime
        +\int_{t_{0,p}}^{t_p}    \Big[
             \frac{\tilde{z}^{-1}y(P)-V_{22}(\tilde{z},x_0,t^\prime)}
        {V_{21}(\tilde{z},x_0,t^\prime)}
       \nonumber \\
       && \qquad \qquad  \times
           \widetilde{V}_{21}(\tilde{z},x_0,t^\prime)
         + \Big(\widetilde{V}_{23}(\tilde{z},x_0,t^\prime)
   - \frac{\widetilde{V}_{21}(\tilde{z},x_0,t^\prime)}{V_{21}(\tilde{z},x_0,t^\prime)}
         V_{23}(\tilde{z},x_0,t^\prime) \Big) \tilde{z}^{-1}
         \phi(P,x_0,t^\prime)
       \nonumber \\
   &&   \qquad \qquad
      +\widetilde{V}_{22}(\tilde{z},x_0,t^\prime)\Big] dt^\prime
      \Bigg),
    \end{eqnarray}
    \begin{eqnarray}\label{5.32}
      \psi_2(P,x,x_0,t_p,t_{0,p})\psi_2(P^\ast,x,x_0,t_p,t_{0,p})
        \psi_2(P^{\ast\ast},x,x_0,t_p,t_{0,p})=
      \frac{E_r(\tilde{z},x,t_p)}{E_r(\tilde{z},x_0,t_{0,p})},
      \quad ~~~
    \end{eqnarray}
    \begin{eqnarray}\label{5.33}
        \psi_{2,x}(P,x,x_0,t_p,t_{0,p})\psi_{2,x}(P^\ast,x,x_0,t_p,t_{0,p})
        \psi_{2,x}(P^{\ast\ast},x,x_0,t_p,t_{0,p})=
      - \frac{F_r(\tilde{z},x,t_p)}{E_r(\tilde{z},x_0,t_{0,p})},
      \nonumber \\
    \end{eqnarray}
    \begin{eqnarray}\label{5.34}
    &&
      \psi_2(P,x,x_0,t_p,t_{0,p})=\Big(\frac{E_r(\tilde{z},x,t_p)}
     {E_r(\tilde{z},x_0,t_{0,p})} \Big)^{1/3}
      \nonumber \\
     && \times ~
     \mathrm{exp}\Bigg\{ \int_{x_0}^x \Big(\frac{y(P)^2 V_{21}(\tilde{z},x^\prime,t_p)
     -y(P)A_r(\tilde{z},x^\prime,t_p)
     +\frac{2}{3}S_r(\tilde{z})V_{21}(\tilde{z},x^\prime,t_p)}
      {E_r(\tilde{z},x^\prime,t_p)} \Big) dx^\prime
     \nonumber \\
     &&
     + \int_{t_{0,p}}^{t_p}
      \Big(\frac{y(P)^2 V_{21}(\tilde{z},x_0,t^\prime)
     -y(P)A_r(\tilde{z},x_0,t^\prime)
     +\frac{2}{3}S_r(\tilde{z})V_{21}(\tilde{z},x_0,t^\prime)}
      {E_r(\tilde{z},x_0,t^\prime)} \Big)
      \nonumber \\
     && \times ~
       \Big(\widetilde{V}_{23}(\tilde{z},x_0,t^\prime)
   - \frac{\widetilde{V}_{21}(\tilde{z},x_0,t^\prime)}{V_{21}(\tilde{z},x_0,t^\prime)}
         V_{23}(\tilde{z},x_0,t^\prime) \Big)
    +\tilde{z}^{-1}y(P)
    \frac{\widetilde{V}_{21}(\tilde{z},x_0,t^\prime)}{V_{21}(\tilde{z},x_0,t^\prime)}
       dt^\prime \Bigg\}. \nonumber \\
    \end{eqnarray}
    \end{the5.3}
  \textbf{Proof.}~~Relation (\ref{5.30}) can be proved as follows. Using
  (\ref{5.1}) and (\ref{5.6}), we have
       \begin{eqnarray}\label{5.35}
         \psi_{2,t_p}(P,x,x_0,t_p,t_{0,p}) &=& \widetilde{V}_{21}\psi_1
          +\widetilde{V}_{22}\psi_2+\widetilde{V}_{23}\psi_3
          \nonumber \\
         &=&
         \widetilde{V}_{21} \Big(\frac{\psi_2-\psi_{2,xx}}{m} \Big) \tilde{z}^2
   +\widetilde{V}_{22}\psi_2+\widetilde{V}_{23}\tilde{z}^{-1}\phi\psi_2
         \nonumber \\
        &=&
      \widetilde{V}_{21}
      \Big(\frac{\tilde{z}^2-\tilde{z}\phi_x-\phi^2}{m}\Big)\psi_2
      +\widetilde{V}_{22}\psi_2+\widetilde{V}_{23}\tilde{z}^{-1}\phi\psi_2
        \nonumber \\
       &=& \Big[
        \widetilde{V}_{21}\Big(\frac{\tilde{z}^2-\tilde{z}\phi_x-\phi^2}{m}\Big)
        + \widetilde{V}_{22} + \widetilde{V}_{23}\tilde{z}^{-1}\phi
        \Big]\psi_2. \nonumber \\
       \end{eqnarray}
   Then using ({5.22}), we obtain
       \begin{eqnarray}\label{5.36}
        \psi_2(P,x,x_0,t_p,t_{0,p})&=& \mathrm{exp} \Bigg( \int_{x_0}^x
         \tilde{z}^{-1}\phi(P,x^\prime,t_p) dx^\prime
        +
        \int_{t_{0,p}}^{t_p} \Big[ \widetilde{V}_{21}(\tilde{z},x_0,t^\prime)
        \nonumber \\
       && \times ~
       \Big(\frac{\tilde{z}^2-\tilde{z}\phi_x(P,x_0,t^\prime)
             -\phi^2(P,x_0,t^\prime)}{m}\Big)
       \nonumber \\
       &&
        + \widetilde{V}_{22}(\tilde{z},x_0,t^\prime)
        + \widetilde{V}_{23}(\tilde{z},x_0,t^\prime)\tilde{z}^{-1}\phi(P,x_0,t^\prime)
        \Big]dt^\prime \Bigg)
        \nonumber \\
       &=&
        \mathrm{exp} \Bigg( \int_{x_0}^x
         \tilde{z}^{-1}\phi(P,x^\prime,t_p) dx^\prime
        +
        \int_{t_{0,p}}^{t_p} \Big[ \widetilde{V}_{21}(\tilde{z},x_0,t^\prime)
          \nonumber \\
        && \times ~
        \Big( \frac{\tilde{z}^{-1}y(P)-V_{22}(\tilde{z},x_0,t^\prime)}
        {V_{21}(\tilde{z},x_0,t^\prime)} \Big)
        + \widetilde{V}_{22}(\tilde{z},x_0,t^\prime)
                \nonumber \\
        &&
                  +\Big(\widetilde{V}_{23}(\tilde{z},x_0,t^\prime)
   - \frac{\widetilde{V}_{21}(\tilde{z},x_0,t^\prime)}{V_{21}(\tilde{z},x_0,t^\prime)}
         V_{23}(\tilde{z},x_0,t^\prime) \Big)
           \nonumber \\
       && \times ~
         \tilde{z}^{-1}
         \phi(P,x_0,t^\prime) \Big]dt^\prime \Bigg),
    \end{eqnarray}
  which is (\ref{5.31}). \\
  Hence
    \begin{eqnarray*}
       &&
        \psi_2(P,x,x_0,t_p,t_{0,p}) \psi_2(P^\ast,x,x_0,t_p,t_{0,p})
         \psi_2(P^{\ast\ast},x,x_0,t_p,t_{0,p})
        \nonumber \\
       &&
       = \mathrm{exp} \Bigg(\tilde{z}^{-1}
       \int_{x_0}^x \phi(P,x^\prime,t_p) dx^\prime
        +\int_{t_{0,p}}^{t_p}    \Big[
             \frac{\tilde{z}^{-1}y(P)-V_{22}(\tilde{z},x_0,t^\prime)}
        {V_{21}(\tilde{z},x_0,t^\prime)}
        \widetilde{V}_{21}(\tilde{z},x_0,t^\prime)
       \nonumber \\
       && \qquad
       + \Big(\widetilde{V}_{23}(\tilde{z},x_0,t^\prime)
   - \frac{\widetilde{V}_{21}(\tilde{z},x_0,t^\prime)}{V_{21}(\tilde{z},x_0,t^\prime)}
         V_{23}(\tilde{z},x_0,t^\prime) \Big) \tilde{z}^{-1}
         \phi(P,x_0,t^\prime)
       \nonumber \\
       && \qquad
    +\widetilde{V}_{22}(\tilde{z},x_0,t^\prime)\Big] dt^\prime
      \Bigg)
       \nonumber \\
    &&  \times ~
    \mathrm{exp} \Bigg(\tilde{z}^{-1}
       \int_{x_0}^x \phi(P^\ast,x^\prime,t_p) dx^\prime
        +\int_{t_{0,p}}^{t_p}    \Big[
             \frac{\tilde{z}^{-1}y(P^\ast)-V_{22}(\tilde{z},x_0,t^\prime)}
        {V_{21}(\tilde{z},x_0,t^\prime)}
        \widetilde{V}_{21}(\tilde{z},x_0,t^\prime)
       \nonumber \\
       && \qquad
       + \Big(\widetilde{V}_{23}(\tilde{z},x_0,t^\prime)
   - \frac{\widetilde{V}_{21}(\tilde{z},x_0,t^\prime)}{V_{21}(\tilde{z},x_0,t^\prime)}
         V_{23}(\tilde{z},x_0,t^\prime) \Big) \tilde{z}^{-1}
         \phi(P^\ast,x_0,t^\prime)
       \nonumber \\
       && \qquad
    +\widetilde{V}_{22}(\tilde{z},x_0,t^\prime)\Big] dt^\prime
      \Bigg)
    \end{eqnarray*}
  \begin{eqnarray}\label{5.37}
    &&  \times ~
    \mathrm{exp} \Bigg(\tilde{z}^{-1}
       \int_{x_0}^x \phi(P^{\ast\ast},x^\prime,t_p) dx^\prime
        +\int_{t_{0,p}}^{t_p}    \Big[
             \frac{\tilde{z}^{-1}y(P^{\ast\ast})-V_{22}(\tilde{z},x_0,t^\prime)}
        {V_{21}(\tilde{z},x_0,t^\prime)}
        \widetilde{V}_{21}(\tilde{z},x_0,t^\prime)
       \nonumber \\
       && \qquad
       + \Big(\widetilde{V}_{23}(\tilde{z},x_0,t^\prime)
   - \frac{\widetilde{V}_{21}(\tilde{z},x_0,t^\prime)}{V_{21}(\tilde{z},x_0,t^\prime)}
         V_{23}(\tilde{z},x_0,t^\prime) \Big) \tilde{z}^{-1}
         \phi(P^{\ast\ast},x_0,t^\prime)
       \nonumber \\
       && \qquad
    +\widetilde{V}_{22}(\tilde{z},x_0,t^\prime)\Big] dt^\prime
      \Bigg)
     \nonumber \\
    &&
    =\mathrm{exp} \Bigg( \int_{x_0}^x \tilde{z}^{-1}
      [\phi(P,x^\prime,t_p)+\phi(P^\ast,x^\prime,t_p)
       +\phi(P^{\ast\ast},x^\prime,t_p)] dx^\prime
       \nonumber \\
    && \qquad +
     \int_{t_{0,p}}^{t_p}
     \Big[3\Big(\widetilde{V}_{22}(\tilde{z},x_0,t^\prime)
      -\frac{\widetilde{V}_{21}(\tilde{z},x_0,t^\prime)}{V_{21}(\tilde{z},x_0,t^\prime)}
          V_{22}(\tilde{z},x_0,t^\prime)\Big)
      \nonumber \\
     && \qquad +
     \tilde{z}^{-1}
     \Big(\widetilde{V}_{23}(\tilde{z},x_0,t^\prime)
      -\frac{\widetilde{V}_{21}(\tilde{z},x_0,t^\prime)}{V_{21}(\tilde{z},x_0,t^\prime)}
          V_{23}(\tilde{z},x_0,t^\prime) \Big)
       \nonumber \\
      && \qquad \times ~
       [\phi(P,x_0,t^\prime)+\phi(P^\ast,x_0,t^\prime)
       +\phi(P^{\ast\ast},x_0,t^\prime)] \Big] dt^\prime \Bigg)
       \nonumber \\
      &&
      =\mathrm{exp} \Bigg( \int_{x_0}^x
        \frac{E_{r,x^\prime}(\tilde{z},x^\prime,t_p)}{E_r(\tilde{z},x^\prime,t_p)}
         dx^\prime
      +
     \int_{t_{0,p}}^{t_p}
     \Big[3\Big(\widetilde{V}_{22}(\tilde{z},x_0,t^\prime)
      -\frac{\widetilde{V}_{21}(\tilde{z},x_0,t^\prime)}{V_{21}(\tilde{z},x_0,t^\prime)}
          V_{22}(\tilde{z},x_0,t^\prime)\Big)
      \nonumber \\
     && \qquad +
      \Big(\widetilde{V}_{23}(\tilde{z},x_0,t^\prime)
      -\frac{\widetilde{V}_{21}(\tilde{z},x_0,t^\prime)}{V_{21}(\tilde{z},x_0,t^\prime)}
          V_{23}(\tilde{z},x_0,t^\prime) \Big)
      \Big(
      \frac{E_{r,x}(\tilde{z},x_0,t^\prime)}{E_r(\tilde{z},x_0,t^\prime)}
      \Big) \Big] dt^\prime \Bigg)
       \nonumber \\
      &&
      =\mathrm{exp} \Bigg( \int_{x_0}^x
      \partial_{x^\prime} (\mathrm{ln}E_r(\tilde{z},x^\prime,t_p)) dx^\prime
      +\int_{t_{0,p}}^{t_p}
      \partial_{t^\prime} (\mathrm{ln}E_r(\tilde{z},x_0,t^\prime)) dt^\prime
       \Bigg)
       \nonumber \\
      &&
     =\frac{E_r(\tilde{z},x,t_p)}{E_r(\tilde{z},x_0,t_{0,p})}.
     \end{eqnarray}
  Then the relation (\ref{5.33}) is follows from (\ref{5.37}) and
  (\ref{5.15}), that is
      \begin{eqnarray}\label{5.38}
        &&
         \psi_{2,x}(P,x,x_0,t_p,t_{0,p}) ~\times ~\psi_{2,x}(P^\ast,x,x_0,t_p,t_{0,p})
          ~\times~\psi_{2,x}(P^{\ast\ast},x,x_0,t_p,t_{0,p})
           \nonumber \\
        &&
         =\tilde{z}^{-1}\phi(P,x,t_p)\psi_2(P,x,x_0,t_p,t_{0,p})
         ~ \times ~
         \tilde{z}^{-1}\phi(P^\ast,x,t_p)\psi_2(P^\ast,x,x_0,t_p,t_{0,p})
            \nonumber \\
         && ~~
           \times ~
  \tilde{z}^{-1}\phi(P^{\ast\ast},x,t_p)\psi_2(P^{\ast\ast},x,x_0,t_p,t_{0,p})
           \nonumber \\
       &&
       =-\frac{F_r(\tilde{z},x,t_p)}{E_r(\tilde{z},x_0,t_{0,p})}.
      \end{eqnarray}
  Moreover, using (\ref{5.20}), we arrive at
      \begin{eqnarray*}
       &&
        \psi_2(P,x,x_0,t_p,t_{0,p})
        =\mathrm{exp} \Bigg( \int_{x_0}^x
         \tilde{z}^{-1}\phi(P,x^\prime,t_p) dx^\prime
        +
        \int_{t_{0,p}}^{t_p} \Big[
        \widetilde{V}_{23}(\tilde{z},x_0,t^\prime)\tilde{z}^{-1}\phi(P,x_0,t^\prime)
      \end{eqnarray*}
      \begin{eqnarray*}
       &&  \qquad
       +\widetilde{V}_{21}(\tilde{z},x_0,t^\prime)
       \Big(\frac{\tilde{z}^2-\tilde{z}\phi_x(P,x_0,t^\prime)
             -\phi^2(P,x_0,t^\prime)}{m}\Big)
       + \widetilde{V}_{22}(\tilde{z},x_0,t^\prime)
        \Big]dt^\prime \Bigg)
        \\
       &&
        =\mathrm{exp} \Bigg( \int_{x_0}^x
         \tilde{z}^{-1}\phi(P,x^\prime,t_p) dx^\prime
        +
        \int_{t_{0,p}}^{t_p} \Big[ \widetilde{V}_{21}(\tilde{z},x_0,t^\prime)
        \Big( \frac{\tilde{z}^{-1}y(P)-V_{22}(\tilde{z},x_0,t^\prime)}
        {V_{21}(\tilde{z},x_0,t^\prime)} \Big)
        \\
        && \qquad
         + \widetilde{V}_{22}(\tilde{z},x_0,t^\prime)
         +\Big(\widetilde{V}_{23}(\tilde{z},x_0,t^\prime)
   - \frac{\widetilde{V}_{21}(\tilde{z},x_0,t^\prime)}{V_{21}(\tilde{z},x_0,t^\prime)}
         V_{23}(\tilde{z},x_0,t^\prime) \Big)
          \tilde{z}^{-1}
         \phi(P,x_0,t^\prime) \Big]dt^\prime \Bigg)
      \\
     &&
     = \mathrm{exp} \Bigg( \int_{x_0}^x \Big(
       \frac{y(P)^2V_{21}(\tilde{z},x^\prime,t_p)-y(P)A_r(\tilde{z},x^\prime,t_p)
       +B_r(\tilde{z},x^\prime,t_p)}{E_r(\tilde{z},x^\prime,t_p)}
       \Big)
       dx^\prime
        \\
     && \qquad
       + \int_{t_{0,p}}^{t_p} M(\tilde{z},x_0,t^\prime) dt^\prime
       \Bigg),
    \end{eqnarray*}
  where
      \begin{eqnarray}\label{5.39}
      M(\tilde{z},x_0,t^\prime)&=&
        \widetilde{V}_{21}(\tilde{z},x_0,t^\prime)
        \Big( \frac{\tilde{z}^{-1}y(P)-V_{22}(\tilde{z},x_0,t^\prime)}
        {V_{21}(\tilde{z},x_0,t^\prime)} \Big)
        + \widetilde{V}_{22}(\tilde{z},x_0,t^\prime)
        \nonumber \\
        &&
         +\Big(\widetilde{V}_{23}(\tilde{z},x_0,t^\prime)
   - \frac{\widetilde{V}_{21}(\tilde{z},x_0,t^\prime)}{V_{21}(\tilde{z},x_0,t^\prime)}
         V_{23}(\tilde{z},x_0,t^\prime) \Big)
          \tilde{z}^{-1}
         \phi(P,x_0,t^\prime). \nonumber \\
      \end{eqnarray}
  From (\ref{5.20}), it is easy to see that
      \begin{eqnarray}\label{5.40}
        &&
        \Big(\widetilde{V}_{22}(\tilde{z},x_0,t^\prime)
        -\frac{\widetilde{V}_{21}(\tilde{z},x_0,t^\prime)}
         {V_{21}(\tilde{z},x_0,t^\prime)}V_{22}(\tilde{z},x_0,t^\prime)\Big)
   =\frac{1}{3}\frac{E_{r,t^\prime}(\tilde{z},x_0,t^\prime)}{E_r(\tilde{z},x_0,t^\prime)}
           \nonumber \\
         && \qquad
          -\frac{1}{3}
          \Big(\widetilde{V}_{23}(\tilde{z},x_0,t^\prime)
          -\frac{\widetilde{V}_{21}(\tilde{z},x_0,t^\prime)}
         {V_{21}(\tilde{z},x_0,t^\prime)}V_{23}(\tilde{z},x_0,t^\prime)\Big)
         \frac{E_{r,x}(\tilde{z},x_0,t^\prime)}{E_r(\tilde{z},x_0,t^\prime)}.
          \nonumber \\
      \end{eqnarray}
  Inserting (\ref{5.40}) into (\ref{5.39}), we arrive at
       \begin{eqnarray}\label{5.41}
                  M(\tilde{z},x_0,t^\prime)&=&
         \Big(\widetilde{V}_{23}(\tilde{z},x_0,t^\prime)
   - \frac{\widetilde{V}_{21}(\tilde{z},x_0,t^\prime)}{V_{21}(\tilde{z},x_0,t^\prime)}
         V_{23}(\tilde{z},x_0,t^\prime) \Big)
                      \nonumber \\
          &\times& \Big( \tilde{z}^{-1}\phi(P,x_0,t^\prime)
    -\frac{1}{3}\frac{E_{r,x}(\tilde{z},x_0,t^\prime)}
    {E_r(\tilde{z},x_0,t^\prime)} \Big)
     +\frac{1}{3}\frac{E_{r,t^\prime}(\tilde{z},x_0,t^\prime)}
     {E_r(\tilde{z},x_0,t^\prime)}
     \nonumber \\
     &+&
     \tilde{z}^{-1}y(P)
       \frac{\widetilde{V}_{21}(\tilde{z},x_0,t^\prime)}{V_{21}(\tilde{z},x_0,t^\prime)}.
       \end{eqnarray}
  Substituting (\ref{5.41}) into the above representation of
  $\psi_2$, we have
       \begin{eqnarray*}
       &&
         \psi_2(P,x,x_0,t_p,t_{0,p})=
         \Big( \frac{E_r(\tilde{z},x,t_p)}{E_r(\tilde{z},x_0,t_p)}
         \Big)^{1/3}
          \\
       && \times ~
        \mathrm{exp} \Bigg( \int_{x_0}^x
        \Big(
       \frac{y(P)^2V_{21}(\tilde{z},x^\prime,t_p)-y(P)A_r(\tilde{z},x^\prime,t_p)
       +\frac{2}{3}V_{21}(\tilde{z},x^\prime,t_p)S_r(\tilde{z})}
       {E_r(\tilde{z},x^\prime,t_p)}
       \Big) dx^\prime \Bigg)  \\
       && \times ~
       \Big( \frac{E_r(\tilde{z},x_0,t_p)}{E_r(\tilde{z},x_0,t_{0,p})}
         \Big)^{1/3}  \\
       && \times ~
        \mathrm{exp} \Bigg(
       \int_{t_{0,p}}^{t_p}
      \Big(\frac{y(P)^2 V_{21}(\tilde{z},x_0,t^\prime)
     -y(P)A_r(\tilde{z},x_0,t^\prime)
     +\frac{2}{3}S_r(\tilde{z})V_{21}(\tilde{z},x_0,t^\prime)}
      {E_r(\tilde{z},x_0,t^\prime)} \Big)
       \\
     && \times ~
     \Big(\widetilde{V}_{23}(\tilde{z},x_0,t^\prime)
   - \frac{\widetilde{V}_{21}(\tilde{z},x_0,t^\prime)}{V_{21}(\tilde{z},x_0,t^\prime)}
         V_{23}(\tilde{z},x_0,t^\prime) \Big)
    +\tilde{z}^{-1}y(P)
    \frac{\widetilde{V}_{21}(\tilde{z},x_0,t^\prime)}{V_{21}(\tilde{z},x_0,t^\prime)}
       dt^\prime \Bigg),
       \end{eqnarray*}
  which implies (\ref{5.34}). \quad $\square$ \\

  The stationary Dubrovin-type equations in Lemma 3.3 have analogs
  for each DP$_p$ flow (indexed by the parameter $t_p$), which
  govern the dynamics of $\mu_j(x,t_p)$ and $\nu_j(x,t_p)$ with
  respect to variations of $x$ and $t_p$. In this context the
  stationary case simply corresponds to the special case $p=0$ as
  described in the following result.

   \newtheorem{lem5.4}[the5.1]{Lemma}
    \begin{lem5.4}
         Assume $(\ref{5.1})-(\ref{5.7})$.\\
     $(\mathrm{i})$
    Suppose the zeros $\{\mu_j(x,t_p)\}_{j=1,\ldots,r-5}$
     of $E_r(\tilde{z},x,t_p)$ remain distinct for $(x,t_p) \in
     \Omega_\mu,$ where $\Omega_\mu \subseteq \mathbb{C}^2$ is open
     and connected. Then
     $\{\mu_j(x,t_p)\}_{j=1,\ldots,r-5}$ satisfy the system of
     differential equations,
      \begin{eqnarray} \label{5.42}
        \mu_{j,x}(x,t_p)=-
   \frac{[S_r(\mu_j(x,t_p))+3y(\hat{\mu}_j(x,t_p))^2]V_{21}(\mu_j(x,t_p),x,t_p)}
            {u(x,t_p)
            \prod_{\scriptstyle k=1 \atop \scriptstyle k \neq j }^{r-5}
            (\mu_j(x,t_p)-\mu_k(x,t_p))},
        \nonumber \\
            \quad j=1,\ldots,r-5,
      \end{eqnarray}
       \begin{eqnarray}\label{5.43}
        \mu_{j,t_p}(x,t_p)&=& -
        [V_{21}(\mu_j(x,t_p),x,t_p)\widetilde{V}_{23}(\mu_j(x,t_p),x,t_p)
         \nonumber \\
       &&
        -\widetilde{V}_{21}(\mu_j(x,t_p),x,t_p)V_{23}(\mu_j(x,t_p),x,t_p)]
         \nonumber \\
       && \times ~
  \frac{[S_r(\mu_j(x,t_p))+3y(\hat{\mu}_j(x,t_p))^2]}
            {u(x,t_p)
            \prod_{\scriptstyle k=1 \atop \scriptstyle k \neq j }^{r-5}
            (\mu_j(x,t_p)-\mu_k(x,t_p))},
      \nonumber \\
      &&    ~~~~~~~~~~~~~~~~~~~~~~~~~~~~~~~
            \quad j=1,\ldots,r-5,
       \end{eqnarray}
   with initial conditions
       \begin{equation}\label{5.44}
         \{\hat{\mu}_j(x_0,t_{0,p})\}_{j=1,\ldots,r-5}
         \in \mathcal{K}_{r-2},
       \end{equation}
   for some fixed $(x_0,t_{0,p}) \in \Omega_\mu$. The initial value
   problem $(\ref{5.43})$, $(\ref{5.44})$ has a unique solution
   satisfying
        \begin{equation}\label{5.45}
         \hat{\mu}_j \in C^\infty(\Omega_\mu,\mathcal{K}_{r-2}),
         \quad j=1,\ldots,r-5.
        \end{equation}

   $\mathrm{(ii)}$
     Suppose the zeros $\{\nu_j(x,t_p)\}_{j=1,\ldots,r-3}$
     of $F_r(\tilde{z},x,t_p)$ remain distinct for $(x,t_p) \in
     \Omega_\nu,$ where $\Omega_\nu \subseteq \mathbb{C}^2$ is open
     and connected. Then
     $\{\nu_j(x,t_p)\}_{j=1,\ldots,r-3}$ satisfy the system of
     differential equations,
        \begin{eqnarray}\label{5.46}
        \nu_{j,x}(x,t_p)&=&
       \nu_j(x,t_p)^{2}\Big([S_r(\nu_j(x,t_p))
        +3y(\hat{\nu}_j(x,t_p))^2]
        \nonumber \\
       &&  \times ~
           V_{31}(\nu_j(x,t_p),x,t_p)
            +m(x,t_p)J_r(\nu_j(x,t_p),x,t_p) \Big)
         \nonumber \\
       &&  \times ~
       \frac{1}
       {u(x,t_p)u_x^2(x,t_p)
            \prod_{\scriptstyle k=1 \atop \scriptstyle k \neq j }^{r-3}
            (\nu_j(x,t_p)-\nu_k(x,t_p))},
       \nonumber \\
       &&  ~~~~~~~~~~~~~~~~~~~~~~~~~~~~~~
                    j=1,\ldots,r-3,
        \end{eqnarray}
      \begin{eqnarray}\label{5.47}
         \nu_{j,t_p}(x,t_p)&=&
         \nu_j(x,t_p)^{2}\Big([S_r(\nu_j(x,t_p))
        +3y(\hat{\nu}_j(x,t_p))^2]
        \nonumber \\
       &&  \times ~
           V_{31}(\nu_j(x,t_p),x,t_p)
           \widetilde{V}_{32}(\nu_j(x,t_p),x,t_p)
         \nonumber \\
       &&
          -\nu_j(x,t_p)^{2}  J_r(\nu_j(x,t_p),x,t_p)
           \widetilde{V}_{31}(\nu_j(x,t_p),x,t_p)\Big)
         \nonumber \\
       &&  \times ~
       \frac{1}
       {u(x,t_p)u_x^2(x,t_p)
            \prod_{\scriptstyle k=1 \atop \scriptstyle k \neq j }^{r-3}
            (\nu_j(x,t_p)-\nu_k(x,t_p))},
       \nonumber \\
       &&  ~~~~~~~~~~~~~~~~~~~~~~~~~~~~~~
                    j=1,\ldots,r-3,
      \end{eqnarray}
   with initial conditions
       \begin{equation}\label{5.48}
         \{\hat{\nu}_j(x_0,t_{0,p})\}_{j=1,\ldots,r-3}
         \in \mathcal{K}_{r-2},
       \end{equation}
   for some fixed $(x_0,t_{0,p}) \in \Omega_\nu$. The initial value
   problem $(\ref{5.47})$, $(\ref{5.48})$ has a unique solution
   satisfying
        \begin{equation}\label{5.49}
         \hat{\nu}_j \in C^\infty(\Omega_\nu,\mathcal{K}_{r-2}),
         \quad j=1,\ldots,r-3.
        \end{equation}
    \end{lem5.4}
  \textbf{Proof.}~~For obvious reasons in suffices to focus on (\ref{5.42})
  and
  (\ref{5.43}). But the proof of (\ref{5.42}) is
  identical to that in Lemma 3.3. We now prove (\ref{5.43}). From
  (\ref{5.8}), we have
        \begin{equation}\label{5.50}
          E_{r,t_p}(\tilde{z},x,t_p)|_{\tilde{z}=\mu_j(x,t_p)}=
           -u(x,t_p)\mu_{j,t_p}(x,t_p)
           \prod_{\scriptstyle k=1 \atop \scriptstyle k \neq j }^{r-5}
            (\mu_j(x,t_p)-\mu_k(x,t_p)).
        \end{equation}
  On the other hand, using (\ref{5.20}) and (\ref{5.42}), one
  computes
      \begin{eqnarray}\label{5.51}
       E_{r,t_p}(\tilde{z},x,t_p)|_{\tilde{z}=\mu_j(x,t_p)}&=&
         E_{r,x}(\mu_j(x,t_p),x,t_p) \Big(\widetilde{V}_{23}-
         \frac{\widetilde{V}_{21}}{V_{21}}V_{23}\Big)
          \nonumber \\
         &=&
          -u(x,t_p)\mu_{j,x}(x,t_p)
           \prod_{\scriptstyle k=1 \atop \scriptstyle k \neq j }^{r-5}
            (\mu_j(x,t_p)-\mu_k(x,t_p))
           \nonumber \\
          &&  \times ~
           \Big(\widetilde{V}_{23}-
         \frac{\widetilde{V}_{21}}{V_{21}}V_{23}\Big)
           \nonumber \\
         &=&
        V_{21}[S_r(\mu_j(x,t_p))+3y(\hat{\mu}_j(x,t_p))^2]
        \Big(\widetilde{V}_{23}-
         \frac{\widetilde{V}_{21}}{V_{21}}V_{23}\Big)
          \nonumber \\
         &=&
         [S_r(\mu_j(x,t_p))+3y(\hat{\mu}_j(x,t_p))^2]
         (V_{21}\widetilde{V}_{23}-\widetilde{V}_{21}V_{23}),
         \nonumber \\
      \end{eqnarray}
  which together with (\ref{5.50}) yields (\ref{5.43}).
  \quad $\square$

The analog of Remark 3.4 directly extends to the current
time-dependent setting.

\section{Time-dependent algebro-geometric solutions }

  In the  final  section,  we extend the results of section 4
  from the
  stationary DP hierarchy to the time-dependent case. In particular,
  we obtain Riemann theta function representations for the
  Baker-Akhiezer function, the meromorphic function $\phi$ and the
 algebro-geometric solutions for  the DP hierarchy.

  We start with the theta function representation of the meromorphic
  function $\phi(P,x,t_p)$.

  \newtheorem{the6.1}{Theorem}[section]
    \begin{the6.1}
       Assume that the curve $\mathcal{K}_{r-2}$ is nonsingular.
        Let $P=(\tilde{z},y) \in \mathcal{K}_{r-2} \setminus
        \{P_{\infty_1},P_0\}$ and let $(x,t_p),(x_0,t_{0,p}) \in \Omega_\mu$,
        where
        $\Omega_\mu \subseteq \mathbb{C}^2$ is open and connected.
        Suppose that $\mathcal{D}_{\underline{\hat{\mu}}(x,t_p)}$, or
        equivalently, $\mathcal{D}_{\underline{\hat{\nu}}(x,t_p)}$ is
        nonspecial for $(x,t_p) \in \Omega_\mu$. Then

      \begin{equation}\label{6.1}
      \begin{split}
      &  \phi(P,x,t_p)=-m^{\frac{1}{3}}(x,t_p)\frac{\theta(\underline{\tilde{z}}(P,\underline{\hat{\nu}}(x,t_p)))
            \theta(\underline{\tilde{z}}(P_{0},\underline{\hat{\mu}}(x,t_p)))}
            {\theta(\underline{\tilde{z}}(P_{0},\underline{\hat{\nu}}(x,t_p)))
            \theta(\underline{\tilde{z}}(P,\underline{\hat{\mu}}(x,t_p)))} \\
          &~~~~~~~~~~~~~~~~\times  \mathrm{exp}\left(e^{(3)}(Q_0)
            -\int_{Q_0}^P \omega_{P_{\infty_1} P_0}^{(3)}\right).   \\
      \end{split}
      \end{equation}
    \end{the6.1}
  \textbf{Proof.}~~The proof of (\ref{6.1}) is analogous to the stationary
  case in Theorem \ref{them4.30}. \quad $\square$  \\

Motivated by (\ref{5.30}), we define the meromorphic function
$I_s(P,x,t_p)$ on $\mathcal{K}_{r-2} \times \mathbb{C}^2$ by
     \begin{eqnarray}\label{6.2}
        I_s(P,x,t_p)&=&
        \frac{\widetilde{V}_{21}(\tilde{z},x,t_p)}{m(x,t_p)}
        (\tilde{z}^2-\tilde{z}\phi_x(P,x,t_p)-\phi^2(P,x,t_p))
         +\widetilde{V}_{22}(\tilde{z},x,t_p)
       \nonumber \\
     &&
        +\widetilde{V}_{23}(\tilde{z},x,t_p)
        \tilde{z}^{-1}\phi(P,x,t_p).
     \end{eqnarray}
The asymptotic properties of $I_s(P,s,t_p)$ are summarized as
follows.
  \newtheorem{the6.2}[the6.1]{Theorem}
    \begin{the6.2}\label{them6.2}
      Let $s=4p+2$, $p \in \mathbb{N}_0$, $(x,t_p) \in
      \mathbb{C}^2$. Then
    \footnote{Here sums with upper limits strictly less than their
    lower limits are interpreted as zero.}
        \begin{eqnarray}
        I_s(P,x,t_p) & \underset{\zeta \rightarrow 0}{=} &
         \frac{2}{3}\zeta^{-s}+\sum_{j=0}^{\frac{s-4}{2}}
         \tilde{\alpha}_{j}\zeta^{-(s-2j-2)}+ \chi_{\frac{s-2}{2}}+O(\zeta^2), \label{6.3}
            \\
         && ~~~~~~~~~~~~~~~~~~~~~~
         \zeta=\tilde{z}^{-1},
         \quad \textrm{as $P \rightarrow P_{\infty_1}$},
            \nonumber \\
        I_2(P,x,t_0)& \underset{\zeta \rightarrow 0}{=}&
        u(x,t_0)m^{1/3}(x,t_0)\zeta^{-2}+O(\zeta^2), \label{6.3c0}
           \\
        && ~~~~~~~~~~~~~~~~~~~~~~
        \zeta=\tilde{z}^{1/3},
        \quad \textrm{as $P\rightarrow P_0$}, \nonumber
        \end{eqnarray}
        where
        $
          \{ \tilde{\alpha}_j \}_{j=0,\ldots,\frac{s-4}{2}}\in\mathbb{C},
        $
        and
        \begin{eqnarray*}
        \chi_0&=&-\frac{u}{m}\left(\kappa_2+2\kappa_0\kappa_{2,x}\right)-u\kappa_0,
        \\
       \chi_{\frac{s-2}{2}} &=&m^{-1} \sum_{k=-2}^{s-2}
       \vartheta_k V_{21}^{([\frac{2j-k+4}{4}],
       \frac{s+2-k}{4}-[\frac{s+2-k}{4}])}
       + \sum_{\ell=0}^{s-2}\kappa_{\ell}V_{23}^{([\frac{s-\ell}{4}],
       \frac{s-\ell}{4}-[\frac{s-\ell}{4}])}, \quad s>2,
        \end{eqnarray*}
    the function $[\, \cdot \, ]$ returns the value of a number rounded downwards to the nearest integer.
    \end{the6.2}
\textbf{Proof.}~~Treating $t_p$ as a parameter, we note that the
asymptotic expansions of $\phi(P)$ near $P_{\infty_1}$ and near
$P_0$ in (\ref{4.1}) and (\ref{4.2a}) still apply in the present
time-dependent context. In terms of local coordinate
$\zeta=\tilde{z}^{-1}$ near $P_{\infty_1}$, from (\ref{4.1}),
(\ref{5.2}) and (\ref{6.2}), we easily get
    \begin{align}\label{6.4}
        I_2(P,x,t_0)&=
        \frac{ \widetilde{V}_{21}(\tilde{z},x,t_0)}{m(x,t_0)}
        \left(\tilde{z}^2-\tilde{z}\phi_x(P,x,t_0)-\phi^2(P,x,t_0)\right)
         + \widetilde{V}_{22}(\tilde{z},x,t_0)
         \nonumber \\
      & ~~~~
      + \widetilde{V}_{23}(\tilde{z},x,t_0)
        \tilde{z}^{-1}\phi(P,x,t_0)\nonumber\\
     &= \frac{u(x,t_0)}{m(x,t_0)}\Big[\left(1-\kappa_{0,x}-
         \kappa_0^2\right)\zeta^{-2}-\left(\kappa_{2,x}
       +2\kappa_0\kappa_{2,x}\right)\Big]-\frac{1}{3}\zeta^{-2}
         \nonumber\\
     & ~~~~
     -u(x,t_0)\kappa_0+O(\zeta^2)\nonumber\\
     &=\frac{2}{3}\zeta^{-2}+\chi_0+O(\zeta^2),
     \qquad \textrm{as $P \rightarrow P_{\infty_1}$},
    \end{align}
where
    \begin{equation*}
      \chi_0=-\frac{u}{m}\left(\kappa_2+2\kappa_0\kappa_{2,x}\right)-u\kappa_0,
    \end{equation*}
    and
    \begin{equation*}
      \kappa_0=\frac{u_x}{u},\quad \kappa_{0,x}
      =\frac{u_{xx}}{u}-\left(\frac{u_x}{u}\right)^2.
    \end{equation*}
Therefore (\ref{6.3}) holds for $s=2$. For $s>2$, recall the
definitions of $\widetilde{V}_{21},\widetilde{ V}_{22},\widetilde{
V}_{23}$ in (\ref{5.2}), we may write
    \begin{align*}
    \widetilde{V}_{21}&=V_{21}^{(0,1)}\tilde{z}^{4p}
     +V_{21}^{(1,0)}\tilde{z}^{4p-2}+V_{21}^{(1,1)}\tilde{z}^{4p-4}+\ldots+
      V_{21}^{(p,0)}\tilde{z}^2+V_{21}^{(p,1)}\\
      &= \sum_{j=0}^{2p}V_{21}^{([\frac{j+1}{2}],
          j+1-[\frac{j+1}{2}])}\zeta^{-(4p-2j)}\\
      &=\sum_{j=0}^{\infty} V_{21}^{([\frac{j+1}{2}],
          j+1-[\frac{j+1}{2}])}\zeta^{-(4p-2j)}, \\
    \widetilde{V}_{22}&=V_{22}^{(0,0)}\tilde{z}^{4p+2}+V_{21}^{(1,0)}
         \tilde{z}^{4p-2}+V_{22}^{(2,0)}\tilde{z}^{4p-6}+\ldots+
      V_{22}^{(p,0)}\tilde{z}^2 ,\\
      &=\sum_{j=0}^{2p}V_{22}^{([\frac{j}{2}],j-[\frac{j}{2}])}\zeta^{-(4p+2-2j)}\\
      &=\sum_{j=0}^{\infty}V_{22}^{([\frac{j}{2}],j-[\frac{j}{2}])}\zeta^{-(4p+2-2j)},\\
    \widetilde{V}_{23}&=V_{23}^{(0,1)}\tilde{z}^{4p}+
       V_{23}^{(1,0)}\tilde{z}^{4p-2}+V_{23}^{(1,1)}\tilde{z}^{4p-4}+\ldots+
        V_{21}^{(p,0)}\tilde{z}^2+V_{23}^{(p,1)}\\
      &= \sum_{j=0}^{2p}V_{23}^{([\frac{j+1}{2}],j+1-[\frac{j+1}{2}])}\zeta^{-(4p-2j)},
      \end{align*}
where
    \begin{eqnarray*}
     &&  V_{21}^{(\beta_1, \beta_2)}=
     V_{23}^{(\beta_1, \beta_2)}=0
     \quad \text{for}\quad \beta_1\geq p+1,
     \quad \beta_1,\beta_2\in\mathbb{N}, \\
     && V_{22}^{(\beta_1, \beta_2)}=0
     \quad \text{for}\quad \beta_1\geq p+1\quad
     \text{or}\quad \beta_2=1,\quad \beta_1,\beta_2\in\mathbb{N}.
    \end{eqnarray*}
Moreover, from (\ref{4.1}), we find
    \begin{eqnarray*}
     &&
     \tilde{z}^{2}-\tilde{z}\phi_x(P,x,t_p)-\phi^2(P,x,t_p)
     \underset{\zeta\rightarrow 0}{=}
     \sum_{j=-1}^{\infty}\vartheta_{2j}\zeta^{2j},\\
    &&~~~~~~~~~~~~~~~~~~~~~~~~
    \underset{\zeta\rightarrow 0}{=}
    \sum_{j=-1}^{\infty}\left(\vartheta_{2j}\zeta^{2j}
    +\vartheta_{2j+1}\zeta^{2j+1}\right),
    \quad \textrm{as $P \rightarrow P_{\infty_1}$},
    \end{eqnarray*}
with
    \begin{eqnarray*}
    && \vartheta_{-2}=1-\kappa_{0,x}-\kappa_0^2=\frac{m(x,t_p)}{u(x,t_p)},\\
    && \vartheta_{2j}=-\kappa_{2j+2}-\sum_{i=0}^{2j+2}\kappa_i\kappa_{2j+2-i},\\
    && \vartheta_{2j+1}=0 ,\quad j\in\mathbb{N}_0.
    \end{eqnarray*}
Therefore, in terms of local coordinate $\zeta=\tilde{z}^{-1}$ near
$P_{\infty_1}$, we obtain
    \begin{eqnarray*}
        I_s(P,x,t_p)&=&
        \frac{\widetilde{V}_{21}(\tilde{z},x,t_p)}{m(x,t_p)}
        (\tilde{z}^2-\tilde{z}\phi_x(P,x,t_p)-\phi^2(P,x,t_p))
         +\widetilde{V}_{22}(\tilde{z},x,t_p)
       \nonumber \\
     &&
        +\widetilde{V}_{23}(\tilde{z},x,t_p)
        \tilde{z}^{-1}\phi(P,x,t_p)\nonumber\\
     &=& m^{-1}(x,t_p)\Big(\sum_{j=0}^{2p}V_{21}^{([\frac{j+1}{2}],
         j+1-[\frac{j+1}{2}])}\zeta^{-(4p-2j)}\Big)\Big(
     \sum_{s=-1}^{\infty}\vartheta_{2s}\zeta^{2s}
     \Big)\nonumber\\
     &&
     +\sum_{j=0}^{2p}V_{22}^{([\frac{j}{2}],
          j-[\frac{j}{2}])}\zeta^{-(4p+2-2j)}
     +\Big(\sum_{j=0}^{2p}V_{23}^{([\frac{j+1}{2}],
          j+1-[\frac{j+1}{2}])}\zeta^{-(4p-2j)}\Big)\nonumber\\
     &&
      \times ~ \zeta ~ \Big(\frac{1}{\zeta}
     \sum_{j=0}^{\infty}\kappa_j\zeta^{j}\Big)\nonumber\\
     &=&m^{-1}\sum_{j=-1}^{\infty}\Big(\sum_{k=-2}^{2j}\vartheta_k
     V_{21}^{([\frac{2j-k+4}{4}],\frac{2j-k+4}{4}-[\frac{2j-k+4}{4}])}\Big)
     \zeta^{-4p+2j}\nonumber\\
     &&
     +\sum_{j=-1}^{2p-1}V_{22}^{([\frac{j+1}{2}],
     j+1-[\frac{j+1}{2}])}\zeta^{-4p+2j}
         \nonumber \\
  \end{eqnarray*}
  \begin{eqnarray}
   &&
     +\sum_{j=0}^{\infty}
     \left(\sum_{\ell=0}^{2j}\kappa_{\ell}V_{23}^{([\frac{2j-\ell+2}{4}],
     \frac{2j-\ell+2}{4}-[\frac{2j-\ell+2}{4}])}
     \right)\zeta^{-4p+2j}
     \nonumber\\
     &=&\frac{2}{3}\zeta^{-(4p+2)}+\sum_{j=0}^{2p-1}\chi_j\zeta^{-4p+2j}+\chi_{2p}
     +\sum_{j=2p+1}^{\infty}\chi_j\zeta^{-4p+2j},\label{aoteman}
  \end{eqnarray}
where
     \begin{eqnarray*}
      \chi_j &=& m^{-1}\sum_{k=-2}^{2j}\vartheta_k
      V_{21}^{([\frac{2j-k+4}{4}],\frac{2j-k+4}{4}-[\frac{2j-k+4}{4}])}
       +V_{22}^{([\frac{j+1}{2}],j+1-[\frac{j+1}{2}])}\\
       &&
       + \sum_{\ell=0}^{2j}\kappa_{\ell}
       V_{23}^{([\frac{2j-\ell+2}{4}],\frac{2j-\ell+2}{4}-[\frac{2j-\ell+2}{4}])},
       \quad\text{for} \quad 0\leq j\leq 2p-1,\quad j\in\mathbb{N}_0,\\
       \chi_j &=&m^{-1} \sum_{k=-2}^{2j}\vartheta_k
       V_{21}^{([\frac{2j-k+4}{4}],\frac{2j-k+4}{4}-[\frac{2j-k+4}{4}])}
       + \sum_{\ell=0}^{2j}\kappa_{\ell}
       V_{23}^{([\frac{2j-\ell+2}{4}],\frac{2j-\ell+2}{4}-[\frac{2j-\ell+2}{4}])},
       \\
       &&~~~~~~~~~~~~~~~~~~~~~~~~~~~~~~~~~~~~~~~~~~~~~~~
       \text{for}\quad j\geq 2p,\quad j\in\mathbb{N}_0.
     \end{eqnarray*}
Then inserting (\ref{aoteman}) into (\ref{5.14}) and comparing the
coefficients of the same powers of $\zeta^{\ell} ~(\ell<0)$ yields
\begin{equation*}
  \chi_{j,x}=0, \quad \textrm{for~~
  $ 0\leq j\leq 2p-1, ~~j\in\mathbb{N}_0$.}
\end{equation*}
Hence, we conclude that
\begin{eqnarray*}
  && \chi_0=\gamma_0(t_p),\\
  && \chi_1=\gamma_1(t_p),\\
  && \ldots\\
  && \chi_{2p-1}=\gamma_{2p-1}(t_p),
\end{eqnarray*}
where $\gamma_j(t_p)~(j=1,2,\ldots)$ are integration constants. Next
we note that the coefficients $\kappa_j ~(j=0,1,\ldots)$ of the power
series for $\phi(P,x,t_p)$ in the coordinate $\zeta$ near
$P_{\infty_1}$ are the ratios of two functions closely related to $u$.
Meanwhile, the coefficients of the homogeneous polynomials
$\widetilde{V}_{ij} ~(i,j=1,2,3)$ are differential polynomials in
$u$. From these considerations it follows that
$\gamma_j=\tilde{\alpha}_j\in\mathbb{C}$. Hence, we obtain
(\ref{6.3}). Finally, (\ref{6.3c0})
follows from (\ref{4.2a}) and (\ref{6.2}). \quad $\square$ \\

Let $\omega_{P_{\infty_1},j}^{(2)}$, $j=4l+2,$ $l \in \mathbb{N}_0$,
be the Abel differentials of the second kind normalized by vanishing
of all their $a$-periods
$$\int_{a_k}\omega_{P_{\infty_1},j}^{(2)}=0,\quad k=1,\ldots,r-2$$
and holomorphic on $\mathcal{K}_{r-2} \setminus \{P_{\infty_1}\}$, with
a pole of order $j$ at $P_{\infty_1}$,
      \begin{equation}\label{6.15}
       \omega_{P_{\infty_1},j}^{(2)}(P)
          \underset{\zeta \rightarrow 0}{=}
           (\zeta^{-j}+O(1) )d \zeta,
             \quad \textrm{as $P \rightarrow P_{\infty_1}$}.
      \end{equation}
Furthermore, define the normalized differential of the second kind
by
      \begin{equation}\label{6.16}
         \widetilde{\Omega}_{P_{\infty_1},s+1}^{(2)}=
           \frac{2}{3}s\omega_{P_{\infty_1},s+1}^{(2)}
           +\sum_{j=0}^{\frac{s-4}{2}}(s-2j-2)\tilde{\alpha}_j
           \omega_{P_{\infty_1},s-2j-3}^{(2)}
      \end{equation}
and
      \begin{equation}\label{6.16b0}
        \widetilde{\Omega}_{P_{0},3}^{(2)}=2\omega_{P_0,3}^{(2)},
      \end{equation}
where $s=4p+2$, $p \in \mathbb{N}_0$. Thus, one infers
$$\int_{a_k}\widetilde{\Omega}_{P_{\infty_1},s+1}^{(2)}=0, \qquad
 \int_{a_k} \widetilde{\Omega}_{P_{0},3}^{(2)} =0,
 ~~~~~~ k=1,\ldots,r-2.$$
In addition, we define the vector of $b$-periods of the differential
of the second kind
  $\widetilde{\Omega}_{P_{\infty_1},s+1}^{(2)}$,
     \begin{equation}\label{6.17}
      \underline{\widetilde{U}}_{s+1}^{(2)}=
       (\widetilde{U}_{s+1,1}^{(2)},\ldots,
       \widetilde{U}_{s+1,r-2}^{(2)}),
     \quad
      \widetilde{U}_{s+1,j}^{(2)}=\frac{1}{2 \pi i} \int_{b_j}
       \widetilde{\Omega}_{P_{\infty_1},s+1}^{(2)},
       \quad   j=1,\ldots, r-2
     \end{equation}
     with $s=4p+2$, $p \in \mathbb{N}_0$. Integrating (\ref{6.16}) and
(\ref{6.16b0}) yields
   \begin{eqnarray}\label{6.18}
       \int_{Q_0}^P \widetilde{\Omega}_{P_{\infty_1},s+1}^{(2)}
    & \underset{\zeta \rightarrow 0}{=} &
    \frac{2}{3}s\int_{\zeta_0}^{\zeta}\omega_{P_{\infty_1},s+1}^{(2)}+
       \sum_{j=0}^{\frac{s-4}{2}}(s-2j-2)\tilde{\alpha}_{j} \int_{\zeta_0}^\zeta
            \omega_{P_{\infty_1},s-2j-3}^{(2)}
        \nonumber \\
    & \underset{\zeta \rightarrow 0}{=} &
      \frac{2}{3}s\int_{\zeta_0}^{\zeta}\frac{1}{\zeta^{s+1}}d\zeta+
       \sum_{j=0}^{\frac{s-4}{2}}(s-2j-2)\tilde{\alpha}_{j} \int_{\zeta_0}^\zeta
            \frac{1}{\zeta^{s-2j-3}}d\zeta\nonumber\\
            &&+ O(1)
        \nonumber \\
    & \underset{\zeta \rightarrow 0}{=} & -\frac{2}{3}\zeta^{-s}
        -\sum_{j=0}^{\frac{s-4}{2}} \tilde{\alpha}_{j}  \frac{1}{\zeta^{s-2j-2}}
        +\hat{e}_{s+1}^{(2)}(Q_0) +O(\zeta),
       \nonumber \\
      &&  ~~~~~~~~~~~~~~~~~~~~~~~~~~~
        \textrm{as $P \rightarrow P_{\infty_1}$},
     \end{eqnarray}
    and
    \begin{eqnarray}\label{6.18a0}
     \int_{Q_0}^P \widetilde{\Omega}_{P_{0},3}^{(2)}
     \underset{\zeta \rightarrow 0}{=}
     -\zeta^{-2}+\tilde{e}_3^{(2)}(Q_0)+O(\zeta),
     \quad \textrm{as $P \rightarrow P_{0}$},
    \end{eqnarray}
where $\hat{e}_{s+1}^{(2)}(Q_0), \tilde{e}_3^{(2)}(Q_0)$ are
constants that arise from evaluating all the integrals at their
lowers limits $Q_0$, and summing accordingly. Combining (\ref{6.3}),
(6.4), (\ref{6.18}) and  (\ref{6.18a0}) yields
   \begin{eqnarray}\label{6.19}
         && \int_{t_{0,p}}^{t_p} I_s(P,x,\tau) d\tau
     \underset{\zeta \rightarrow 0}{=}  (t_p-t_{0,p})
      \Bigg(\hat{e}_{s+1}^{(2)}(Q_0)-\int_{Q_0}^P
      \widetilde{\Omega}_{P_{\infty_1},s+1}^{(2)}
        \Bigg) \nonumber\\
      &&~~~~~~~~~~~~~~~~~
      +\int_{t_{0,p}}^{t_p}\chi_{\frac{s-2}{2}}(x,\tau)d\tau+O(\zeta),
      \quad \textrm{as $P \rightarrow P_{\infty_1}$},
      \end{eqnarray}
and
   \begin{eqnarray}\label{6.19a0}
     && \int_{t_{0,0}}^{t_0} I_2(P,x,\tau) d\tau
      \underset{\zeta \rightarrow  0}{=}
       \int_{t_{0,0}}^{t_0}\left(u(x,\tau)m^{\frac{1}{3}}(x,\tau)
       \left(\tilde{e}_3^{(2)}(Q_0)-\int_{Q_0}^{P}
       \widetilde{\Omega}_{P_0,3}^{(2)}\right)\right)d\tau
         \nonumber \\
       &&~~~~~~~~~~~~~~~~~~~~~~~+O(\zeta)~~~~~~~~~~~~~~~
       \textrm{as $P\rightarrow P_{0}$}.
   \end{eqnarray}
Given these preparations, the theta function representation of
$\psi_2(P,x,x_0,t_p,t_{0,p})$ reads as follows.

  \newtheorem{the6.3}[the6.1]{Theorem}
   \begin{the6.3}\label{them6.3}
        Assume that the curve $\mathcal{K}_{r-2}$ is nonsingular.
        Let $P=(\tilde{z},y) \in \mathcal{K}_{r-2} \setminus
        \{P_{\infty_1}, P_0\}$ and let $(x,t_p),(x_0,t_{0,p}) \in \Omega_\mu$,
        where
        $\Omega_\mu \subseteq \mathbb{C}^2$ is open and connected.
        Suppose that $\mathcal{D}_{\underline{\hat{\mu}}(x,t_p)}$, or
        equivalently, $\mathcal{D}_{\underline{\hat{\nu}}(x,t_p)}$ is
        nonspecial for $(x,t_p) \in \Omega_\mu$. Then
        for $s=2$
         \begin{eqnarray}\label{6.20b0}
        \psi_2(P,x,x_0,t_0,t_{0,0})&=&\frac
        {\theta\left(\underline{\tilde{z}}(P,\underline{\hat{\mu}}(x,t_0))\right)
        \theta(\underline{\tilde{z}}(P_{0},\underline{\hat{\mu}}(x_0,t_{0,0})))}
        {\theta(\underline{\tilde{z}}(P_{0},\underline{\hat{\mu}}(x,t_0)))
        \theta(\underline{\tilde{z}}(P,\underline{\hat{\mu}}(x_0,t_{0,0})))}
          \\
     & \times&
        \mathrm{exp} \Bigg( \int_{x_0}^x 2m^{1/3}(x^\prime,t_0) dx^\prime
       \Big( \int_{Q_0}^P
       \omega_{P_0,3}^{(2)}-e_3^{(2)}(Q_0)\Big)\Bigg)
         \nonumber \\
       &+&
       (t_0-t_{0,0})
      \Big(\hat{e}_{3}^{(2)}(Q_0)-\int_{Q_0}^P \widetilde{\Omega}_{P_{\infty_1},3}^{(2)}
      \Big)+
      \int_{t_{0,0}}^{t_0}\chi_{0}(x_0,\tau)d\tau \Bigg)
      \nonumber \\
       &\times&
        \int_{t_{0,0}}^{t_0}\left(u(x_0,\tau)
       m^{1/3}(x_0,\tau)\left(\tilde{e}_3^{(2)}(Q_0)-\int_{Q_0}^{P}
       \widetilde{\Omega}_{P_0,3}^{(2)}\right)\right)d\tau,
       \nonumber
       \end{eqnarray}
and  for $s>2$
       \begin{eqnarray}\label{6.20}
       \psi_2(P,x,x_0,t_p,t_{0,p})&=&\frac
        {\theta\left(\underline{\tilde{z}}(P,\underline{\hat{\mu}}(x,t_p))\right)
        \theta(\underline{\tilde{z}}(P_{0},\underline{\hat{\mu}}(x_0,t_{0,p})))}
        {\theta(\underline{\tilde{z}}(P_{0},\underline{\hat{\mu}}(x,t_p)))
        \theta(\underline{\tilde{z}}(P,\underline{\hat{\mu}}(x_0,t_{0,p})))}
          \\
    & \times &~
        \mathrm{exp} \Bigg( \int_{x_0}^x 2m^{1/3}(x^\prime,t_p) dx^\prime
       \Big( \int_{Q_0}^P
       \omega_{P_0,3}^{(2)}-e_3^{(2)}(Q_0)\Big)\Bigg)
         \nonumber \\
      &+&
        (t_p-t_{0,p})
      \Big(\hat{e}_{s+1}^{(2)}(Q_0)-\int_{Q_0}^P
      \widetilde{\Omega}_{P_{\infty_1},s+1}^{(2)}
      \Big)+
      \int_{t_{0,p}}^{t_p}\chi_{\frac{s-2}{2}}(x_0,\tau)d\tau
      \Bigg).
      \nonumber
     \end{eqnarray}
  \end{the6.3}
  \textbf{Proof.}~~We present only the proof of the time variation here, since
  the proof of the space variation is analogous to the stationary
  case in Theorem \ref{them4.40}. Let $\psi_2(P,x,x_0,t_p,t_{0,p})$ be defined
  as in (\ref{5.31}). For $s>2$,  we denote the right-hand side of (\ref{6.20})
  by $\Psi(P,x,x_0,t_p,t_{0,p})$. While $s=2$, for our convenience,
  we also denote the right-hand side of (\ref{6.20b0}) by
  $\Psi(P,x,x_0,t_p,t_{0,p})$.
  Temporarily assume that
    \begin{equation}\label{6.21}
             \mu_j(x,t_p) \neq \mu_k(x,t_p),
             \quad \textrm{ for $j \neq k$ and $(x,t_p) \in \widetilde{\Omega}_\mu
             \subseteq \Omega_\mu $ },
          \end{equation}
    where $\widetilde{\Omega}_\mu$ is open and connected. In order
    to prove that $\psi_2=\Psi$, by using (\ref{5.20}) and
    (\ref{5.22}), we compute
        \begin{eqnarray}\label{6.22}
    I_s(P,x,t_p)&=&\frac{\widetilde{V}_{21}}{m}(\tilde{z}^2-\tilde{z}\phi_x
          -\phi^2)+\widetilde{V}_{22}+\widetilde{V}_{23}\tilde{z}^{-1}\phi
           \nonumber \\
    &=&
    \widetilde{V}_{21}\frac{\tilde{z}^{-1}y-V_{22}}{V_{21}}+\widetilde{V}_{22}
      +( \widetilde{V}_{23}-\frac{\widetilde{V}_{21}}{V_{21}}V_{23})
      \tilde{z}^{-1}\phi
           \nonumber \\
    &=&
      ( \widetilde{V}_{23}-\frac{\widetilde{V}_{21}}{V_{21}}V_{23})
      \tilde{z}^{-1}\phi
    +\widetilde{V}_{22}-\frac{\widetilde{V}_{21}}{V_{21}}V_{22}
    +\tilde{z}^{-1}y \frac{\widetilde{V}_{21}}{V_{21}}
          \nonumber \\
    &=&
      ( \widetilde{V}_{23}-\frac{\widetilde{V}_{21}}{V_{21}}V_{23})
      \Big(\frac{y^2V_{21}-yA_r+\frac{2}{3}S_rV_{21}+\frac{1}{3}E_{r,x}}{E_r}
      \Big) \nonumber \\
    &&
      +\widetilde{V}_{22}-\frac{\widetilde{V}_{21}}{V_{21}}V_{22}
        +\tilde{z}^{-1}y \frac{\widetilde{V}_{21}}{V_{21}}
            \nonumber \\
    &=&
       \frac{1}{3}\frac{E_{r,t_p}}{E_r}
       +( \widetilde{V}_{23}-\frac{\widetilde{V}_{21}}{V_{21}}V_{23})
      \Big(\frac{y^2V_{21}-yA_r+\frac{2}{3}S_rV_{21}}{E_r}
      \Big)
            \nonumber \\
    &&
       +\tilde{z}^{-1}y \frac{\widetilde{V}_{21}}{V_{21}}
            \nonumber \\
    &=&
     \frac{1}{3}\frac{E_{r,t_p}}{E_r}
       +( \widetilde{V}_{23}-\frac{\widetilde{V}_{21}}{V_{21}}V_{23})
       \Bigg[\frac{2}{3}\frac{V_{21}(3y^2+S_r)}{E_r}-
              \frac{yV_{21}(y+\frac{A_r}{V_{21}})}{E_r} \Bigg]
               \nonumber \\
     &&
       +\tilde{z}^{-1}y \frac{\widetilde{V}_{21}}{V_{21}}.
        \end{eqnarray}
  Hence,
        \begin{eqnarray}\label{6.23}
          I_s(P,x,t_p)&=& -\frac{1}{3}\frac{\mu_{j,t_p}}{\tilde{z}-\mu_j}
          -\frac{2}{3}\frac{\mu_{j,t_p}}{\tilde{z}-\mu_j} +O(1)
           \nonumber \\
          &=&
           -\frac{\mu_{j,t_p}}{\tilde{z}-\mu_j}+O(1),
           \qquad \textrm{as $\tilde{z} \rightarrow \mu_j(x,t_p)$}.
        \end{eqnarray}
  More concisely,
        \begin{equation}\label{6.24}
          I_s(P,x_0,\tau)=\frac{\partial}{\partial \tau}
           \mathrm{ln} (\tilde{z}-\mu_j(x_0,\tau)) +O(1)
           \quad
          \textrm{for $P$ near $\hat{\mu}_j(x_0,t_p)$}.
        \end{equation}
  Therefore
      \begin{eqnarray}\label{6.25}
        && \mathrm{exp} \left( \int_{t_{0,p}}^{t_p} d \tau
          \left(\frac{\partial}{\partial \tau}\mathrm{ln}(\tilde{z}-\mu_j(x_0,\tau))
           +O(1)\right)\right) \nonumber \\
        && \qquad
          = \frac{\tilde{z}-\mu_j(x_0,t_p)}{\tilde{z}-\mu_j(x_0,t_{0,p})}O(1)
          \nonumber \\
          && \qquad
            =
            \begin{cases}
            (\tilde{z}-\mu_j(x_0,t_p))O(1) \quad \textrm{for $P$ near
                     $\hat{\mu}_j(x_0,t_p) \neq \hat{\mu}_j(x_0,t_{0,p})$},\\
            O(1) \quad \textrm{for $P$ near
                     $\hat{\mu}_j(x_0,t_p)=\hat{\mu}_j(x_0,t_{0,p})$},\\
            (\tilde{z}-\mu_j(x_0,t_{0,p}))^{-1}O(1) \quad \textrm{for $P$ near
                     $\hat{\mu}_j(x_0,t_{0,p}) \neq \hat{\mu}_j(x_0,t_p)$},
            \end{cases}
       \end{eqnarray}
  where $O(1) \neq 0$ in (\ref{6.25}). Consequently, all zeros and
  poles of $\psi_2$ and $\Psi$ on $\mathcal{K}_{r-2} \setminus
  \{P_{\infty_1}, P_0\}$ are simple and coincident. It remains to identify
  the essential singularity of $\psi_2$ and $\Psi$ at $P_{\infty_1}$ and $P_0$
  with respect to the time variation. By (\ref{6.19}) and (\ref{6.19a0}), we see that
  the singularities in the exponential terms of $\psi_2$ and $\Psi$
  with respect to the time variation coincide. The uniqueness result
  for Baker-Akhiezer functions (\cite{12,14,14.0,15.0}) completes the proof that
  $\psi_2=\Psi$ on $\widetilde{\Omega}_\mu$.
   The extension of this result from $(x,t_p) \in
   \widetilde{\Omega}_\mu$ to $(x,t_p) \in   \Omega_\mu$ then simply
   follows from the continuity of $\underline{\alpha}_{Q_0}$ and the
   hypothesis of $\mathcal{D}_{\underline{\hat{\mu}}(x,t_p)}$ being
   nonspecial for $(x,t_p) \in \Omega_\mu$. \quad $\square$ \\

  \newtheorem{rem6.33}[the6.1]{Remark}
  \begin{rem6.33}
   We provided two explicit representations for the Baker-Akhiezer
   function $\psi_2$ in terms of the Riemann theta function,
   corresponding to the case $s=2$ and $s>2$, respectively.
   By $(\ref{6.3c0})$, $I_2=O(\zeta^{-2})$, $P_0$ is a
   essential singularity of $\psi_2$ for $s=2$. However,
   for $s>2$, $I_s=O(\zeta^2)$ near $P_0$, and there are no singularities
   in this case.
   Thus, we investigated these two situations respectively in Theorem
   $\ref{them6.2}$ and Theorem $\ref{them6.3}$. What we want to emphasize is that
   these results will not take any trouble for us to obtain the
   solution $u(x,t_p)$. We can deal with the two
   expressions $(\ref{6.20b0})$ and $(\ref{6.20})$ uniformly.
   The more details will be given in Theorem $\ref{them6.6}$.
  \end{rem6.33}

   The straightening out of the DP flows by the Abel map is
   contained in our next result.

   \newtheorem{the6.4}[the6.1]{Theorem}
    \begin{the6.4}
      Assume that the curve $\mathcal{K}_{r-2}$ is nonsingular, and
      let $(x,t_p)$, $(x_0,t_{0,p}) \in \mathbb{C}^2$. Then for $s>2,$
       \begin{eqnarray}\label{6.26}
       \underline{\alpha}_{Q_0}(\mathcal{D}_{\underline{\hat{\mu}}(x,t_p)})
      &=&
\underline{\alpha}_{Q_0}(\mathcal{D}_{\underline{\hat{\mu}}(x_0,t_{0,p})})
      - \left( \int_{x_0}^x2m^{\frac{1}{3}}(x^\prime,t_{p})dx^\prime\right)
      \hat{\underline{U}}_{3}^{(2)}
       \nonumber \\
      &&
       +~ \underline{\widetilde{U}}_{s+1}^{(2)}(t_p-t_{0,p}),
          \end{eqnarray}
          \begin{eqnarray}\label{6.27}
\underline{\alpha}_{Q_0}(\mathcal{D}_{\underline{\hat{\nu}}(x,t_p)})
      &=&
      \underline{\alpha}_{Q_0}(\mathcal{D}_{\underline{\hat{\nu}}(x_0,t_{0,p})})
       -\left( \int_{x_0}^x2m^{\frac{1}{3}}(x^\prime,t_{p})dx^\prime\right)
       \hat{\underline{U}}_{3}^{(2)}
          \nonumber \\
      &&
       +~ \underline{\widetilde{U}}_{s+1}^{(2)}(t_p-t_{0,p}),
          \end{eqnarray}
    and for $s=2,$
     \begin{eqnarray}\label{6.26d001}
       \underline{\alpha}_{Q_0}(\mathcal{D}_{\underline{\hat{\mu}}(x,t_0)})
        &=&
       \underline{\alpha}_{Q_0}(\mathcal{D}_{\underline{\hat{\mu}}(x_0,t_{0,0})})
      - \left( \int_{x_0}^x2m^{\frac{1}{3}}(x^\prime,t_{0})dx^\prime\right)
       \hat{\underline{U}}_{3}^{(2)}
       \\
       &+&
       \underline{\widetilde{U}}_{s+1}^{(2)}(t_0-t_{0,0})
       +\left(\int_{t_{0,0}}^{t_0}2u(x_0,\tau)m^{\frac{1}{3}}(x_0,\tau)d\tau\right)\hat{\underline{U}}_3^{(2)},\nonumber
          \end{eqnarray}
     \begin{eqnarray}\label{6.26d002}
       \underline{\alpha}_{Q_0}(\mathcal{D}_{\underline{\hat{\nu}}(x,t_0)})
        &=&
       \underline{\alpha}_{Q_0}(\mathcal{D}_{\underline{\hat{\nu}}(x_0,t_{0,0})})
      - \left( \int_{x_0}^x2m^{\frac{1}{3}}(x^\prime,t_{0})dx^\prime\right)
       \hat{\underline{U}}_{3}^{(2)}
       \\
       &+&
       \underline{\widetilde{U}}_{s+1}^{(2)}(t_0-t_{0,0})
       +\left(\int_{t_{0,0}}^{t_0}2u(x_0,\tau)m^{\frac{1}{3}}(x_0,\tau)d\tau\right)\hat{\underline{U}}_3^{(2)}.\nonumber
          \end{eqnarray}

    \end{the6.4}
  \textbf{Proof.}~~As in the context of Theorem \ref{them4.70}, it suffices to prove
  (\ref{6.26}). Temporarily assume that
   $\mathcal{D}_{\underline{\hat{\mu}}(x,t_p)}$ is nonspecial for
   $(x,t_p) \in \Omega_\mu \subseteq \mathbb{C}^2$, where
   $\Omega_\mu$ is open and connected. We introduce the meromorphic
   differential
       \begin{equation}\label{6.28}
         \Omega(x,x_0,t_p,t_{0,p})=\frac{\partial}{\partial \tilde{z}}
         \mathrm{ln} (\psi_2(\cdot,x,x_0,t_p,t_{0,p})) d\tilde{z}.
       \end{equation}
  From the representation (\ref{6.20}), one infers
       \begin{eqnarray}\label{6.29}
         \Omega(x,x_0,t_p,t_{0,p})&=&\left( \int_{x_0}^x2m^{\frac{1}{3}}
         (x^\prime,t_{p})dx^\prime\right)\omega_{P_0,3}^{(2)}
         -(t_p-t_{0,p})\widetilde{\Omega}_{P_{\infty_1},s+1}^{(2)}
          \nonumber \\
         &&
         -\sum_{j=1}^{r-5} \omega^{(3)}_{\hat{\mu}_j(x_0,t_{0,p}),\hat{\mu}_j(x,t_p)}
         + \hat{\omega},
       \end{eqnarray}
  where $\hat{\omega}$ denotes a holomorphic differential on
  $\mathcal{K}_{r-2}$, that is, $\hat{\omega}=\sum_{j=1}^{r-2}e_j
  \omega_j$ for some $e_j \in \mathbb{C}$ and $\omega_j$ $(j=1,\ldots,r-2)$
  denote the normalized holomorphic differentials
  (see (\ref{4.25})). Since
  $\psi_2(\cdot,x,x_0,t_p,t_{0,p})$ is single-valued on
  $\mathcal{K}_{r-2}$, all $a$- and $b$-periods of $\Omega$ are
  integer multiples of $2 \pi i$ and hence
      \begin{equation}\label{6.30}
        2 \pi i m_k = \int_{a_k} \Omega(x,x_0,t_p,t_{0,p})
        = \int_{a_k} \hat{\omega}=e_k, \quad k=1,\ldots,r-2,
      \end{equation}
  for some $m_k \in \mathbb{Z}$. Similarly, for some $n_k \in
  \mathbb{Z}$,
      \begin{eqnarray}\label{6.31}
        2 \pi i n_k &=&\int_{b_k} \Omega(x,x_0,t_p,t_{0,p})
         \nonumber \\
        &=&
         \left( \int_{x_0}^x2m^{\frac{1}{3}}(x^\prime,t_{p})dx^\prime\right)
         \int_{b_k}\omega_{P_0,3}^{(2)}
         -(t_p-t_{0,p})\int_{b_k}\widetilde{\Omega}_{P_{\infty_1},s+1}^{(2)}
          \nonumber \\
           &&
         -\sum_{j=1}^{r-5} \int_{b_k}
        \omega^{(3)}_{\hat{\mu}_j(x_0,t_{0,p}),\hat{\mu}_j(x,t_p)}
        +2 \pi i \sum_{j=1}^{r-2} m_j \int_{b_k} \omega_j
          \nonumber \\
        &=&
  2\pi i\left( \int_{x_0}^x2m^{\frac{1}{3}}(x^\prime,t_{p})dx^\prime\right)\hat{U}_{3,k}^{(2)}
            - 2\pi i(t_p-t_{0,p})
       \widetilde{U}_{s+1,k}^{(2)}
          \nonumber \\
       &&
        -2 \pi i \sum_{j=1}^{r-5} \int_{\hat{\mu}_j(x,t_p)}^{\hat{\mu}_j(x_0,t_{0,p})}
         \omega_k
         +2 \pi i \sum_{j=1}^{r-2} m_j \int_{b_k} \omega_j
          \nonumber \\
       & =& 2\pi i\left( \int_{x_0}^x2m^{\frac{1}{3}}(x^\prime,t_{p})
       dx^\prime\right)\hat{U}_{3,k}^{(2)}
            -  2\pi i(t_p-t_{0,p})
       \widetilde{U}_{s+1,k}^{(2)}\nonumber\\
       && +2 \pi i \alpha_{Q_0,k}(\mathcal{D}_{\underline{\hat{\mu}}(x,t_p)})-
       2 \pi i \alpha_{Q_0,k}(\mathcal{D}_{\underline{\hat{\mu}}(x_0,t_{0,p})})
       +2 \pi i \sum_{j=1}^{r-2}m_j \Gamma_{j,k},\nonumber\\
      \end{eqnarray}
      where we have used the formula
      \begin{equation}\label{6.32}
        \int_{b_k} \omega_{Q_1,Q_2}^{(3)}= 2 \pi i \int_{Q_2}^{Q_1}
         \omega_k,  \qquad k=1,\ldots,r-2.
      \end{equation}
      By symmetry of $\Gamma$ this is equivalent to
       \begin{eqnarray}\label{6.33}
       \underline{\alpha}_{Q_0}(\mathcal{D}_{\underline{\hat{\mu}}(x,t_p)})
      &=&
      \underline{\alpha}_{Q_0}(\mathcal{D}_{\underline{\hat{\mu}}(x_0,t_{0,p})})
      - \left( \int_{x_0}^x2m^{\frac{1}{3}}(x^\prime,t_{p})dx^\prime\right)
      \hat{\underline{U}}_{3}^{(2)}
       \nonumber \\
      &&
       +~ \underline{\widetilde{U}}_{s+1}^{(2)}(t_p-t_{0,p}),
          \end{eqnarray}
for $(x,t_p) \in \Omega_\mu$, which leads to (\ref{6.26}). Since
$\mathcal{D}_{P_0\underline{\hat{\nu}}}$ and
$\mathcal{D}_{P_{\infty_1}\underline{\hat{\mu}}}$ are linearly
equivalent, that is
  \begin{equation*}
    \underline{A}_{Q_0}(P_0)
    +\underline{\alpha}_{Q_0}(\mathcal{D}_{\underline{\hat{\nu}}(x,t_p)})
    =\underline{A}_{Q_0}(P_{\infty_1})
    +\underline{\alpha}_{Q_0}(\mathcal{D}_{\underline{\hat{\mu}}(x,t_p)}),
  \end{equation*}
hence (\ref{6.27}) holds. Similarly, one can prove (\ref{6.26d001}) and (\ref{6.26d002}).
Finally, this result extends from $(x,t_p)
\in \Omega_\mu$ to $(x,t_p) \in \mathbb{C}^2$ using the continuity
of $\underline{\alpha}_{Q_0}$ and the fact that positive nonspecial
divisors are dense in the space of divisors. \quad $\square$ \\

Our main result, the theta function representation of time-dependent
algebro-geometric solutions for the DP hierarchy now quickly follows
from the materials prepared above.
  \newtheorem{the6.5}[the6.1]{Theorem}
    \begin{the6.5}\label{them6.6}
        Assume that $u$ satisfies the $p$-th DP equation
        $(\ref{7})$, that is, $DP_p(u)=m_{t_p}-X_p=0$, and
        the curve $\mathcal{K}_{r-2}$ is nonsingular
        Let $(x,t_p) \in \Omega_\mu$,    where
        $\Omega_\mu \subseteq \mathbb{C}^2$ is open and connected.
        Suppose also that $\mathcal{D}_{\underline{\hat{\mu}}(x,t_p)}$, or
        equivalently, $\mathcal{D}_{\underline{\hat{\nu}}(x,t_p)}$ is
        nonspecial for $(x,t_p) \in \Omega_\mu$. Then
   \begin{eqnarray}\label{6.34}
         u(x,t_p)=u(x_0,t_{0,p})\frac{\theta(\underline{\tilde{z}}(P_0,\underline{\hat{\mu}}(x_0,t_{0,p})))
          \theta(\underline{\tilde{z}}
          (P_{\infty_1},\underline{\hat{\mu}}(x,t_p)))}
           {\theta(\underline{\tilde{z}}
          (P_{\infty_1},\underline{\hat{\mu}}(x_0,t_{0,p})))
          \theta(\underline{\tilde{z}}(P_0,\underline{\hat{\mu}}(x,t_p)))}.
       \end{eqnarray}
   \end{the6.5}
\textbf{Proof.}~~In the time-dependent context, we will use the same
strategy as was used in Theorem \ref{them4.80} in the stationary case, treating
$t_p$ as a parameter. A closer look at Theorem \ref{them6.3}, we note that the
two expressions of $\psi_2$ in (\ref{6.20b0}) ($p=0, s=2)$ and
(\ref{6.20}) ($p>0, s>2$) can be written uniformly as the following
form near $P_{\infty_1}$,
 \begin{eqnarray}\label{6.100}
   \psi_2&\underset{\zeta \rightarrow 0}{=}&
   \left(\sigma_0+\sigma_1\zeta+\sigma_2\zeta^2+O(\zeta^3)\right)\nonumber\\
   &\times &
         \mathrm{exp}\left(\left(\int_{x_0}^{x}
         2m^{\frac{1}{3}}(x^\prime,t_p)dx^\prime\right)\left(f_3^{(2)}(Q_0)
          \zeta^2+O(\zeta^4)\right)\right)\nonumber\\
          &\times&\exp\Big(\left(t_p-t_{0,p}\right)
          \Big(\frac{2}{3}\zeta^{-s}+\sum_{j=0}^{\frac{s-4}{2}}
          \tilde{\alpha}_{j}  \frac{1}{\zeta^{s-2j-2}}+\int_{t_{0,p}}^{t_p}
          \chi_{\frac{s-2}{2}}(x_0,t_{p}^{\prime})dt_p^{\prime}\Big)
         \nonumber \\
         &&+O(\zeta^2)\Big),~~~~~ \zeta=\tilde{z}^{-1}, \quad
         \textrm{as $P\rightarrow P_{\infty_1},$}
    \end{eqnarray}
where the terms $\sigma_i=\sigma_i(x,t_p) ~(i=1,2,3)$ come from the
Taylor expansion about $P_{\infty_1}$ of the ratios of the theta
functions in (\ref{6.20b0}) ($p=0$) or (\ref{6.20}) ($p>0$) (see
(\ref{aoteman01})). That is
  \begin{eqnarray*}
   \frac{\theta\left(\underline{\tilde{z}}
          (P,\underline{\hat{\mu}}(x,t_p))\right)}
     {\theta\left(\underline{\tilde{z}}
          (P_{0},\underline{\hat{\mu}}(x,t_p))\right)}
     \underset{\zeta \rightarrow 0}{=}
    \frac{\theta_0}{\theta_1}-
    \frac{\partial_x  \, \theta_0 }{\theta_1} \, \zeta +
      \frac{\frac{1}{2}\partial_x^2 \theta_0 -
        \partial_{\underline{U}_3^{(2)}}\theta_0}{\theta_1}
    \zeta^2 + O(\zeta^3),
     \quad \textrm{as $ P \rightarrow P_{\infty_1}$}, \\
     \end{eqnarray*}
with
    \begin{equation*}
   \theta_0=\theta_0(x,t_p)=
    \theta(\underline{\tilde{z}}
          (P_{\infty_1},\underline{\hat{\mu}}(x,t_p)))=
   \theta\left(\underline{\Xi}_{Q_0}-\underline{A}_{Q_0}(P_{\infty_1})
  +\underline{\alpha}_{Q_0}(\mathcal{D}_{\underline{\hat{\mu}}(x,t_p)})\right),
    \end{equation*}
     \begin{equation*}
  \theta_1=\theta_1(x,t_p)= \theta(\underline{\tilde{z}}
          (P_{0},\underline{\hat{\mu}}(x,t_p)))
  =\theta\left(\underline{\Xi}_{Q_0}-\underline{A}_{Q_0}(P_{0})
  +\underline{\alpha}_{Q_0}(\mathcal{D}_{\underline{\hat{\mu}}(x,t_p)})\right),
    \end{equation*}
    and
    \begin{equation*}
   \partial_{\underline{U}_3^{(2)}}= \sum_{j=1}^{r-2}
    U_{3,j}^{(2)}\frac{\partial}{\partial \tilde{z}_j}.
   \end{equation*}
  Similarly, we have
      \begin{eqnarray*}
      \frac{\theta\left(\underline{\tilde{z}}(P_0,\underline{\hat{\mu}}(x_0,t_{0,p}))\right)}
      {\theta\left(\underline{\tilde{z}}(P,\underline{\hat{\mu}}(x_0,t_{0,p}))\right)}
      &=& \left(\frac{\theta\left(\underline{\tilde{z}}(P,\underline{\hat{\mu}}(x,t_p))\right)}
      {\theta\left(\underline{\tilde{z}}(P_0,\underline{\hat{\mu}}(x,t_p))\right)} \right)^{-1}
       \Big |_{(x,t_p)=(x_0,t_{0,p})}    \\
      &\underset{\zeta\rightarrow 0}{=}&
      \left(\frac{\theta_0}{\theta_1}
      \left(  1- \frac{\partial_x\theta_0}{\theta_0}\zeta + O(\zeta^2)  \right)\right)^{-1}
      \Big |_{(x,t_p)=(x_0,t_{0,p})}\\
      &\underset{\zeta \rightarrow 0}{=}&
      \frac{\theta_1}{\theta_0}
      \left(1+ \partial_x\ln\theta_0\,\zeta+ O(\zeta^2)\right)
      \Big |_{(x,t_p)=(x_0,t_{0,p})}  \\
       &\underset{\zeta \rightarrow 0}{=}&
       \frac{\theta_1(x_0,t_{0,p})}{\theta_0(x_0,t_{0,p})}
       \left(1+ \partial_x\ln\theta_0(x,t_p)
       \Big |_{(x,t_p)=(x_0,t_{0,p})}\,\zeta+ O(\zeta^2)\right) , \\
       &&~~~~~~~~~~~~~~~~~~~~~~~~~~~~~~~~~~~~~~~~~~~~~~~~~~
        \quad \textrm{as $P \rightarrow P_{\infty_1}$}.
   \end{eqnarray*}
   Then we will give the Taylor expansion about $\psi_2$,
   \begin{eqnarray}\label{6.101}
     \psi_2
        &\underset{\zeta \rightarrow 0}{=}&
            \frac
        {\theta(\underline{\tilde{z}}(P,\underline{\hat{\mu}}(x,t_p)))
        \theta(\underline{\tilde{z}}(P_{0},\underline{\hat{\mu}}(x_0,t_{0,p})))}
        {\theta(\underline{\tilde{z}}(P_{0},\underline{\hat{\mu}}(x,t_p)))
        \theta(\underline{\tilde{z}}(P,\underline{\hat{\mu}}(x_0,t_{0,p})))}
          \nonumber \\
        &&  \times ~
          \mathrm{exp}\left(\left(\int_{x_0}^{x}2m^{\frac{1}{3}}(x^\prime,t_p)
          dx^\prime\right)\left(f_3^{(2)}(Q_0)
          \zeta^2+O(\zeta^4)\right)\right)
              \nonumber  \\
        &&\times~
        \exp\Big(\left(t_p-t_{0,p}\right)
          \Big(\frac{2}{3}\zeta^{-s}+\sum_{j=0}^{\frac{s-4}{2}} \tilde{\alpha}_{j}
           \frac{1}{\zeta^{s-2j-2}}+\int_{t_{0,p}}^{t_p}
           \chi_{\frac{s-2}{2}}(x_0,t_{p}^{\prime})dt_p^{\prime}\Big)
         \nonumber \\
         &&+O(\zeta^2)\Big) \nonumber \\
        & \underset{\zeta \rightarrow 0}{=} &
          \Big[  \frac{\theta_1(x_0,t_{0,p})}{\theta_0(x_0,t_{0,p})} \,
          \frac{\theta_0(x,t_p)}{\theta_1(x,t_p)} \, + \,
          \frac{\theta_1(x_0,t_{0,p})}{\theta_0(x_0,t_{0,p})} \,
          \frac{\theta_0(x,t_p)}{\theta_1(x,t_p)} \, \nonumber\\
         &&\times~
         \Big(\partial_x\ln \theta_0(x,t_p)
         \Big |_{(x,t_p)=(x_0,t_{0,p})}- \partial_x\ln \theta_0(x,t_p)\Big) \zeta
         + O(\zeta^2)\Big] \nonumber \\
         &&   \times~
         \mathrm{exp}\left(\left(\int_{x_0}^{x}2m^{\frac{1}{3}}(x^\prime,t_p)
         dx^\prime\right)\left(f_3^{(2)}(Q_0)
          \zeta^2+O(\zeta^4)\right)\right)\nonumber\\
          &&\times~
          \exp\Big(\left(t_p-t_{0,p}\right)
          \Big(\frac{2}{3}\zeta^{-s}+\sum_{j=0}^{\frac{s-4}{2}} \tilde{\alpha}_{j}
           \frac{1}{\zeta^{s-2j-2}}+\int_{t_{0,p}}^{t_p}
           \chi_{\frac{s-2}{2}}(x_0,t_{p}^{\prime})dt_p^{\prime}\Big)
         \nonumber \\
         &&+O(\zeta^2)\Big),~~~~~~~~~~~~~~~~~~~~~~~~~~~
         \textrm{as $P\rightarrow P_{\infty_1}.$}
         \end{eqnarray}
         Hence, comparing the same powers of $\zeta$ in (\ref{6.100}) and
         (\ref{6.101}) gives
         \begin{eqnarray}
           \sigma_0(x,t_p) &=& \frac{\theta_1(x_0,t_{0,p})}
           {\theta_0(x_0,t_{0,p})} \frac{\theta_0(x,t_p)}{\theta_1(x,t_p)}, \label{6.102}   \\
           \sigma_1(x,t_p) &=&  \Big(\partial_x\ln \theta_0(x,t_p)
           \Big |_{(x,t_p)=(x_0,t_{0,p})}- \partial_x\ln \theta_0(x,t_p)\Big)\nonumber\\
           &&\times\frac{\theta_1(x_0,t_{0,p})}{\theta_0(x_0,t_{0,p})} \,
           \frac{\theta_0(x,t_p)}{\theta_1(x,t_p)}.
         \end{eqnarray}
         If we set
         \begin{eqnarray*}
          \psi_2 \underset{\zeta \rightarrow 0}{=}\left( \sigma_0(x,t_p)
          +\sigma_1(x,t_p)\zeta+\sigma_2(x,t_p)\zeta^2+ O(\zeta^2) \right)
          \exp\left(\Delta\right)\exp\left(\tilde{\Delta}\right), \\
              \textrm{as $P\rightarrow P_{\infty_1},$}
          \end{eqnarray*}
          with
          \begin{eqnarray*}
           \exp\left(\Delta\right) = \exp\left(\left(\int_{x_0}^{x}2m^{\frac{1}{3}}
           (x^\prime,t_p)dx^\prime\right)\left(f_3^{(2)}(Q_0)
           \zeta^2+O(\zeta^4) \right)\right)
         \end{eqnarray*}
         and
         \begin{eqnarray*}
            \exp(\tilde{\Delta}) =\exp\left(\left(t_p-t_{0,p}\right)
            \Big(\frac{2}{3}\zeta^{-s}+\sum_{j=0}^{\frac{s-4}{2}} \tilde{\alpha}_{j}
             \frac{1}{\zeta^{s-2j-2}}+\int_{t_{0,p}}^{t_p}
             \chi_{\frac{s-2}{2}}(x_0,t_{p}^{\prime})dt_p^{\prime}\Big)+O(\zeta^2)\right),
          \end{eqnarray*}
          then we can show as ($P\rightarrow P_{\infty_1}$)
          \begin{eqnarray}\label{6.50}
            \psi_{2,x}&\underset{\zeta \rightarrow 0}{=}&
             \left(\sigma_{0,x}+\sigma_{1,x}\zeta+ O(\zeta^2)\right)
             \exp\left(\tilde{\Delta}\right),\nonumber\\
            \psi_{2,xx}&\underset{\zeta \rightarrow 0}{=}&
            \left(\sigma_{0,xx}+\sigma_{1,xx}\zeta+ O(\zeta^2)\right)\exp
            \left(\tilde{\Delta}\right),\\
            \psi_{2,xxx}&\underset{\zeta \rightarrow 0}{=}&
            \left(\sigma_{0,xxx}+\sigma_{1,xxx}\zeta+ O(\zeta^2)\right)\exp
            \left(\tilde{\Delta}\right).\nonumber
          \end{eqnarray}
          By eliminating $\psi_1$ and $\psi_3$ in (\ref{1}), we
          arrive at
      \begin{equation}\label{6.51}
         \psi_{2,xxx}=-m\tilde{z}^{-2}+\frac{m_x}{m}\psi_{2,xx}
         -\frac{m_x}{m}\psi_2+\psi_{2,x}.
       \end{equation}
  Substituting (\ref{6.50}) into (\ref{6.51}) and comparing the
  coefficients of $\zeta^0$ yields
  \begin{equation*}
    \sigma_{0,xxx}=\frac{m_x}{m}(\sigma_{0,xx}-\sigma_0)+\sigma_{0,x},
  \end{equation*}
  namely
  \begin{equation*}
  \frac{( \sigma_{0,xx}-\sigma_{0})_x}{\sigma_{0,xx}-\sigma_{0}}=\frac{m_x}{m}
                 =\frac{(u(x,t_p)-u_{xx}(x,t_p))_x}{u(x,t_p)-u_{xx}(x,t_p)},
  \end{equation*}
  which together with (\ref{6.102}) leads to (\ref{6.34}). \quad $\square$

\section*{Acknowledgments}
The authors would like to express their sincerest thanks to both
referees for  useful and valuable
suggestions. The work described in this paper was supported by
grants from the National Science Foundation of China (Project
No.{\it10971031}), and the Shanghai Shuguang Tracking Project
(Project No.{\it08GG01}). Qiao was partially supported by the US
Department of Defense Army Research Office under grant number {\it
W911NF-08-1-0511}, and by the Norman Hackerman Advanced Research
Program under grant number {\it 003599-0001-2009}.


\begin{thebibliography}{99}
\bibitem{1} M.J. Ablowitz, D.J. Kaup, A.C. Newell and H. Segur, The inverse
 scattering transform--Fourier analysis for nonlinear
 problems, Stud.Appl.Math. 53 (1974) 249--315.
\bibitem{2} A. Degasperis and M. Procesi, Asymptotic integrability,
 in A. Degasperis and G. Gaeta (eds), Symmetry and Perturbation Theory,
 World Scientific, pp. 23--37, 1999.
\bibitem{3} A. Degasperis, D.D. Holm and A.N.W. Hone, A new
integrable equation with peakon solutions, Theor.Math.Phys. 133
(2002) 1463--1474.
\bibitem{4} E.D. Belokolos, A.I. Bobenko, V.Z. Enol'skii, A.R. Its,
and V.B. Matveev, Algebro-Geometric Approach to Nonlinear Integrable
Equations, Springer, Berlin, 1994.
\bibitem{5} I.M. Krichever, Algebraic-geometric construction of the
Zaharov-Sabat equations and their periodic solutions, Dokl.Akad.Nauk
SSSR 227 (1976) 394--397.
\bibitem{6} I.M. Krichever, Integration of nonlinear equations by
the methods of algebraic geometry, Funct.Anal.Appl. 11 (1977) 12--26

\bibitem{8} S.P. Novikov, S.V. Manakov, L.P. Pitaevskii, V.E.
Zakharov, Theory of Solitons, the Inverse Scattering Methods,
Concultants Bureau, New York, 1984.
\bibitem{9} B.A. Dubrovin, Completely integrable Hamiltonian systems
associated with matrix operators and Abelian varieties,
Funct.Anal.Appl. 11 (1977) 265--277.
\bibitem{10} B.A. Dubrovin, Theta functions and nonlinear equations,
Russian Math.Surveys. 36 (1981) 11--80.
\bibitem{11} B.A. Dubrovin, Matrix finite-gap operators,
Revs.Sci.Tech. 23 (1983) 33--78.
\bibitem{11a}D. Mumford, Tata Lectures on Theta II, Birkh\textrm{$\ddot{a}$}user, Boston, 1984.
\bibitem{11b} H. M. Farkas, I. Kra, Riemann Surfaces. Second ed. Springer, New York,
(1992).
\bibitem{12} F. Gesztesy and R. Ratneseelan, An alternative approach
to algebro-geometric solutions of the AKNS hierarchy, Rev.Math.Phys.
10 (1998) 345--391.
\bibitem{13}  F. Gesztesy and H. Holden, Algebro-geometric solutions
of the Camassa-Holm hierarchy, Rev.Mat.Iberoam. 19 (2003) 73--142.
\bibitem{14}F. Gesztesy, H. Holden,
 Soliton Equations and Their Algebro-Geometric Solutions,
 Volume I: (1+1)-Dimensional Continuous Models,
 Cambridge Studies in Advanced Mathematics, Vol. 79, Cambridge University Press, (2003).

\bibitem{14.0} F. Gesztesy, H. Holden,  J. Michor, and G. Teschl,
Soliton Equations and Their Algebro-Geometric Solutions, Volume II:
(1+1)-Dimensional Discrete Models, Cambridge Studies in Advanced
Mathematics, Vol. 114, Cambridge University Press, Cambridge, (2008).
\bibitem{15} J.S. Geronimo, F. Gesztesy and H. Holden, Algebro-geometric solutions
of the Baxter-Szeg$\mathrm{\ddot{o}}$ difference equation,
Comm.Math.Phys. 258 (2005) 149--177.


\bibitem{15.0}W. Bulla, F. Gesztesy, H. Holden, G. Teschl,
Algebro-Geometric Quasi-Periodic Finite-Gap Solutions of the
Toda and Kac-van Moerbeke Hierarchies, Amer. Math. Soc, 135 (1998) 1-79.


\bibitem{17} A. Constantin, R.I. Ivanov and J. Lenells,
Inverse scattering transform for the Degasperis--Procesi euqation,
Nonlinearity, 23 (2010) 2559--2575.

\bibitem{18} A. Constantin and D. Lannes, The hydrodynamical relevance
of the Camassa-Holm and Degasperis-Procesi equations,
Arch.Ration.Mech.Anal, 192 (2009) 165--186.

\bibitem{19} E.G. Fan, The positive and negative Camassa-Holm-$\gamma$ hierarchies,
zero curvature representations, bi-Hamiltonian structures, and
algebro-geometric solutions, J.Math.Phys, 50 (2009) 013525 1--23.

\bibitem{20} Y.C. Hon and E.G. Fan, Uniformly constructing finite-band solutions
for a family of derivative nonlinear Schr$\mathrm{\ddot{o}}$dinger
equations, Chaos Solitons and Fractals, 24 (2005) 1087--1096.

\bibitem{21} R. Dickson, F. Gesztesy and K. Unterkofler, A new
approach to the Boussinesq hierarchy, Math.Nachr, 198 (1999)
51--108.

\bibitem{22} R. Dickson, F. Gesztesy and K. Unterkofler,
Algebro-geometric solutions of the Boussinesq hierarchy,
Rev.Math.Phys, 11 (1999) 823--879.
\bibitem{23} X.G. Geng, L.H. Wu and G.L. He, Algebro-geometric
constructions of the modified Boussinesq flows and quasi-periodic
solutions, Physica D, 240 (2011) 1262--1288.

\bibitem{28} R.I. Ivanov, Water waves and integrability,
Phil.Trans.R.Soc.Lond.A 365 (2007) 2267--2280.

\bibitem{31} Y. Matsuno, The $N$-soliton solution of the
Degasperis--Procesi equation, Inverse Problems, 21 (2005)
2085--2101.

\bibitem{34} V.B. Matveev and M.I. Yavor, Solutions presque
p$\mathrm{\acute{e}}$riodiques et $\mathrm{\grave{a}}$ $N$-solitons
de l'$\mathrm{\acute{e}}$quation hydrodynamique non
lin$\mathrm{\acute{e}}$aire de Kaup, Ann.Inst.
H.Poincar$\mathrm{\acute{e}}$ Sect. A 31 (1979) 25--41.

\bibitem{38} Z.J. Qiao, Integrable Hierarchy, $3 \times 3$
Constrained Systems and Parametric Solutions, Acta.Appl.Math, 83
(2004) 199--220.
\bibitem{39} Z.J. Qiao, The Camassa-Holm Hierarchy, $N$-Dimensional
Integrable Systems, and Algebro-Geometric Solution on a Symplectic
Submanifold, Commun.Math.Phys. 239 (2003) 309--341.


\bibitem{55} R.S. Johnson, Camassa-Holm, Korteweg-de Vries and
related models for water waves, J.Fluid.Mech, 455 (2002) 63--82.
\bibitem{56} Y.C. Ma, M.J. Ablowitz, The periodic cubic
Schr$\mathrm{\ddot{o}}$dinger equation, Stud.Appl.Math, 65 (1981)
11--80.
\bibitem{57} E. Date and S. Tanaka, Periodic multi-soliton solutions
of Korteweg-de Vries equation and Toda lattice,
Progr.Theoret.Phys.Suppl, 59 (1976) 107--125.
\bibitem{58} J. L. Burchnall and T. W. Chaundy, Commutative ordinary differential operators, Proc.Roy.Soc.
London A118 (1928) 557--583.
\end{thebibliography}
\end{document}